\documentclass[journal,onecolumn,11pt]{IEEEtran}
\usepackage{epsfig,rotating,setspace,latexsym,amsmath,epsf,amssymb,amsfonts,bm,epstopdf,cite,authblk,bbm,mathtools,amsthm,algorithm,color,systeme,mathtools,amsmath,derivative,bigints,mathrsfs,tabularx}
\usepackage[short,c2]{optidef}
\usepackage[utf8]{inputenc}
\usepackage[T1]{fontenc}
\usepackage[thinc]{esdiff}
\usepackage[noend]{algpseudocode}
\usepackage[caption=false,font=footnotesize]{subfig}

\theoremstyle{plain}
\newtheorem{theorem}{Theorem}
\newtheorem{lemma}{Lemma}
\newtheorem{corollary}{Corollary}

\newenvironment{Proof}[1]{\medskip\par\noindent{\bf Proof:\,}\,#1}{{\mbox{\,$\blacksquare$}\par}}
\newenvironment{Proof Sketch}[1]{\medskip\par\noindent{\bf Proof Sketch:\,}\,#1}{{\mbox{\,$\blacksquare$}\par}}

\algrenewcommand\algorithmicforall{\textbf{foreach}}
\algrenewcommand\algorithmicindent{.8em}

\allowdisplaybreaks

\begin{document}

\title{Fundamental Tradeoffs in Sampling, Transmission and Information Freshness: Delayed Feedback}

\author{Subhankar Banerjee \qquad Sennur Ulukus\\
\normalsize Department of Electrical and Computer Engineering\\
\normalsize University of Maryland, College Park, MD 20742\\
\normalsize \emph{sbanerje@umd.edu} \qquad \emph{ulukus@umd.edu} }

\maketitle

\begin{abstract}
   We study a status update system where a sampler-transmitter pair aims to deliver fresh information to a monitor over an erasure channel. We characterize the freshness of the monitor with the age of information metric. We consider separate sampling and transmission constraints to model the different resource budgets in such systems. Due to the separate constraints, tradeoffs among sampling, transmission, and freshness arise. Sampling and transmitting a new packet may improve the age of the monitor significantly but will consume both resource budgets, whereas transmitting a stale packet may improve the age moderately but will consume the transmission budget; on the contrary, when the sampler-transmitter pair stays idle, it degrades the freshness of the monitor but saves the budget for later usage. We consider a delayed feedback system, i.e., after a transmission, the sampler-transmitter pair receives the corresponding feedback from the monitor after a delay. The aforementioned tradeoffs together with the delayed feedback impose challenges in implementing an effective joint sampling and transmission strategy. We first propose a minimum mean-squared error (MMSE)-based $(W,R)_d$ policy class, inspired by the $(W,R)$ policy class introduced in \cite{banerjee2026Fundamental}. We then propose a feasible policy by randomizing among four carefully chosen $(W,R)_d$ policies. We show that this policy achieves a universal $4$-approximation guarantee for the monitor's average age under arbitrary feedback delay. Furthermore, for fixed delay, we characterize this average age using countable-state Markov chains. We establish an $O(T^{-\frac{1}{2}})$ bound on the absolute difference between the corresponding finite-horizon and long-term averages, where $T$ is the time horizon. For a given $\epsilon>0$, we provide an explicit time horizon $T_\epsilon$ and show that the finite-horizon average age computed over any horizon of length at least $T_\epsilon$ lies within $\epsilon$ of the long-term average age. Moreover, we use a finite embedded Markov chain to compute the long-term average age and both rates in finitely many steps. Finally, we obtain an optimal mixture over all $(W,R)_d$ policies with $W,R\in\mathbb{N}$.
\end{abstract}

\section{Introduction}
Information freshness is an important aspect of applications such as mission-critical control, automated transportation, AR/VR, and industrial IoT \cite{abbas2023comprehensive,YatesSurvey}. These applications can be abstractly formulated with information sources, a sampler-transmitter pair, and monitors with a feedback channel. The goal of the sampler is to acquire fresh information packets from the sources, whereas the transmitter aims to deliver them to the intended monitors through communication channels. In the literature, for such status-update systems, the metric age of information has been used to characterize the information freshness of the monitors \cite{YatesSurvey}. When the sampler can actively decide when to acquire information from the source and can sample it accordingly, we call such systems generate-at-will \cite{YatesSurvey}. In the status update systems mentioned above, though the goal of both the sampler and the transmitter is to maximize the information freshness at the monitors, they jointly control two complementary actions to achieve it. Consequently, sampling and transmission operations may require different resources. For example, the sampling operation may depend on the computing capacity required to process and encode a measurement, available sensing energy, and the availability of sensing equipment, whereas the transmission of a sampled packet may depend on channel uses, transmission power, or access to a shared communication channel. Thus, separate average sampling- and transmission-rate constraints provide a simple abstraction of the resource limitations associated with acquiring and communicating fresh information, respectively.
   
\begin{figure}[t]
    \centerline{\includegraphics[width = 0.5\columnwidth]{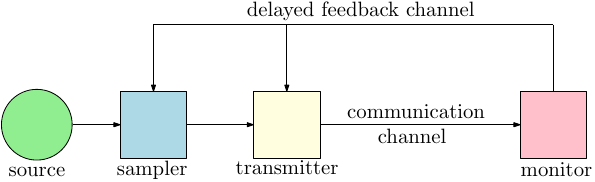}}
    \caption{Different network components for the considered delayed feedback status-update system.}
    \label{fig:1}
\end{figure}

In this work, we consider a time-slotted generate-at-will status-update system shown in Fig.~\ref{fig:1}, with a single source and monitor, where the communication channel is erroneous with probability of successful transmission $0 <q\leq 1$. Moreover, we consider a transmission and a sampling constraint $0<f_{r}\leq 1$ and $0<f_s\leq 1$, respectively. The sampling constraint limits the average number of acquired samples per slot, whereas the transmission constraint restricts the average number of transmission attempts per slot, and both the actions are coupled: if the transmitter has no packet and decides to transmit, the sampler has to sample a packet, whereas if a sampler samples a packet and the transmitter does not transmit it, it results in wastage of sampling resources. Thus, the sampler-transmitter pair has to jointly decide when to stay idle, when to retransmit a stale packet at the transmitter, and when to sample and transmit a fresh packet, either after discarding an old available packet or when the transmitter is empty. Due to this three-action control problem, a tradeoff arises between sampling, transmission, and the age of the monitor. Intuitively, when the age of the monitor is large, the sampler-transmitter pair may choose to sample and transmit a fresh packet, whereas if the age is moderate, the pair may choose to retransmit a previously sampled packet; on the contrary, if the age is low, then the pair may choose to stay idle.

Due to practical limitations, it is possible that when the sampler-transmitter pair transmits a packet, the pair receives the corresponding ACK/NACK from the monitor with a delay. We model this as a delayed feedback channel with arbitrary delay. The three-action control problem described above becomes substantially more challenging under the delayed feedback, as the pair may not even know the current age of the monitor when making its decisions.   

In the literature, the resource-constrained age of information problems for various system models have been studied extensively. The works in \cite{sun2019sampling,sun2019sampling2,tang2023age,banerjee2025equidistant} studied systems with only a sampling budget constraint and for instantaneous feedback. In \cite{banerjee2025equidistant}, we have proposed the \emph{equidistant-sampling policy}, which samples fresh information periodically, where the period is determined by the sampling constraint, and we showed that either an equidistant-sampling policy or a randomized mixture of two such policies is optimal for the average age of information minimization problem. However, this policy may become infeasible with the inclusion of a separate transmission constraint. The works in \cite{sun2019sampling,sun2019sampling2,pan2023optimal,tang2023age} considered a non-preemptive system, i.e., once the transmission of a packet begins, it cannot be preempted before that transmission finishes; thus, the control problem in those works was how long to stay idle before sampling a fresh packet. Thus, these works fundamentally differ from our three-action control problem. 

In \cite{hatami2022demand,tang2020minimizing,chen2023minimizing,wang2019minimizing,ceran2019average}, the works considered resource constraints related only to transmission. In \cite{hatami2022demand,tang2020minimizing,chen2023minimizing}, the authors considered either that the packet is coming from an outside process or that a fresh sample can be generated whenever the transmitter decides to transmit. Thus, essentially, they studied when to stay idle and when to transmit a fresh available packet without taking into account the retransmission of a stored packet at the transmitter, which makes those works different than ours. In \cite{wang2019minimizing,ceran2019average}, the works considered retransmission of a stored packet; however, their motivation to retransmit a stored packet was not due to a separate sampling constraint, but rather due to the unavailability of a fresh packet \cite{wang2019minimizing} or the improvement in decoding reliability through HARQ \cite{ceran2019average}, which makes those works different from ours. Moreover, those works considered instantaneous feedback, which also makes those problems fundamentally different. In \cite{zakeri2022query}, the authors considered both sampling and transmission constraints; however, they studied the query age minimization problem for their system, in which a freshness-related cost appears only when a query occurs, which is fundamentally different than the average age of information, which penalizes the system continuously.  

Age of information minimization with delayed or imperfect feedback has been studied in systems without explicit resource constraints \cite{tsai2020age}, under sampling-rate constraints \cite{pan2023optimal,liyanaarachchi2023early}, and under transmission-side resource constraints \cite{ji2024ageoptimal,zhu2026age}. These works optimize sampling or transmission decisions in settings where the exact age of the monitor may be unavailable, but they do not address the joint sampling-and-transmission control problem considered here. Without additional average sampling- or transmission-rate constraints, \cite{zhao2025optimizing} constructed minimum mean-squared error (MMSE) estimates of receiver ages and incorporated them into a Max-Weight scheduling policy. In our setting, however, the age estimate must guide the three-action control problem while accounting for separate sampling- and transmission-rate constraints.

In \cite{banerjee2026Fundamental}, we considered a similar system model with separate sampling- and transmission-rate constraints and instantaneous feedback. There, we proposed a policy set $\bar{\Pi}$ that contains policies with specific structural properties. We named these policies as $(W,R)$ policies, where $W$ identifies a waiting threshold and $R$ identifies a packet replacement threshold. We showed that randomizing among at most three different $(W,R)$ policies provides an optimal policy for the instantaneous feedback problem. However, due to the feedback delay, the sampler-transmitter pair may not have accurate knowledge of the age of the monitor. Thus, we cannot employ $(W,R)$ policies in this work immediately. Thus, we first construct the MMSE estimate of the age of the monitor based on the available delayed feedback information. Then we propose MMSE-based $(W,R)$ policies, denoted as $(W,R)_d$. We develop a feasible policy, $\pi_{d,4}$, that randomizes among four carefully chosen $(W,R)_{d}$ policies. We then establish a $4$-approximation optimality guarantee for this policy under any feedback-delay process. The approximation guarantee for $\pi_{d,4}$ holds even when feedback delays vary across transmissions, with feedback potentially arriving out of order. To implement $\pi_{d,4}$, the sampler-transmitter pair only uses all the available information at each decision time and does not require knowledge of the distribution of the feedback delays.

For a fixed finite feedback delay, we develop a Markov-chain representation for evaluating the average age of the monitor for an arbitrary $(W,R)_d$ policy. We identify an observable state that determines the MMSE age estimate and consequently the sampling and transmission actions. We show that this state evolves as a countable-state time-homogeneous Markov chain. We express the long-term average age in terms of a stationary distribution of this chain. We prove an $O(T^{-1/2})$ bound on the absolute difference between the finite-horizon and long-term average ages. For a given $\epsilon>0$, we obtain a sufficient horizon $T_\epsilon$ such that for every $T\geq T_\epsilon$, the expected average age over $T$ slots differs from its long-term value by less than $\epsilon$. We also use a finite embedded chain to compute the long-term average age and sampling- and transmission-rates using finitely many recursion steps and finite linear systems. We derive finite cutoffs $K$ and $R_*$ for the waiting and packet replacement thresholds, respectively. Then, we solve a finite linear program (LP) over mixtures of policies with $1\leq W<K$ and $1\leq R\leq R_*$. We show that the resulting mixture is optimal among all feasible mixtures of $(W,R)_d$ policies with $W,R\in\mathbb{N}$.

\section{System Model and Problem Formulation}
We assume that a sampling action occurs at the beginning of a slot. Similarly, transmission of a freshly sampled packet or retransmission of a stored packet occurs at the beginning of a slot, while the transmission takes the entire time slot; i.e., if the transmission is successful, the monitor receives the packet at the end of the same slot. Note that, in a slot, if a fresh packet gets sampled, that packet can be transmitted immediately. For a policy $\pi$, we denote the sampling action at time $t$ by $a_{s}^{\pi}(t) \in\{0,1\}$, where $a_{s}^{\pi}(t)=0$ implies that the sampler does not sample a fresh packet and $a_{s}^{\pi}(t)=1$ implies otherwise. Similarly, the transmission action is denoted by $a^{\pi}_r(t)\in\{0,1\}$, where $a_r^\pi(t)=0$ denotes no transmission and $a_r^{\pi}(t)=1$ denotes otherwise. We denote the channel realization at time $t$ by $Z(t)$, where $Z(t)=1$ implies that a transmission at time $t$ is successful and $Z(t)=0$ implies otherwise. Thus, $\mathbb{P}(Z(t)=1)=q$. We assume that $Z(t)$ is independent and identically distributed (i.i.d.) over time, and is independent of policy $\pi$. 

We denote the age of the monitor at time $t$ by $v_{1}^{\pi}(t)$, which evolves as follows
\begin{align}
    v_{1}^{\pi}(t+1) = \begin{cases}
        1, & a_{s}^{\pi}(t)=1  \ \textrm{and} \  a_{r}^{\pi}(t)Z(t)=1,\\ 
        v_{2}^{\pi}(t)+1, & a_{s}^{\pi}(t)=0  \ \textrm{and} \  a_{r}^{\pi}(t)Z(t)=1,\\ 
        v_{1}^{\pi}(t)+1, &a_{r}^{\pi}(t)Z(t)=0.
    \end{cases}
\end{align}
At the beginning of time $t$, the age of the packet stored at the transmitter is denoted by $v_{2}^{\pi}(t)$, with $v_{2}^{\pi}(t)=\infty$, if the transmitter has no packet. We assume that if $v_2^\pi(t)=\infty$ and $a_s^\pi(t)=0$, then $a_r^\pi(t)=0$.

If a transmission occurs in slot $t$, let $d(t)\in\mathbb{N}_{0}$ denote the random delay of the corresponding ACK/NACK. Thus, the ACK/NACK, or equivalently the channel outcome $Z(t)$, is received by the sampler-transmitter pair at the end of slot $t+d(t)$. We define $d(t)$ for all $t\geq1$, but it affects the system only when a transmission occurs in slot $t$. We assume that the feedback-delay process $\{d(t)\}_{t\geq1}$ is independent of the channel-realization process $\{Z(t)\}_{t\geq1}$. 

Let $\mathcal{I}_{d}^{\pi}(t)$ denote the information available to the sampler-transmitter pair at the beginning of slot $t$ under policy $\pi$. It contains the initial state, all sampling and transmission actions taken before slot $t$, the generation times of the transmitted packets, and all ACK/NACKs received by the beginning of slot $t$, together with the transmission slots to which they correspond. We denote the set of policies adapted to ${\mathcal{I}_{d}^{\pi}(t)}_{t\geq1}$ by $\Pi_d$. For every policy $\pi\in\Pi_d$, we initialize the system with $v_{1}^{\pi}(1)=1$ and $v_{2}^{\pi}(1)=\infty$. A packet remains stored at the transmitter until an ACK for that packet is received, unless it is replaced by a fresher sampled packet. Feedback corresponding to a packet that has already been replaced does not affect the packet currently stored at the transmitter. Without loss of optimality, we consider $a_s^\pi(t)\leq a_r^\pi(t)$, since sampling a packet can be postponed until its first transmission.

For $\pi\in\Pi_d$, we denote the average age, sampling and transmission rates over the time horizon $T$ with $\Delta_d(\pi;T)$, $\nu_{s,d}(\pi;T)$, and $\nu_{r,d}(\pi;T)$, respectively,
\begin{align}
    & \Delta_d(\pi;T) = \frac{1}{T}  \sum_{t=1}^{T} \mathbb{E}_{\pi}[v_{1}^{\pi}(t)],\label{eq:nn3} \\ 
    &\nu_{s,d}(\pi;T) = \frac{1}{T}\sum_{t=1}^{T}\mathbb{E}_{\pi}[a_{s}^{\pi}(t)],\label{eq:nn4} \\
    & \nu_{r,d}(\pi;T) = \frac{1}{T}\sum_{t=1}^{T}\mathbb{E}_{\pi}[a_{r}^{\pi}(t)].\label{eq:nn5}
\end{align}
We denote the corresponding limsups of (\ref{eq:nn3})-(\ref{eq:nn5}) by $\Delta_d(\pi)$, $\nu_{s,d}(\pi)$, and $\nu_{r,d}(\pi)$, respectively. Thus,
\begin{align}
     &\Delta_d(\pi) = \limsup_{T\rightarrow\infty} \Delta_d(\pi;T) , \\ 
     &\nu_{s,d}(\pi) = \limsup_{T\rightarrow\infty} \nu_{s,d}(\pi;T), \\ 
     &\nu_{r,d}(\pi)= \limsup_{T\rightarrow\infty} \nu_{r,d}(\pi;T).
\end{align}

In this work, we aim to minimize the long-term average age of the monitor subject to the long-term average sampling and transmission resource constraints under delayed feedback, i.e., 
\begin{align}\label{eq:df_opt_problem}
    \Delta_{d}^{*}=\inf_{\pi\in\Pi_{d}}&\quad \Delta_d(\pi)\nonumber\\ 
    \textrm{s.t.}& \quad\nu_{s,d}(\pi)\leq f_s,\nonumber\\
    &\quad \nu_{r,d}(\pi)\leq f_r.
\end{align}

\begin{figure}[t]
    \centerline{\includegraphics[width =0.5 \columnwidth]{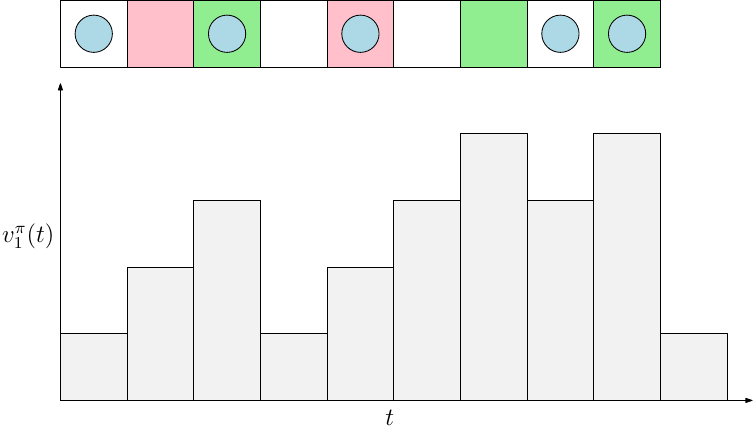}}
    \caption{An illustration of the age evolution. The cyan circles represent sampling of fresh packets. Pink boxes correspond to failed transmission attempts, whereas green boxes indicate successful delivery of the most recently sampled packet.}
    \label{fig:age_evo}
\end{figure} 

\section{Delayed Feedback Policy Design}\label{sec:delayed}
Let $\Gamma^\pi(t)$ denote the generation time of the freshest packet received by the monitor by the beginning of slot $t$. If no packet has been received, we set $\Gamma^\pi(t)=0$. Thus, $v_1^\pi(t)=t-\Gamma^\pi(t)$. Since the outcomes of some previous transmissions may be unavailable at time $t$, $v_{1}^{\pi}(t)$ is not, in general, measurable with respect to $\mathcal{I}_{d}^{\pi}(t)$. We define the conditional estimate of $v_{1}^{\pi}(t)$ as $\hat{v}_{1,d}^{\pi}(t)$,
\begin{align}\label{eq:estimator}
    \hat{v}_{1,d}^{\pi}(t) = \mathbb{E}_{\pi}[v_{1}^{\pi}(t) | \mathcal{I}_{d}^{\pi}(t)].
\end{align}

Next, we obtain an explicit expression for the estimator in (\ref{eq:estimator}). Let $e_{d,0}^{\pi}(t)$ denote the generation time of the freshest packet whose successful delivery is known to the sampler-transmitter pair by the beginning of time $t$. If by the beginning of time $t$, the transmitter does not receive feedback for any successful transmission, we set $e_{d,0}^{\pi}(t)=0$. Let $J_{d}^{\pi}(t)$ denote the number of distinct packets whose generation times are strictly greater than $e_{d,0}^{\pi}(t)$ and for which the outcomes of at least one transmission have not been received by the sampler-transmitter pair by the beginning of slot $t$. We denote the generation time of those packets with $e_{d,j}^{\pi}(t)$, which satisfy
\begin{align}
    e_{d,1}^{\pi}(t) >e_{d,2}^{\pi}(t)>\cdots>e_{d,J_{d}^{\pi}(t)}^{\pi}(t)>e_{d,0}^{\pi}(t).
\end{align}
For $j=1,\ldots,J_{d}^{\pi}(t)$, let $n_{d,j}^{\pi}(t)$ denote the number of transmissions of the packet generated at $e_{d,j}^{\pi}(t)$ for which the feedback has not been received by the sampler-transmitter pair by the beginning of slot $t$. We define
\begin{align}
    p_{d,j}^{\pi}(t) = 1-(1-q)^{n_{d,j}^{\pi}(t)}.
\end{align}
By the i.i.d.~channel model and the independence of the feedback delays and the channel realizations, the arrival or nonarrival of feedback does not change the distribution of the unknown transmission outcomes. Thus, for $1\leq j\leq J_{d}^{\pi}(t)$, we have
\begin{align}
    \mathbb{P}(\Gamma^{\pi}(t)=e_{d,j}^{\pi}(t) | \mathcal{I}_{d}^{\pi}(t))&= p_{d,j}^{\pi}(t) \prod_{k=1}^{j-1} (1-p_{d,k}^{\pi}(t)),\label{eq:110}\\ 
    \mathbb{P}(\Gamma^{\pi}(t)=e_{d,0}^{\pi}(t) | \mathcal{I}_{d}^{\pi}(t))&= \prod_{k=1}^{J_{d}^{\pi}(t)}  (1-p_{d,k}^{\pi}(t)).\label{eq:111}
\end{align}
Thus,
\begin{align}\label{eq:estimated_age}
    \hat{v}_{1,d}^{\pi}(t) = t- e_{d,0}^{\pi}(t)\prod_{k=1}^{J_{d}^{\pi}(t)}  (1-p_{d,k}^{\pi}(t)) -\sum_{j=1}^{J_{d}^{\pi}(t)} e_{d,j}^{\pi}(t) p_{d,j}^{\pi}(t) \prod_{k=1}^{j-1} (1-p_{d,k}^{\pi}(t)).
\end{align}
When $J_{d}^{\pi}(t)=0$, the products in (\ref{eq:estimated_age}) are interpreted as one and we obtain
\begin{align}
    \hat{v}_{1,d}^{\pi}(t) = t-e_{d,0}^{\pi}(t).
\end{align}

From the literature, we know that the conditional expectation minimizes the mean-squared error (MSE) over all estimators measurable with respect to the conditioning information, e.g., see~\cite{poor2013introduction}. Thus, from (\ref{eq:estimator}), $\hat{v}_{1,d}^{\pi}(t)$ is the minimum mean-squared error (MMSE) estimator of $v_{1}^{\pi}(t)$ based on $\mathcal{I}_{d}^{\pi}(t)$. Moreover, by the tower property of conditional expectation,
\begin{align}\label{eq:516}
    \mathbb{E}_{\pi}[\hat{v}_{1,d}^{\pi}(t)] = \mathbb{E}_{\pi}[v_{1}^{\pi}(t)].
\end{align}

\begin{figure}[t]
    \centerline{\includegraphics[width=0.5\columnwidth]{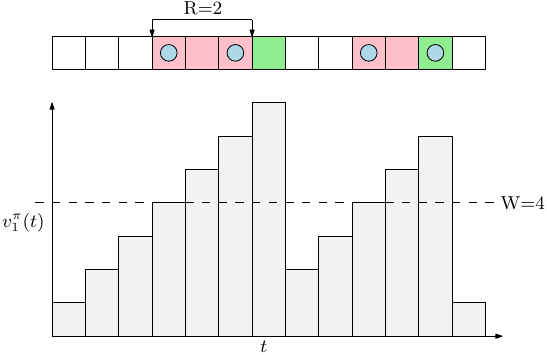}}
    \caption{Sampling and transmission actions under an
    instantaneous-feedback $(W,R)$ policy with $W=4$ and $R=2$
    \cite{banerjee2026Fundamental}. The color coding is the same as in Fig.~\ref{fig:age_evo}.}
    \label{fig:wr_policy}
\end{figure}

We first recall the $(W,R)$ policies considered in \cite{banerjee2026Fundamental}, where $W\in\mathbb{N}$ and $R\in\mathbb{N}\cup\{\infty\}$. Under instantaneous feedback, a $(W,R)$ policy keeps the sampler-transmitter pair idle when $v_1^\pi(t)<W$. Otherwise, a transmission occurs. If the transmitter has no packet or the age of the stored packet is at least $R$, the sampler samples a fresh packet, which is immediately transmitted. Otherwise, the transmitter retransmits the stored packet. A successfully delivered packet is removed from the transmitter. Thus, $W$ determines when to transmit, whereas $R$ determines when to replace a stored packet with a fresh packet. For $R=\infty$, a stored packet is retransmitted until it is successfully delivered, without being replaced. These policies form the policy set $\bar{\Pi}$ in \cite{banerjee2026Fundamental}. Fig.~\ref{fig:wr_policy} illustrates an instantaneous-feedback $(W,R)$ policy.

We next introduce the MMSE-based $(W,R)$ policies, denoted by $(W,R)_d$, where $W,R\in\mathbb{N}$. These policies use the MMSE age estimate instead of the actual age of the monitor in the waiting-threshold decision. At time $t$, the sampler-transmitter pair computes $\hat{v}_{1,d}^{\pi}(t)$. Under a $(W,R)_{d}$ policy $\pi$, the transmitter and the sampler remain idle, if $\hat{v}_{1,d}^{\pi}(t) < W$. If $\hat{v}_{1,d}^{\pi}(t)\geq W$, a transmission occurs. In this case, if the transmitter has no packet or the age of the stored packet is at least $R$, the sampler samples a fresh packet, and the transmitter immediately transmits it. Otherwise, the transmitter retransmits the stored packet. 

\begin{lemma}\label{lemma:df_policy_bounds}
    For every $W,R\in\mathbb{N}$ with $W\geq R$, the policy $\pi=(W,R)_{d}$ satisfies
    \begin{align}
        \nu_{s,d}(\pi)&\leq \frac{1}{R},\label{eq:df_sampling_bound}\\
        \nu_{r,d}(\pi)&\leq \frac{1}{q(W-R+1)},\label{eq:df_transmission_bound}\\
        \Delta_d(\pi)&\leq W+\frac{1}{q}.\label{eq:df_age_bound}
    \end{align}
\end{lemma}

The proof of this lemma is provided in Appendix~\ref{appen:th:df_six_approx}.

For the sampling constraint, we use the same definitions of $R_1$, $R_2$, and $\theta_s$ as in \cite{banerjee2026Fundamental}. We also introduce $Q_1$, $Q_2$, and $\theta_q$ for the transmission constraint. We define
\begin{align}
    R_1 = \left\lfloor\frac{1}{f_s}\right\rfloor,\quad  R_2 = \left\lceil\frac{1}{f_s}\right\rceil,\quad Q_1 = \left\lfloor\frac{1}{qf_r}\right\rfloor, \quad  Q_2 = \left\lceil\frac{1}{qf_r}\right\rceil.
\end{align}
We choose
\begin{align}\label{eq:theta_s}
    \theta_s =
    \begin{cases}
        1, & R_1=R_2,\\
        \frac{f_s-\frac{1}{R_2}}{\frac{1}{R_1}-\frac{1}{R_2}}, & R_1<R_2.
    \end{cases}
\end{align}
Thus, $\frac{\theta_s}{R_1}+\frac{1-\theta_s}{R_2}=f_s$. Similarly, if $Q_1=Q_2$, we set $\theta_q=1$. Otherwise, we choose $\theta_q\in[0,1]$, such that
\begin{align}\label{eq:theta_q}
    \frac{\theta_q}{Q_1}+\frac{1-\theta_q}{Q_2}=qf_r.
\end{align}
For $i,j\in\{1,2\}$, let us consider the following policies
\begin{align}\label{eq:346}
    \pi_{d}^{i,j} = (R_{i}+Q_{j}-1,R_{i})_{d}.
\end{align}
We define the following probabilities
\begin{align}
    \theta_{1,1} = \theta_{s} \theta_{q}, \quad  \theta_{1,2}=\theta_{s}(1-\theta_{q}), \quad \theta_{2,1} = (1-\theta_{s})\theta_{q}, \quad \theta_{2,2} = (1-\theta_{s}) (1-\theta_{q}). 
\end{align}
Now, we define a policy $\pi_{d,4}$ under which, at time $t=1$, the sampler-transmitter pair selects policy $\pi_d^{i,j}$ with probability $\theta_{i,j}$ and follows the selected policy for the entire time horizon.

\begin{theorem}\label{th:df_six_approx}
    For any feedback-delay process ${d(t)}_{t\geq1}$, the policy $\pi_{d,4}$ satisfies both constraints in (\ref{eq:df_opt_problem}). Moreover,
    \begin{align}
       \Delta_d(\pi_{d,4}) \leq 4\Delta_d^{*}- \frac{1}{q} - \frac{(1-q)f_{r}}{q}\leq 4 \Delta_{d}^{*}.
    \end{align}
\end{theorem}

The proof of this theorem is provided in Appendix~\ref{appen:th:df_six_approx}.

\section{Analysis for Fixed Feedback Delay}
The results in Section~\ref{sec:delayed} hold for an arbitrary feedback-delay process. Here, we consider a deterministic fixed finite feedback delay, i.e., $d(t)=d \in\mathbb{N}$, for all $t\geq1$, and evaluate the average age of the monitor. Thus, if a transmission occurs in slot $t$, its corresponding feedback is received at the end of slot $t+d$. Recall that $e_{d,0}^{\pi}(t)$ denotes the generation time of the freshest packet whose successful delivery is known to the sampler-transmitter pair at the beginning of slot $t$. For $t\geq 1$, we define 
\begin{align}\label{eq:364}
    c_d^{\pi}(t)=t-e_{d,0}^{\pi}(t).
\end{align}
Thus, $c_d^{\pi}(t)$ is the age corresponding to the freshest confirmed successful delivery. Next, for $t\geq 1$, we define
\begin{align}
    b_d^{\pi}(t) =
        \begin{cases}
        t-v_{2}^{\pi}(t)-e_{d,0}^{\pi}(t), &v_{2}^{\pi}(t)<\infty,\\
        0,&v_{2}^{\pi}(t)=\infty.
        \end{cases}\label{eq:365}
\end{align}
Thus, whenever a packet is available at the transmitter, $b_d^{\pi}(t)$ is the difference between its generation time and $e_{d,0}^{\pi}(t)$. Consequently, when $b_d^{\pi}(t)>0$, we have, $v_{2}^{\pi}(t)=c_d^{\pi}(t)-b_d^{\pi}(t)$. Under a fixed feedback delay $d$, the only transmission outcomes that can remain unknown at the beginning of slot $t$ are those corresponding to the slots $t-d,t-d+1,\ldots,t-1$. For $k\in\{1,\ldots,d\}$, consider slot $t-d+k-1$. If a transmission occurred in this slot, let $e_r^\pi(t-d+k-1)$ denote the generation time of the transmitted packet, and we define
\begin{align}
    u_{d,k}^{\pi}(t) = \left( e_r^\pi(t-d+k-1)-e_{d,0}^{\pi}(t) \right), \label{eq:366}
\end{align}
for $t\geq 1$. If no transmission occurred in that slot, we set $u_{d,k}^{\pi}(t)=0$. We define
\begin{align}
    \chi_d^{\pi}(t)=\left(c_d^{\pi}(t),b_d^{\pi}(t),u_{d,1}^{\pi}(t),\cdots,u_{d,d}^{\pi}(t)\right),\label{eq:367}
\end{align}
and denote the corresponding countable state space by $\mathcal{X}_d$. Throughout this section, we fix an arbitrary MMSE-based $(W,R)$ policy $\pi=(W,R)_d$, where $W,R\in\mathbb{N}$.

\begin{theorem}\label{th:12}
    For every $t\geq1$, $\chi_d^\pi(t)$ is uniquely determined by $\mathcal I_d^\pi(t)$, and
    \begin{align}\label{eq:368}
        \hat{v}_{1,d}^\pi(t)=\mathbb E_\pi\left[v_1^\pi(t)\,\middle|\,\chi_d^\pi(t)\right].
    \end{align}
    Moreover, $v_2^\pi(t)$ is uniquely determined by $\chi_d^\pi(t)$. Consequently, the action pair $\big(a_s^\pi(t),a_r^\pi(t)\big)$ selected by $\pi$ is uniquely determined by $\chi_d^\pi(t)$. Furthermore, for every $B\subseteq\mathcal X_d$,
    \begin{align}\label{eq:369}
        \mathbb P\left(\chi_d^\pi(t+1)\in B\,\middle|\,\mathcal I_d^\pi(t)\right)=\mathbb P\left(\chi_d^\pi(t+1)\in B\,\middle|\, \chi_d^\pi(t)\right),
    \end{align}
    where the conditional probability on the right-hand side does not depend explicitly on $t$. Consequently, $\{\chi_d^\pi(t)\}_{t\geq1}$ is a time-homogeneous Markov chain on the countable state space $\mathcal X_d$. We denote the one-step transition kernel of this Markov chain by $ P_{\pi}$, i.e., 
    \begin{align}  
        P_{\pi}(x,x') = \mathbb P \left( \chi_d^\pi(t+1)=x' \,\middle|\, \chi_d^\pi(t)=x \right), \qquad x,x'\in\mathcal X_d, \label{eq:370} 
    \end{align}
\end{theorem}

The proof of this theorem is provided in Appendix~\ref{appen:th:12}.

We next explicitly characterize the transition kernel $P_{\pi}(\cdot,\cdot)$. Let $\chi_{d}^{\pi}(t)=x$, where $x=(c,b,u_1,\cdots,u_d)\in\mathcal X_d$. Now, from Theorem~\ref{th:12}, we know that $(a_{s}^{\pi}(t),a_{r}^{\pi}(t))$ is uniquely determined by $x$. Thus, $\bar{b}_{d}^{\pi}(t)$ defined in (\ref{eq:n375}) is also uniquely determined by $x$. Thus, we write the corresponding $\bar{b}_{d}^{\pi}(t)$ as $\bar{b}(x)$, i.e.,
\begin{align}
    \bar b(x) = a_s^\pi(x)c + \left(1-a_s^\pi(x)\right)b.
\end{align}
Now, we determine the possible states in the next slot. First, let us assume that $u_{1}>0$ and the transmission of the corresponding slot, i.e., slot $t-d$, was not successful, i.e., $D(t)=0$. Therefore, from (\ref{eq:n377})-(\ref{eq:379d}), we have 
\begin{align}
    x^{(0)} = \left(c+1,\,\bar b(x),\,u_2,\cdots,u_d,\,a_r^\pi(x)\bar b(x)\right). \label{eq:543}
\end{align}
Now, if $u_{1}>0$, and the transmission at slot $t-d$ was successful then we have $D(t)=u_1$, and the next state is
\begin{align}
    x^{(1)} = \left(c+1-u_1,\,\left(\bar b(x)-u_1\right)^+,\,\left(u_2-u_1\right)^+,\cdots,\left(u_d-u_1\right)^+,\left(a_r^\pi(x)\bar b(x)-u_1\right)^+\right). \label{eq:544}
\end{align}
Note that, when $u_1=0$, from (\ref{eq:543}) and (\ref{eq:544}), we get $x^{(0)}=x^{(1)}$. Thus, we have 
\begin{align}
    P_\pi(x,x') = (1-q)\bm{1}_{\{x'=x^{(0)}\}} + q\bm{1}_{\{x'=x^{(1)}\}}, \quad x,x'\in\mathcal X_d. \label{eq:545}
\end{align}

\begin{lemma} \label{lemma:n7}
    Starting from any state $\chi_{d}^{\pi}(t)$, at the beginning of any slot $t$, at least one transmission occurs during the slots $t,t+1,\ldots,t+d+W-1$. 
\end{lemma}

The proof of this lemma is provided in Appendix~\ref{appen:lemma:n7}.

Next, we define,
\begin{align}
    \bar{\tau}(t) = \inf \left\{m\geq1: e_{d,0}^\pi(t+m) > e_{d,0}^\pi(t+m-1) \right\}.
\end{align}
Note that, $\bar{\tau}(t)$ is a random variable which denotes the required time slots to update the generation time of the freshest packet whose successful delivery is known. For slot $\tau\geq 2$, if we have $e_{d,0}^{\pi}(\tau)>e_{d,0}^{\pi}(\tau-1)$, then from (\ref{eq:364}), equivalently we have
\begin{align}
    c_d^\pi(\tau)<c_d^\pi(\tau-1)+1.
\end{align}

\begin{lemma}\label{lemma:n8}
    We define $B_{d}=W+2d$, then for every $t$ and every integer $n\geq0$, the following relation holds
    \begin{align}
        \mathbb P\left(\bar{\tau}(t)>nB_{d} \mid\mathcal I_d^\pi(t)\right)\leq(1-q)^n\label{eq:388}
    \end{align}
    almost surely. Moreover, for every integer $m\geq1$, there exists a finite constant $C_m$, independent of $t$ and the current information state $\mathcal{I}_{d}^{\pi}(t)$, such that
    \begin{align}
        \mathbb E_\pi\left[\bar{\tau}(t)^m\mid\mathcal I_d^\pi(t)\right]\leq C_m. \label{eq:389}
    \end{align}
\end{lemma}

The proof of this lemma is provided in Appendix~\ref{appen:lemma:n7}.

\begin{lemma}\label{lemma:n9}
    If at slot $\tau\geq 2$, we get $e_{d,0}^\pi({\tau})>e_{d,0}^{\pi}(\tau-1)$, then $\chi_d^\pi(\tau)$ belongs to the set $\mathcal C_{W,R,d}$, where
    \begin{align}
        \mathcal C_{W,R,d} = \Big\{ (c,b,u_1,\cdots,u_d)\in\mathcal X_d: 1\leq c\leq R+d, 0\leq b,u_k\leq R+d, \ k=1,\cdots,d\Big\}.\label{eq:409}
    \end{align}
\end{lemma}

The proof of this lemma is provided in Appendix~\ref{appen:lemma:n7}.

We now consider $\chi_d^\pi(t)$ at the successive slots at which $e_{d,0}^\pi(t)$ increases. Using the definition of $\bar{\tau}(t)$, we recursively define these slots as 
\begin{align}
    \bar{\tau}_0 &=1+\bar{\tau}(1),\label{eq:423}\\
    \bar{\tau}_{n+1}&=\bar{\tau}_n+\bar{\tau}(\bar{\tau}_n),\quad n\geq0.\label{eq:n424}
\end{align}
Thus, $\bar{\tau}_n$ is the slot at which $e_{d,0}^\pi(t)$ increases for the $(n+1)$th time.

\begin{lemma}\label{lemma:n10}
    The process $\{\chi_d^\pi(\bar{\tau}_n)\}_{n\geq0}$ is a finite-state, discrete-time, time-homogeneous Markov chain whose state space is contained in $\mathcal C_{W,R,d}$. We denote its one-step transition kernel by $\tilde P_\pi(\cdot,\cdot)$.
\end{lemma}

The proof of this lemma is provided in Appendix~\ref{appen:lemma:n10}.

\begin{lemma} \label{lemma:n11}
    Every recurrent class of $\{\chi_d^\pi(t)\}_{t\geq1}$ is positive recurrent. Moreover, there are only finitely many recurrent classes, and starting from any initial state $x\in\mathcal X_d$, the chain eventually enters one of them in finite time with probability one.
\end{lemma}

The proof of this lemma is provided in Appendix~\ref{appen:lemma:n10}.

Let $\mathcal R_1^{\pi}, \ldots, \mathcal R_{M_{\pi}}^{\pi}$ denote all recurrent classes of $\{\chi_d^\pi(t)\}_{t\geq1}$. By Lemma~\ref{lemma:n11}, $M_{\pi}<\infty$, and every $\mathcal R_\ell^{\pi}$ is positive recurrent. For every $\ell\in\{1,\cdots,M_{\pi}\}$, we denote the unique stationary probability distribution associated with $\mathcal R_\ell^{\pi}$ by $\psi_\ell^{\pi}$. We consider $\psi_\ell^{\pi}$ as a probability mass function on the entire state space $\mathcal X_d$ by setting $\psi_\ell^{\pi}(y)=0$, for all $y\notin\mathcal R_\ell^{\pi}$. For every initial state $x\in\mathcal X_d$, we define
\begin{align}
    \hat{\psi}_\ell^{\pi}(x) =\mathbb P\left(\exists\, t\in\mathbb N,\ t\geq1: \chi_d^\pi(t)\in\mathcal R_\ell^{\pi} \mid \chi_d^\pi(1)=x \right).
\end{align}
Thus, $\hat{\psi}_\ell^{\pi}(x)$ is the probability that the original chain, when initialized at $x$, eventually enters $\mathcal R_\ell^{\pi}$. Moreover, Lemma~\ref{lemma:n11} implies that, starting from every $x\in\mathcal X_d$, the chain eventually enters one of its recurrent classes with probability one. Therefore,
\begin{align}
    \sum_{\ell=1}^{M_{\pi}} \hat{\psi}_\ell^{\pi}(x) = 1, \quad x\in\mathcal X_d.
\end{align} 
For an initial state $x\in\mathcal X_d$, we define
\begin{align}\label{eq:143}
\bar{\psi}_{\pi}^{x}(y) = \sum_{\ell=1}^{M_{\pi}} \hat{\psi}_\ell^{\pi}(x) \psi_\ell^{\pi}(y), \quad y\in\mathcal X_d.
\end{align}

\begin{lemma}\label{lemma:n12}
    For $x,y\in\mathcal X_d$, we define
    \begin{align}
        p_t^{\pi}(y\mid x) = \mathbb P\left( \chi_d^\pi(t)=y \mid \chi_d^\pi(1)=x \right),
    \end{align}
    and
    \begin{align}\label{eq:nnn145}
        \bar p_T^{\pi}(y\mid x) = \frac{1}{T} \sum_{t=1}^{T} p_t^{\pi}(y\mid x).
    \end{align}
    Then, for every $x,y\in\mathcal X_d$,
    \begin{align}
        \lim_{T\rightarrow\infty}\bar p_T^{\pi}(y\mid x)= \bar{\psi}_{\pi}^{x}(y). \label{eq:467}
    \end{align}
    Moreover, for every $x\in\mathcal X_d$,
    \begin{align}
        \lim_{T\rightarrow\infty}\sum_{y\in\mathcal X_d} \left| \bar p_T^{\pi}(y\mid x) - \bar{\psi}_{\pi}^{x}(y) \right| = 0. \label{eq:n468}
    \end{align}
\end{lemma}

The proof of this lemma is provided in Appendix~\ref{appen:lemma:n10}.

For $k\in\{s,r\}$, let $a_k^\pi(y)$ denote the action $a_k^\pi(t)$ when $\chi_d^\pi(t)=y$.
\begin{theorem} \label{th:13}
    Let $x=(1,0,\ldots,0)$ denote the initial state of $\{\chi_d^\pi(t)\}_{t\geq1}$ corresponding to the initial state of the system, $(1,\infty)$. Then,
    \begin{align}
        &\Delta_d(\pi) = \sum_{y\in\mathcal X_d} \bar{\psi}_{\pi}^{x}(y) \hat v_{1,d}(y), \label{eq:515}\\
        &\nu_{k,d}(\pi) = \sum_{y\in\mathcal X_d} \bar{\psi}_{\pi}^{x}(y) a_{k}^\pi(y), \quad k\in\{s,r\},\label{eq:52}
    \end{align}
    where $\hat v_{1,d}(y)$ denotes the value of $\hat v_{1,d}^\pi(t)$ when $\chi_d^\pi(t)=y$.
\end{theorem}
The proof of this theorem is provided in Appendix~\ref{appen:th:13}. 

For policies $\pi=\pi_{d}^{i,j}$ in (\ref{eq:346}) we denote $\bar{\psi}_{\pi}^{x}(y)$ with $\bar{\psi}_{i,j}^{x}(y)$, where $x=(1,0,\cdots,0)$. Then, the following corollary follows immediately from Theorem~\ref{th:13}.

\begin{corollary}
   The average age of the monitor corresponding to policy $\pi_{d,4}$ is
    \begin{align}\label{eq:nn149}
        \Delta_{d}(\pi_{d,4}) = \sum_{i,j\in\{1,2\}} \theta_{i,j} \sum_{y\in\mathcal X_d} \bar{\psi}_{i,j}^{x}(y) \hat v_{1,d}(y).
    \end{align}
\end{corollary}

\subsection{Finite-Horizon Average Age Analysis}
In this subsection, we analyze the finite-horizon average age of the monitor. In Theorem~\ref{th:15}, we establish a rate of convergence of the finite-horizon average age to the long-term average age. Theorem~\ref{th:16} gives a sufficient horizon $T_\epsilon$ such that the finite-horizon average age over at least this many slots lies within $\epsilon$ of the long-term average age. Theorem~\ref{th:17} gives a finite procedure for computing the kernels needed to obtain this horizon.

To study the convergence of the finite-horizon average age, we first compute the state distribution at each slot. For each constituent policy $\pi=\pi_d^{i,j}$, we initialize
\begin{align}\label{eq:nn332}
    p_1^\pi(y\mid x) = \bm{1}_{\{y=x\}}, \quad y\in\mathcal X_d,
\end{align}
and recursively compute
\begin{align}\label{eq:nn333}
    p_{t+1}^\pi(y\mid x) = \sum_{z\in\mathcal X_d} p_t^\pi(z\mid x)P_\pi(z,y), \quad t\geq1.
\end{align}
The transition kernel $P_\pi$ is explicitly given in (\ref{eq:545}). Since every state has at most two possible successor states, $p_t^\pi(\cdot\mid x)$ has finite support for every finite $t$. Therefore, we can compute $\bar p_T^\pi(\cdot\mid x)$ by a finite recursion for every finite $T$. From the proof of Theorem~\ref{th:13} in Appendix~\ref{appen:th:13}, we see that
\begin{align}\label{eq:nn334}
    \Delta_{d}(\pi;T)=\frac{1}{T} \sum_{t=1}^{T} \mathbb E_{\pi} \left[ v_1^\pi(t) \,\middle|\, \chi_{d}^{\pi}(1)=x \right] = \sum_{y\in\mathcal X_d} \bar p_T^\pi(y\mid x) \hat v_{1,d}(y).
\end{align}
Now, from Theorem~\ref{th:13}, we see that the limit $\lim_{T\rightarrow\infty} \Delta_{d}(\pi;T)$ exists, which is given in (\ref{eq:515}). In the next result, we study the convergence rate of this limit. 

\begin{theorem}\label{th:15}
    For every initial state $x\in\mathcal X_d$, there exist two finite constants $C_\pi(x)$ and $C_{c}(x)$, independent of $T$, such that
    \begin{align}
        &\sum_{y\in\mathcal X_d} \left|\bar p_T^\pi(y\mid x) - \bar\psi_\pi^x(y) \right|  \leq \frac{C_\pi(x)}{T}, \quad T\geq1, \label{eq:600}\\
        &\left|\Delta_d(\pi;T)-\Delta_d(\pi) \right|  \leq 2\sqrt{\frac{2C_c(x)C_\pi(x)}{T}}, \quad T\geq1. \label{eq:601}
    \end{align}
\end{theorem}

The proof of this theorem is provided in Appendix~\ref{appen:th:15}.

Theorem~\ref{th:15} gives the convergence rates for the averaged state distribution and the average age. We next obtain computable bounds on $C_{\pi}(x)$ and $C_{c}(x)$ in Theorem~\ref{th:15}, and consequently a sufficient horizon for a given approximation error. For $j\in\{1,2\}$, we introduce the duration-weighted transition kernels
\begin{align}
    \tilde P_{\pi,j}(z,z')&=\mathbb E_\pi\left[\bar\tau(1)^j\bm1_{\{\chi_d^\pi(1+\bar\tau(1))=z'\}}\,\middle|\,\chi_d^\pi(1)=z\right],\label{eq:155}
\end{align}
where $z\in\mathcal X_d$ and $z'\in\mathcal C_{W,R,d}$. From Lemma~\ref{lemma:n8} we have
\begin{align}\label{eq:nn156}
    \tilde{P}_{\pi,j}(z,z')<\infty.
\end{align}
In the proof of Theorem~\ref{th:15} in Appendix~\ref{appen:th:15}, we have introduced $\bar{\mathcal{R}}_{\pi}$ to denote the union of all recurrent classes of the finite embedded chain $\{\chi_d^\pi(\bar\tau_n)\}_{n\geq0}$ and we have also introduced the reference states 
\begin{align}
    r_\ell\in\mathcal R_\ell^\pi\cap \bar{\mathcal{R}}_{\pi}, \quad \ell\in\{1,2,\cdots,M_{\pi}\}. 
\end{align}
Now, we denote the set of non-reference states of the finite embedded chain with  $\mathcal{C}_{\pi}^{nr}$
\begin{align}
    \mathcal C_\pi^{nr}=\mathcal C_{W,R,d}\setminus\{r_1,\cdots,r_{M_\pi}\}.\label{eq:nn157}
\end{align}
We define the following matrix
\begin{align}
    \bar{P}_\pi&=\bigl(\tilde P_\pi(z,z')\bigr)_{z,z'\in\mathcal C_\pi^{nr}}. \label{eq:nn158}
\end{align}
Later, we show that $(I-\bar{P}_{\pi})^{-1}$ is well-defined. Now, we define two vectors $v_{\pi}^{(1)}$ and $v_{\pi}^{(2)}$
\begin{align}
    v_\pi^{(1)}&=\left(I- \bar{P}_\pi\right)^{-1} \left(\sum_{z'\in\mathcal C_{W,R,d}} \tilde P_{\pi,1}(z,z') \right)_{z\in\mathcal C_\pi^{nr}},\label{eq:nn159}\\
    v_\pi^{(2)} &=\left(I-\bar{P}_\pi\right)^{-1} \Bigg[ \left(\sum_{z'\in\mathcal C_{W,R,d}\cap\mathcal X_d} \tilde P_{\pi,2}(z,z') \right)_{z\in\mathcal C_\pi^{nr}} + 2\bigl(\tilde P_{\pi,1}(z,z')\bigr)_{z,z'\in\mathcal C_\pi^{nr}} v_\pi^{(1)} \Bigg]. \label{eq:nn160}
\end{align}
We denote the $i$th component of $v_{\pi}^{(j)}$ by $v_{\pi}^{(j)}(i)$, and extend
it to the reference states by $v_\pi^{(j)}(r_\ell)=0$. For every $\ell$, let us consider the following iteration 
\begin{align}
    \bar{\zeta}_{\pi,\ell,0}(y) &=\bm1_{\{y=r_\ell\}}, \qquad y\in\mathcal X_d, \label{eq:nn161}\\
    \bar{\zeta}_{\pi,\ell,n+1}(y) &= \bm1_{\{y\ne r_\ell\}} \sum_{z\in\mathcal X_d} \bar\zeta_{\pi,\ell,n}(z)P_\pi(z,y), \qquad n\geq0. \label{eq:nn162}
\end{align}
For every positive integer $\tilde K$, we define
\begin{align}
    \bar m_{\pi,\ell}(\tilde K)&= \sum_{n=0}^{\tilde K-1} \sum_{y\in\mathcal X_d} \bar \zeta_{\pi,\ell,n}(y), \label{eq:nn163}\\
    \bar m_{\pi,\ell}^{(2)} &= \sum_{z\in\mathcal C_{W,R,d}} \tilde P_{\pi,2}(r_\ell,z) + 2\sum_{z\in\mathcal C_\pi^{nr}} \tilde P_{\pi,1}(r_\ell,z)v_\pi^{(1)}(z) + \sum_{z\in\mathcal C_\pi^{nr}} \tilde P_\pi(r_\ell,z)v_\pi^{(2)}(z). \label{eq:nn164}
\end{align}

\begin{theorem}\label{th:16}
     For every initial state $x\in\mathcal X_d$, the constant $C_c(x)$ in
     Theorem~\ref{th:15} can be chosen as
    \begin{align}
        C_c(x) &=\max\{c(x),R+d\}^2+\frac{2\max\{c(x),R+d\}B_d}{q} +\frac{B_d^2(2-q)}{q^2}, \label{eq:nn165}
    \end{align}
    where $c(x)$ is the first coordinate of the state $x$. For every $\tilde K\in\mathbb N$, the constant $C_\pi(x)$ in Theorem~\ref{th:15} satisfies
    \begin{align}
        C_\pi(x) &\leq C_\pi(x;\tilde K), \label{eq:nn166}
    \end{align}
    where
    \begin{align}
         C_\pi(x;\tilde K) =& 2\left[ 1+ \sum_{z\in\mathcal C_{W,R,d}} \tilde P_{\pi,1}(x,z) + \sum_{z\in\mathcal C_{W,R,d}} \tilde P_\pi(x,z)v_\pi^{(1)}(z) \right] \nonumber\\
        &+ 2\sum_{\ell=1}^{M_\pi}\hat\psi_\ell^\pi(x) \left( \frac{\bar m_{\pi,\ell}^{(2)}} {\bar m_{\pi,\ell}(\tilde K)}-1 \right). \label{eq:nn167}
    \end{align}
    The probability with which $\chi_{d}^{\pi}(t)$ enters class $\mathcal{R}_{\ell}^{\pi}$ starting from state $x$, $\hat{\psi}_{\ell}^{\pi}(x)$ is
    \begin{align}
        \hat\psi_\ell^\pi(r_j) &=\bm1_{\{j=\ell\}}, \label{eq:nn168}\\
        \bigl(\hat\psi_\ell^\pi(z)\bigr)_{z\in\mathcal C_\pi^{nr}} &= \left(I-\bm \bar P_\pi\right)^{-1}\bigl(\tilde P_\pi(z,r_\ell)\bigr)_{z\in\mathcal C_\pi^{nr}}, \label{eq:nn169}\\
        \hat\psi_\ell^\pi(x) &= \sum_{z\in\mathcal C_{W,R,d}} \tilde P_\pi(x,z)\hat\psi_\ell^\pi(z).\label{eq:nnn171}
    \end{align}
    Furthermore, for $\epsilon>0$, let
    \begin{align}
        T_\epsilon &= 1+\left\lfloor \max\left\{ \frac{ C_\pi(x;\tilde K)}{\epsilon}, \frac{8C_c(x)  C_\pi(x;\tilde K)}{\epsilon^2}\right\} \right\rfloor. \label{eq:nn170}
    \end{align}
    Then, every integer $T\geq T_\epsilon$ satisfies
    \begin{align}
        &\sum_{y\in\mathcal X_d} \left|\bar p_T^\pi(y\mid x)-\bar\psi_\pi^x(y)\right| <\epsilon, \label{eq:nn171}\\
        &\left|\Delta_d(\pi;T)-\Delta_d(\pi)\right| <\epsilon. \label{eq:nn172}
    \end{align}
\end{theorem}

The proof of this theorem is provided in Appendix~\ref{appen:th:16}.

The bounds in Theorem~\ref{th:16} depend on the transition kernels $\tilde{P}_\pi$, $\tilde{P}_{\pi,1}$, and $\tilde{P}_{\pi,2}$. We next show how to compute these kernels from the transition rules in (\ref{eq:543}), (\ref{eq:544}), and (\ref{eq:545}). In particular, we show that, for every initial state, these kernels can be computed using finitely many recursion steps.

\begin{theorem}\label{th:17}
    For $z\in\mathcal X_d$, the kernels $\tilde P_\pi(z,\cdot)$, $\tilde P_{\pi,1}(z,\cdot)$, and $\tilde P_{\pi,2}(z,\cdot)$ can be computed exactly using finitely many recursion steps. For $0<q<1$, we define
    \begin{align}
        \bar{n}_\pi(z) &= d+2+\max\left\{0,\left\lceil\frac{W}{(1-q)^d}\right\rceil-c(z) \right\}, \label{eq:nn300}
    \end{align}
    where $c(z)$ is the first coordinate of $z$. For every $y\in\mathcal X_d$, we initialize
    \begin{align}
        \mathbb{P}\left( \bar{\tau}(1)>0,\ \chi_d^\pi(1)=y \,\middle|\, \chi_d^\pi(1)=z \right) &=\bm{1}_{\{y=z\}}. \label{eq:nn303}
    \end{align}
    For every $n\geq1$ and $y\in\mathcal X_d$, the probabilities are computed recursively as
    \begin{align}
        \mathbb{P}&\left( \bar{\tau}(1)>n,\ \chi_d^\pi(n+1)=y \,\middle|\, \chi_d^\pi(1)=z \right)\nonumber\\
        &=\sum_{x\in\mathcal{X}_d} \mathbb{P}\left( \bar{\tau}(1)>n-1,\ \chi_d^\pi(n)=x \,\middle|\, \chi_d^\pi(1)=z \right) \left(1-q\bm{1}_{\{u_1>0\}}\right) \bm{1}_{\{y=x^{(0)}\}}, \label{eq:nn304}\\
        \mathbb{P}&\left( \bar{\tau}(1)=n,\ \chi_d^\pi(n+1)=y \,\middle|\, \chi_d^\pi(1)=z \right) \nonumber\\
        &=\sum_{x\in\mathcal{X}_d} \mathbb{P}\left( \bar{\tau}(1)>n-1,\ \chi_d^\pi(n)=x \,\middle|\, \chi_d^\pi(1)=z \right) q\bm{1}_{\{u_1>0\}}\bm{1}_{\{y=x^{(1)}\}}. \label{eq:nn305}
    \end{align}
    At each index $n$, the sums contain at most one nonzero term. If $q=1$, then for every $y\in\mathcal C_{W,R,d}$,
    \begin{align}
        \tilde P_\pi(z,y)&= \sum_{n=1}^{B_d} \mathbb P\left( \bar\tau(1)=n,\ \chi_d^\pi(n+1)=y \,\middle|\,\chi_d^\pi(1)=z\right),\label{eq:nn330}\\
        \tilde P_{\pi,j}(z,y)&= \sum_{n=1}^{B_d}n^j \mathbb P\left(\bar\tau(1)=n,\ \chi_d^\pi(n+1)=y \,\middle|\,\chi_d^\pi(1)=z\right), \quad j\in\{1,2\}, \label{eq:nn331}
    \end{align}
    where $B_d=W+2d$. If $0<q<1$, then, for every $y\in\mathcal C_{W,R,d}$,
    \begin{align}
        \tilde{P}_\pi(z,y) &= \sum_{n=1}^{\bar{n}_\pi(z)} \mathbb{P}\left(\bar{\tau}(1)=n,\ \chi_d^\pi(n+1)=y \,\middle|\, \chi_d^\pi(1)=z \right) \nonumber\\
        &\quad+ \frac{1}{1-(1-q)^R} \sum_{n=\bar{n}_\pi(z)+1}^{\bar{n}_\pi(z)+R}\mathbb{P}\left( \bar{\tau}(1)=n,\ \chi_d^\pi(n+1)=y \,\middle|\, \chi_d^\pi(1)=z \right). \label{eq:nn327}\\
        \tilde{P}_{\pi,1}(z,y)&= \sum_{n=1}^{\bar{n}_\pi(z)} n\,\mathbb{P}\left( \bar{\tau}(1)=n,\ \chi_d^\pi(n+1)=y \,\middle|\, \chi_d^\pi(1)=z \right) + \sum_{n=\bar{n}_\pi(z)+1}^{\bar{n}_\pi(z)+R} \Bigg[ \frac{n}{1-(1-q)^R} \nonumber\\
        &\quad+ \frac{R(1-q)^R}{(1-(1-q)^R)^2} \Bigg] \mathbb{P}\left( \bar{\tau}(1)=n,\ \chi_d^\pi(n+1)=y \,\middle|\,\chi_d^\pi(1)=z \right). \label{eq:nn328}\\
        \tilde{P}_{\pi,2}(z,y) &\!=\! \sum_{n=1}^{\bar{n}_\pi(z)} \!\!n^2\,\mathbb{P}\left(
        \bar{\tau}(1)\!=\!n,\chi_d^\pi(n\!+\!1)\!=\!y \,\middle|\,\chi_d^\pi(1)\!=\!z\right)
        \!+\!\!\sum_{n=\bar{n}_\pi(z)+1}^{\bar{n}_\pi(z)+R} \!\Bigg[\frac{n^2}{1\!-\!(1\!-\!q)^R}\!+\! \frac{2Rn(1\!-\!q)^R}{(1\!-\!(1\!-\!q)^R)^2} \nonumber\\ &\quad+ \frac{R^2(1-q)^R(1+(1-q)^R)} {(1-(1-q)^R)^3} \Bigg] \mathbb{P}\left( \bar{\tau}(1)=n,\ \chi_d^\pi(n+1)=y \,\middle|\, \chi_d^\pi(1)=z\right).\label{eq:nn329}
    \end{align}
    For $y\notin\mathcal C_{W,R,d}$, all three kernels are zero.
\end{theorem}
    
The proof of this theorem is provided in Appendix~\ref{appen:th:17}.

\subsection{Long-Term Average Age Evaluation}
In this subsection, we use the finite embedded chain to compute the long-term average age and both sampling and transmission rates directly. For every $\ell\in\{1,\ldots,M_\pi\}$, we define $\tilde{\mathcal R}_\ell^\pi =\mathcal R_\ell^\pi\cap\bar{\mathcal R}_\pi$. From the proof of Theorem~\ref{th:16} in Appendix~\ref{appen:th:16}, $\tilde{\mathcal R}_\ell^\pi$ is the unique recurrent class of the embedded chain contained in $\mathcal R_\ell^\pi$. Let $\mu_\ell^\pi$ denote its stationary distribution. For $k\in\{s,r\}$, let $a_k^\pi(y)$ denote the action $a_k^\pi(t)$ when $\chi_d^\pi(t)=y$. For $z\in\mathcal X_d$ and $g\in\{\hat v_{1,d},a_s^\pi,a_r^\pi\}$, we define
\begin{align}
    L_\pi(z) &=\mathbb E_\pi\left[\bar\tau(1) \,\middle|\,\chi_d^\pi(1)=z\right],\label{eq:1120}\\
    F_\pi(z;g) &=\mathbb E_\pi\left[ \sum_{t=1}^{\bar\tau(1)}g(\chi_d^\pi(t))\,\middle|\,\chi_d^\pi(1)=z\right].\label{eq:1121}
\end{align}

\begin{theorem}\label{th:18}
    For every $g\in\{\hat v_{1,d},a_s^\pi,a_r^\pi\}$, the following holds
    \begin{align}\label{eq:1122}
        \sum_{y\in\mathcal X_d}\bar\psi_\pi^x(y)g(y) &= \sum_{\ell=1}^{M_\pi}\hat\psi_\ell^\pi(x) \frac{ \sum_{z\in\tilde{\mathcal R}_\ell^\pi} \mu_\ell^\pi(z)F_\pi(z;g)} {\sum_{z\in\tilde{\mathcal R}_\ell^\pi} \mu_\ell^\pi(z)L_\pi(z)}.
    \end{align}
    The right-hand side of (\ref{eq:1122}) can be computed using finitely many recursion steps and finite linear systems.
\end{theorem}

The proof of this theorem is provided in Appendix~\ref{appen:th:18}.

To compute $\Delta_d(\pi_{d,4})$, we take $g=\hat v_{1,d}$ in (\ref{eq:1122}) for each constituent policy and combine the ages using (\ref{eq:nn149}).

\subsection{Optimizing a Mixture of $(W,R)_d$ Policies}\label{sec:7}
We define $\bar\Pi_d=\{(W,R)_d: W,R\in\mathbb N\}$. Let
\begin{align}\label{eq:1105}
    R_0&=\left\lfloor\frac{1}{f_s}\right\rfloor+1,\quad
    W_0=R_0+\left\lfloor\frac{1}{qf_r}\right\rfloor,\quad
    \pi_0=(W_0,R_0)_d.
\end{align}
From Lemma~\ref{lemma:df_policy_bounds}, we have
\begin{align}\label{eq:1106}
    \nu_{s,d}(\pi_0)\leq \frac{1}{R_0}<f_s,\quad \nu_{r,d}(\pi_0)\leq  \frac{1}{q(W_0-R_0+1)}<f_r, \quad \Delta_d(\pi_0)\leq W_0+\frac{1}{q}.
\end{align}
For simplicity of the presentation, we define
\begin{align}\label{eq:101}
    b_s=\frac{1}{R_0}, \quad b_r= \frac{1}{q(W_0-R_0+1)}, \quad B_0=W_0+\frac{1}{q}.
\end{align}
We also define
\begin{align}\label{eq:1109}
    \gamma=\min\{f_s-b_s,f_r-b_r\}>0.
\end{align}
For $K\in\mathbb N$ with $K>W_0$, let $R_*\geq R_0$ be the threshold constructed in the proof of Lemma~\ref{lemma:17} for this $K$. We define
\begin{align}\label{eq:1104}
    \bar\Pi_{d,K} &=\{(W,R)_d\in\bar\Pi_d:W<K,\ R\leq R_*\}.
\end{align}
By the construction in the proof of Lemma~\ref{lemma:17}, we have $R_0\leq R_*$. Since $K>W_0$, it is immediate that $\pi_0\in\bar\Pi_{d,K}$. Moreover, $R_i\leq R_0$ and $Q_j\leq W_0-R_0+1$, thus all the constituent policies of $\pi_{d,4}$, $\pi_{d}^{i,j}$, for $i,j\in\{1,2\}$ in (\ref{eq:346}) also belong to $\bar\Pi_{d,K}$.

We denote the policy that selects $\pi\in\bar{\Pi}_{d,K}$ with probability $w_\pi$ at $t=1$ and follows the selected policy for the entire time horizon by $\pi_w$. Thus, from Theorem~\ref{th:13}, we have
\begin{align}
    \Delta_d(\pi_w)&=\sum_{\pi\in\bar\Pi_{d,K}}w_\pi\Delta_d(\pi),\label{eq:1107}\\
    \nu_{k,d}(\pi_w)&=\sum_{\pi\in\bar\Pi_{d,K}}w_\pi\nu_{k,d}(\pi), \qquad k\in\{s,r\}.
\end{align}
Now, we consider the following linear program (LP)
\begin{align}\label{eq:1108}
    V_K=\min_{w}&\quad \sum_{\pi\in\bar\Pi_{d,K}}w_\pi\Delta_d(\pi)\nonumber\\
    \textrm{s.t.}&\quad \sum_{\pi\in\bar\Pi_{d,K}}w_\pi\nu_{k,d}(\pi)\leq f_k,\quad k\in\{s,r\},\nonumber\\
    &\quad \sum_{\pi\in\bar\Pi_{d,K}}w_\pi=1,\quad w_\pi\geq0,\quad \pi\in\bar\Pi_{d,K}.
\end{align}
The LP is feasible since $\pi_0\in\bar\Pi_{d,K}$ satisfies both rate constraints. Since $\pi_{d,4}$ is feasible by Theorem~\ref{th:df_six_approx} and its constituent policies belong to $\bar\Pi_{d,K}$, we have 
\begin{align}\label{eq:105}
    V_K\leq\Delta_d(\pi_{d,4}).
\end{align}
With a slack variable for each rate constraint, (\ref{eq:1108}) has three equality constraints. Therefore, an optimal basic feasible solution assigns positive probability to at most three policies.

Let $\lambda_s,\lambda_r\geq0$ be the Lagrangian multipliers for the sampling and transmission constraints, and let $\zeta\in\mathbb R$ be the multiplier for the equality constraint for the LP in (\ref{eq:1108}). Let $(\lambda_s^K,\lambda_r^K,\zeta_K)$ be an optimal solution of the dual LP of (\ref{eq:1108}). By dual feasibility and strong duality for the finite LP in (\ref{eq:1108}), respectively, we have
\begin{align}
    \zeta_K &\leq \Delta_d(\pi)+\lambda_s^K\nu_{s,d}(\pi) +\lambda_r^K\nu_{r,d}(\pi), \quad \pi\in\bar\Pi_{d,K},\label{eq:1113}\\
    V_K&=\zeta_K-\lambda_s^K f_s-\lambda_r^K f_r.\label{eq:n115}
\end{align}
Next, we find a lower bound on the average age of the monitor for policy $\pi\in\bar{\Pi}_d$.

\begin{lemma}\label{lemma:16}
For every $\pi=(W,R)_d\in\bar\Pi_d$, we have
\begin{align}\label{eq:n114}
    \Delta_d(\pi)\geq\frac{W+1}{2}.
\end{align}
\end{lemma}

The proof of this lemma is provided in Appendix~\ref{appen:th:8}.

\begin{lemma}\label{lemma:17}
    Let us assume that $K>W_0$. For $\lambda=(\lambda_s,\lambda_r)$ with $\lambda_s,\lambda_r\geq0$, we define
    \begin{align}
        G_\lambda(\pi)=\Delta_d(\pi)+\lambda_s\nu_{s,d}(\pi) +\lambda_r\nu_{r,d}(\pi).
    \end{align}
    There exists a finite integer $R_*\geq R_0$ such that, for every $\lambda$ satisfying $\lambda_s+\lambda_r\leq\frac{B_0}{\gamma}$, we have
    \begin{align}\label{eq:1141}
        G_\lambda((W,R_*)_d) &\leq G_\lambda((W,R)_d), \quad 1\leq W<K,\quad R>R_*.
    \end{align}
\end{lemma}

The proof of this lemma is provided in Appendix~\ref{appen:th:8}.

\begin{theorem}\label{th:8}
    For $\pi=(W,R)_d\in\bar\Pi_d$ and $\lambda^K=(\lambda_{s}^K,\lambda_{r}^K)$, if 
     \begin{align}\label{eq:1117}
        G_{\lambda^K}((W,R)_d)\geq\zeta_K,\quad R\in\mathbb N,\quad K\leq W\leq\lfloor 2\zeta_K-1\rfloor,
    \end{align}
    then $V_K$ is the optimal average age of the monitor among all feasible mixtures of policies in $\bar\Pi_d$. If the inequality in (\ref{eq:1117}) is strict for every policy in the range in (\ref{eq:1117}), then every optimal mixture over $\bar\Pi_d$ assigns zero probability to policies with $W\geq K$.
\end{theorem}

The proof of this theorem is provided in Appendix~\ref{appen:th:8}.

We next provide a finite choice of $K$ for which (\ref{eq:1117}) holds. Let $K>W_0$, then from the previous analysis we know that $\pi_0\in\bar\Pi_{d,K}$, and from (\ref{eq:1113}), we have
\begin{align}\label{eq:n128}
    \zeta_K&\leq \Delta_d(\pi_0) +\lambda_s^K\nu_{s,d}(\pi_0)    +\lambda_r^K\nu_{r,d}(\pi_0)\leq B_0+\lambda_s^K b_s+\lambda_r^K b_r,
\end{align}
where the second inequality follows from the definitions in (\ref{eq:101}). Now, from (\ref{eq:1109}) and (\ref{eq:n115}), we have
\begin{align}
    V_K &=\zeta_K-\lambda_s^K f_s-\lambda_r^K f_r\\
    &\leq B_0-\lambda_s^K(f_s-b_s)-\lambda_r^K(f_r-b_r)\\
    &\leq B_0-\gamma(\lambda_s^K+\lambda_r^K).
\end{align}
Since $V_K\geq0$, we obtain
\begin{align}\label{eq:n129}
    \lambda_s^K+\lambda_r^K\leq \frac{B_0}{\gamma}.
\end{align}
Thus, from (\ref{eq:n128}) we get
\begin{align}\label{eq:1119}
    \zeta_K \leq B_0+\max\{b_s,b_r\}(\lambda_s^K+\lambda_r^K) \leq B_0+\frac{B_0}{\gamma}\max\{b_s,b_r\}.
\end{align}
Note that the right-hand side of the second inequality is independent of $K$. Thus, we choose $K$ as
\begin{align}
    K=\max\left\{ W_0+1, 1+\left\lfloor
    2\left(B_0+\frac{B_0}{\gamma}\max\{b_s,b_r\}\right)-1\right\rfloor \right\}.
\end{align}
Note that this choice gives $K>W_0$ and $K>\lfloor 2\zeta_K-1\rfloor$. Thus, the range of $W$ in (\ref{eq:1117}) is empty. Therefore, by Theorem~\ref{th:8}, solving (\ref{eq:1108}) with this $K$ and the corresponding $R_*$ from Lemma~\ref{lemma:17} gives an optimal mixture over $\bar\Pi_d$, and every optimal mixture assigns zero probability to policies with $W\geq K$. We denote an optimal solution of (\ref{eq:1108}) by $w^*$, and the corresponding policy by $\pi_{w^*}$. Moreover, from (\ref{eq:105}) and Theorem~\ref{th:df_six_approx}, we observe that the policy corresponding to this optimal mixture provides a $4$-approximation guarantee as well. 

\section{Numerical Results}\label{sec:num}
For numerical analysis, we also consider the stationary randomized policy $\pi_{\mathrm{SR}}^*$ introduced in \cite{banerjee2026Fundamental}. In every slot, the sampler-transmitter pair selects idling, retransmission, or sampling and transmission of a fresh packet with probabilities $1-f_r$, $f_r-\min\{f_s,f_r\}$, and $\min\{f_s,f_r\}$, respectively. If retransmission is selected when no packet is stored, the pair stays idle. The selections are independent across slots and of the feedback-delay processes. For the lower bound in the figures, we use the maximum of the universal lower bound $\Delta_{\mathrm{LB}}$ in \cite{banerjee2026Fundamental} and the sampling-only and transmission-only lower bounds in \cite{banerjee2026Fundamental}, given in (\ref{eq:565}) and (\ref{eq:566}), respectively.

\begin{figure}[t]
    \centering
      \subfloat[Average age\label{fig:delayed_age_0.45}]{%
        \includegraphics[width=0.5\columnwidth]{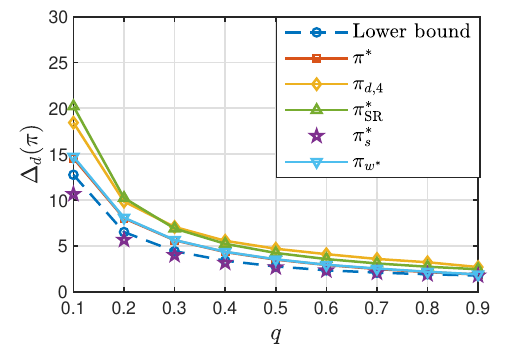}}
        \hfill
        \subfloat[Average sampling rate\label{fig:delayed_usage_0.45}]{%
        \includegraphics[width=0.5\columnwidth]{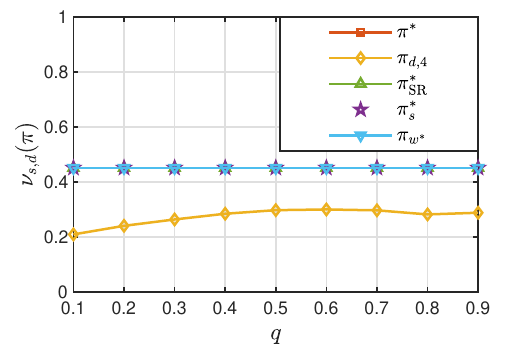}}%
    \caption{Average age of the monitor and sampling rate versus $q$ for $d=3$, $f_s=0.45$, and $f_r=0.5$.}
    \label{fig:comparison_0.45}
\end{figure}

\begin{figure}[t]
    \centering
      \subfloat[Average age\label{fig:delayed_age_0.1}]{%
        \includegraphics[width=0.5\columnwidth]{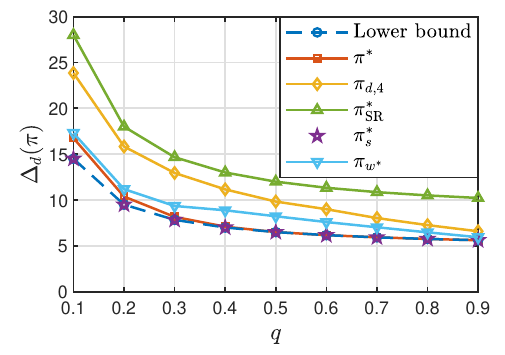}}
        \hfill
        \subfloat[Average sampling rate\label{fig:delayed_usage_0.1}]{%
        \includegraphics[width=0.5\columnwidth]{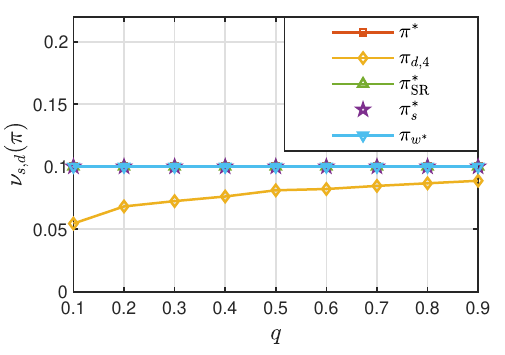}}%
    \caption{Average age of the monitor and sampling rate versus $q$ for $f_s=0.1$, and $f_r=0.5$.}
    \label{fig:comparison_0.1}
\end{figure}

For Fig.~\ref{fig:comparison_0.45} and Fig.~\ref{fig:comparison_0.1}, we truncate the ages by $50$ and enumerate all reachable information states within the resulting bounded state space. We then solve the corresponding occupation-measure LP \cite{altman2021constrained} to obtain the optimal policy of the truncated delayed feedback CMDP, $\pi_{d,\mathrm{tr}}^{*}$. We compute the average age of the monitor under each constituent policy of $\pi_{d,4}$ using Theorem~\ref{th:18}. 

Fig.~\ref{fig:comparison_0.45} presents the average age and sampling rate for $d=3$, $f_s=0.45$, and $f_r=0.5$ as $q$ varies. In Fig.~\ref{fig:delayed_age_0.45}, $\pi_{d,4}$ achieves a lower average age than $\pi_{\mathrm{SR}}^*$ at $q=0.1$ and $q=0.2$ and a higher average age at every plotted $q\geq0.3$. As shown in Fig.~\ref{fig:delayed_usage_0.45}, $\pi_{\mathrm{SR}}^*$ uses the full sampling budget. For the plotted values $q\geq0.3$, $\pi_{d,4}$ uses approximately $58\%$ to $66\%$ of the sampling budget. This lower sampling rate intuitively explains its higher age in  this range. The policy $\pi_{w^*}$ uses nearly the entire sampling budget and achieves a lower average age than both $\pi_{d,4}$ and $\pi_{\mathrm{SR}}^*$ at every plotted $q$. We also plot the equidistant-sampling policy $\pi_s^*$, proposed in \cite{banerjee2026Fundamental} in the instantaneous feedback setting. Under delayed feedback, this policy transmits the most recently sampled packet until its ACK is received or a fresh packet is sampled. For $f_s=0.45$, it chooses at $t=1$ to sample every two or three slots with probabilities $0.7$ and $0.3$, respectively. Since each sampling interval is at most three slots, for $d=3$ a new packet is sampled before the ACK of the current packet arrives. Thus, $\nu_{r,d}(\pi_s^*)=1>f_r$ at every plotted $q$, and we show its curve as a sampling-only benchmark.

Fig.~\ref{fig:comparison_0.1} presents the results for $d=3$, $f_s=0.1$, and $f_r=0.5$. For every plotted $q\geq0.2$, we have $f_s\leq qf_r$, which makes the transmission constraint redundant under instantaneous feedback \cite{banerjee2026Fundamental}. Under delayed feedback, the transmitter may continue sending a
successfully delivered packet while its ACK is pending. Here, $\pi_s^*$ samples a fresh packet every $10$ slots, so its average sampling rate is $f_s=\frac{1}{10}$. Its average age and transmission rate are
\begin{align}
    \Delta_d(\pi_s^*) &=\frac{1}{q}+\frac{9}{2},\\
    \nu_{r,d}(\pi_s^*) &=\frac{3+\frac{1-(1-q)^7}{q}}{10}.
\end{align}
The policy exceeds the transmission budget at $q=0.1,0.2,0.3,0.4$ and satisfies it at $q=0.5,\ldots,0.9$. We show its curve at all plotted values of $q$. The plotted lower bound equals the sampling-only lower-bound in (\ref{eq:565}) for all considered $q$. Hence, at each point where $\pi_s^*$ satisfies the transmission constraint, it attains the lower bound and is optimal for the delayed feedback problem. At these points, the truncated-CMDP curve also provides the same age. As shown in Fig.~\ref{fig:delayed_usage_0.1}, $\pi_{d,4}$ uses between approximately $72\%$ and $89\%$ of the sampling budget at the plotted $q\geq0.3$. Thus, its relative sampling-rate gap to $\pi_{\mathrm{SR}}^*$ is smaller than the gap in Fig.~\ref{fig:delayed_usage_0.45}. In this setting, $\pi_{d,4}$ achieves a lower average age than $\pi_{\mathrm{SR}}^*$ at every plotted $q$. The mixture $\pi_{w^*}$ uses nearly the entire sampling budget and further reduces the average age at every plotted $q$. 

\section{Conclusion}
We studied age of information minimization in a time-slotted generate-at-will status-update system with delayed feedback and separate average sampling- and transmission-rate constraints. We first derived the MMSE estimate of the age of the monitor based on the information available to the sampler-transmitter pair. We then introduced MMSE-based $(W,R)$ policies, denoted by $(W,R)_d$, and constructed a feasible policy $\pi_{d,4}$ that randomizes among four carefully chosen $(W,R)_d$ policies. We showed that $\pi_{d,4}$ achieves a $4$-approximation guarantee for any feedback-delay process in our model, including varying delays with feedback potentially arriving out of order. The policy also does not require knowledge of the distribution of the feedback delays. For a deterministic fixed finite feedback delay, we identified a state that is observable to the sampler-transmitter pair and evolves as a countable-state, time-homogeneous Markov chain under an arbitrary $(W,R)_d$ policy. We expressed the long-term average age in terms of a stationary distribution of this chain. We also established an $O(T^{-\frac{1}{2}})$ bound on the absolute difference between the finite-horizon and long-term average ages. For any prescribed accuracy $\epsilon>0$, we obtained a sufficient horizon $T_\epsilon$ such that the average age over $T$ slots differs from the corresponding long-term average age  by less than $\epsilon$ whenever $T\geq T_\epsilon$. We also evaluated the long-term average age and both sampling and transmission rates directly using a finite embedded Markov chain. Finally, by solving a finite linear program over mixtures of $(W,R)_d$ policies with $1\leq W<K$ and $1\leq R\leq R_*$, we obtained $\pi_{w^*}$, which is optimal among all feasible mixtures of $(W,R)_d$ policies with $W,R\in\mathbb{N}$.
 
\begin{appendices}

\section{Proofs of Lemma~\ref{lemma:df_policy_bounds} and Theorem~\ref{th:df_six_approx}}\label{appen:th:df_six_approx}
\subsection{Proof of Lemma~\ref{lemma:df_policy_bounds}}
\begin{Proof}
First, we show that two consecutive sampling instants are separated by at least $R$ slots. Let $\sigma_k$ and $\sigma_{k+1}$ be two consecutive sampling instants. If the packet sampled at $\sigma_k$ is still available at the transmitter at $\sigma_{k+1}$, then, from the definition of $(W,R)_{d}$ policy, it is replaced only if its age is at least $R$. Hence,
\begin{align}
        \sigma_{k+1}-\sigma_k\geq R.
\end{align}
Otherwise, the feedback corresponding to the successful delivery of the packet sampled at $\sigma_k$ has been received by the beginning of slot $\sigma_{k+1}$. Therefore, $\Gamma^{\pi}(\sigma_{k+1})=\sigma_k$, and
\begin{align}
    v_1^{\pi}(\sigma_{k+1}) = \sigma_{k+1}-\sigma_k
\end{align}
almost surely. Taking the conditional expectation with respect to $\mathcal{I}_d^{\pi}(\sigma_{k+1})$ gives
\begin{align}
    \hat{v}_{1,d}^{\pi}(\sigma_{k+1}) = \sigma_{k+1}-\sigma_k.
\end{align}
Since sampling at $\sigma_{k+1}$ requires $\hat{v}_{1,d}^{\pi}(\sigma_{k+1})\geq W\geq R$, we have
\begin{align}
    \sigma_{k+1}-\sigma_k\geq R.
\end{align}
Thus,
\begin{align}
    &\sum_{k=1}^{\sum_{t=1}^{T}a_{s}^{\pi}(t) -1} \!\!\!\!\!\sigma_{k+1} -\sigma_{k} \geq \left(\sum_{t=1}^{T}a^{\pi}_{s}(t)-1\right) R,\\
    &\frac{\sigma_{\sum_{t=1}^{T}a^{\pi}_{s}(t)} -\sigma_{1}}{R} \geq  \left(\sum_{t=1}^{T}a_{s}^{\pi}(t)-1\right), \\ 
    &\limsup_{T\rightarrow \infty} \frac{1}{T} \left(\sum_{t=1}^{T} a_{s}^{\pi}(t)-1\right) \leq \frac{1}{R},\label{eq:352}
\end{align}
where (\ref{eq:352}) follows from the fact that $\sigma_{\sum_{t=1}^{T}a^{\pi}_{s}(t)}$ is upper bounded by $T$ and $\sigma_{1}$ is lower bounded by $0$. Now, from the definition of $\nu_{s,d}(\pi)$, we have (\ref{eq:df_sampling_bound}). 

Let $\tilde{v}_{2}^{\pi}(t)$ denote the age of the packet \textit{transmitted} in slot $t$. Thus, we have $\tilde{v}_{2}^{\pi}(t)=0$ if no transmission occurs or if a fresh packet is sampled and transmitted, and $\tilde{v}_{2}^{\pi}(t)=v_2^{\pi}(t)$ when the stored packet is retransmitted. From the structure of the $(W,R)_{d}$ policy, we have 
\begin{align}\label{eq:df_transmitted_packet_age}
    0\leq \tilde{v}_{2}^{\pi}(t) \leq R-1, \qquad \textrm{whenever } a_r^{\pi}(t)=1.
\end{align}
Thus, we have the following age evolution,
\begin{align}\label{eq:df_exact_age_evolution}
    v_1^{\pi}(t+1)\!=\! v_1^{\pi}(t)\!+\!1 \!-\!a_r^{\pi}(t)Z(t)\left( v_1^{\pi}(t)\!-\!\tilde{v}_{2}^{\pi}(t)
    \right).
\end{align}
The action $a_r^{\pi}(t)$ and $\tilde{v}_{2}^{\pi}(t)$ are $\mathcal{I}_d^{\pi}(t)$-measurable, while $Z(t)$ is independent of $v_1^{\pi}(t)$, $\tilde{v}_{2}^{\pi}(t)$, and $\mathcal{I}_d^{\pi}(t)$. Thus, from (\ref{eq:df_exact_age_evolution}), we have
\begin{align}\label{eq:df_conditional_age_drift}
    \mathbb{E}_{\pi}\!\left[v_1^{\pi}(t+1) | \mathcal{I}_d^{\pi}(t)\right]=\hat{v}_{1,d}^{\pi}(t)+1 -q a_r^{\pi}(t) \left( \hat{v}_{1,d}^{\pi}(t)-\tilde{v}_{2}^{\pi}(t)\right).
\end{align}
Note that, $\mathcal{I}_d^{\pi}(t)\subseteq\mathcal{I}_d^{\pi}(t+1)$, thus, from the tower property of conditional expectation, we have
\begin{align}\label{eq:df_tower_drift}
    \mathbb{E}_{\pi}\!\left[\hat{v}_{1,d}^{\pi}(t+1)|\mathcal{I}^{\pi}_d(t)\right]=\mathbb{E}_{\pi}\!\left[v_1^{\pi}(t+1)|\mathcal{I}_d^{\pi}(t)\right].
\end{align}
From the construction of the $(W,R)_{d}$ policy and (\ref{eq:df_transmitted_packet_age}), we have
\begin{align}\label{eq:357}
    \hat{v}_{1,d}^{\pi}(t)\geq W\geq  R>R-1\geq\tilde{v}_{2}^{\pi}(t), \qquad
    \textrm{whenever } a_r^{\pi}(t)=1.
\end{align}
From (\ref{eq:df_conditional_age_drift}), (\ref{eq:df_tower_drift}), and (\ref{eq:357}), we have
\begin{align}
    \mathbb{E}_{\pi}\!\left[\hat{v}_{1,d}^{\pi}(t+1)-\hat{v}_{1,d}^{\pi}(t) |\mathcal{I}_d^{\pi}(t)\right] & \leq 1-q a_{r}^{\pi}(t)\left(\hat{v}_{1,d}^{\pi}(t)-R+1\right)\label{eq:df_estimator_drift}\\ & \leq 1- q a_{r}^{\pi}(t) (W-R+1). \label{eq:df_estimator_drift_UB}
\end{align}
Taking expectations in both sides of (\ref{eq:df_estimator_drift_UB}), summing from $t=1$ to $T$, and using $\hat{v}_{1,d}^{\pi}(T+1)\geq0$, we have
\begin{align}\label{eq:telescope}
    q(W-R+1)\sum_{t=1}^{T} \mathbb{E}_{\pi}[a_r^{\pi}(t)] \leq T+\mathbb{E}_{\pi}[\hat{v}_{1,d}^{\pi}(1)].
\end{align}
Now, from the definition of $\nu_{r,d}(\pi)$ and (\ref{eq:telescope}), we get (\ref{eq:df_transmission_bound}). Now,
\begin{align}
    \hat{v}_{1,d}^{\pi}(t) &=\hat{v}_{1,d}^{\pi}(t)(1-a_r^{\pi}(t)) +\hat{v}_{1,d}^{\pi}(t)a_r^{\pi}(t) \label{eq:df_age_decomposition1}\\
    &\leq W(1-a_r^{\pi}(t))+(R-1)a_r^{\pi}(t) +\left(\hat{v}_{1,d}^{\pi}(t)-R+1\right)a_r^{\pi}(t) \label{eq:df_age_decomposition2}\\
    &\leq W+\left(\hat{v}_{1,d}^{\pi}(t)-R+1\right)a_r^{\pi}(t) \label{eq:df_age_decomposition3}\\
    &\leq W+\frac{1}{q}+ \frac{\mathbb{E}_{\pi}\!\left[ \hat{v}_{1,d}^{\pi}(t)-\hat{v}_{1,d}^{\pi}(t+1)\,\middle|\,\mathcal{I}_d^{\pi}(t)\right]}{q}.
    \label{eq:df_age_decomposition4}
\end{align}
Here, (\ref{eq:df_age_decomposition2}) follows from $\hat{v}_{1,d}^{\pi}(t)<W$ whenever $a_r^{\pi}(t)=0$, (\ref{eq:df_age_decomposition3}) follows from $R-1\leq W$, and (\ref{eq:df_age_decomposition4}) follows from (\ref{eq:df_estimator_drift}). Now, taking expectations in both sides of (\ref{eq:df_age_decomposition4}) then summing from $t=1$ to $T$ and using the fact that $\hat{v}_{1,d}^{\pi}(T+1)\geq 0$, we have
\begin{align}
    \sum_{t=1}^{T} \mathbb{E}_{\pi}[\hat{v}_{1,d}^{\pi}(t)] \leq W T + \frac{T}{q} + \frac{\mathbb{E}_{\pi}[\hat{v}_{1,d}^{\pi}(1)]}{q}.
\end{align}
Now, from (\ref{eq:516}) and the definition of $\Delta_{d}(\pi)$, we have (\ref{eq:df_age_bound}).
\end{Proof}

\subsection{Proof of Theorem~\ref{th:df_six_approx}}
\begin{Proof}
From  the construction of the policy $\pi_{d,4}$, we have
\begin{align}
    \nu_{s,d}(\pi_{d,4};T) = \sum_{i,j\in\{1,2\}} \theta_{i,j} \nu_{s,d}(\pi_{d}^{i,j};T),
\end{align}
and therefore, we have
\begin{align}
    \nu_{s,d}(\pi_{d,4}) &\leq \sum_{i,j\in\{1,2\}} \theta_{i,j} \nu_{s,d}(\pi_{d}^{i,j}) \\ 
    &\leq \sum_{i,j\in\{1,2\}} \frac{\theta_{i,j}} {R_{i}}\label{eq:373} \\ 
    &= \frac{\theta_{s}}{R_{1}} + \frac{1-\theta_{s}}{R_{2}}\\
    &=f_{s},\label{eq:375}
\end{align}
where (\ref{eq:373}) follows from  Lemma~\ref{lemma:df_policy_bounds} and (\ref{eq:375}) follows from (\ref{eq:theta_s}). Similarly, we have
\begin{align}
    \nu_{r,d} (\pi_{d,4}) &\leq  \sum_{i,j\in\{1,2\}} \theta_{i,j} \nu_{r,d}(\pi_{d}^{i,j}) \\ 
    &\leq  \sum_{i,j\in\{1,2\}} \frac{\theta_{i,j}} {qQ_{j}}\label{eq:377}\\
    &\leq  f_{r},\label{eq:378}
\end{align}
where (\ref{eq:377}) follows from the construction of the policies $\pi_{d}^{i,j}$ and Lemma~\ref{lemma:df_policy_bounds}, and (\ref{eq:378}) follows from (\ref{eq:theta_q}). 

For the average age of the monitor, we similarly obtain
\begin{align}
    \Delta_d(\pi_{d,4}) &\leq \sum_{i,j\in\{1,2\}}\theta_{i,j}\Delta_d(\pi_d^{i,j})\\
    &\leq \sum_{i,j\in\{1,2\}}\theta_{i,j}\left(R_i+Q_j-1+\frac{1}{q}\right)\label{eq:380}\\
    &\leq \theta_sR_1+(1-\theta_s)R_2+Q_2-1+\frac{1}{q}\\
    &\leq \theta_sR_1+(1-\theta_s)R_2+\frac{1}{qf_r}+\frac{1}{q}.\label{eq:563}
\end{align}
Here, (\ref{eq:380}) follows from (\ref{eq:346}) and Lemma~\ref{lemma:df_policy_bounds}. The last two inequalities follow from the definitions of  $Q_1$ and $Q_2$.

Let $\Delta^*$ denote the optimal average age of the monitor for the corresponding instantaneous-feedback problem with the same $q$, $f_s$, $f_r$, and initial state \cite{banerjee2026Fundamental}. Providing instantaneous feedback gives the sampler-transmitter pair additional information and therefore cannot increase the optimal average age. Hence,
\begin{align}\label{eq:564}
    \Delta_d^*\geq\Delta^*.
\end{align}
The sampling-only and transmission-only lower bounds established in \cite{banerjee2026Fundamental}, together with (\ref{eq:564}), give
\begin{align}
    \Delta_d^{*} &\geq \frac{1}{q} + \frac{ \theta_sR_1+(1-\theta_s)R_2-1 }{2}, \label{eq:565}\\
    \Delta_d^{*} &\geq  \frac{1}{2} +\frac{1}{2qf_r}  +\frac{(1-q)f_r}{2q}. \label{eq:566}
\end{align}
Adding (\ref{eq:565}) and (\ref{eq:566}) and multiplying the resulting inequality by $2$, we have
\begin{align}
    4\Delta_d^{*} &\geq 2\left[ \frac{1}{q} + \frac{ \theta_sR_1+(1-\theta_s)R_2-1}{2} + \frac{1}{2} + \frac{1}{2qf_r} + \frac{(1-q)f_r}{2q} \right]\\
    &= \theta_sR_1+(1-\theta_s)R_2 +\frac{1}{qf_r} +\frac{2}{q}+\frac{(1-q)f_r}{q} \\ 
    & \geq \Delta_d(\pi_{d,4})+\frac{1}{q} +\frac{(1-q)f_r}{q}, \label{eq:567}
\end{align}
where (\ref{eq:567}) follows from (\ref{eq:563}). Moreover,
\begin{align}
    \frac{1}{q}+\frac{(1-q)f_r}{q}\geq0. \label{eq:568}
\end{align}
Therefore, from (\ref{eq:567}) and (\ref{eq:568}), we obtain
\begin{align}
    \Delta_d(\pi_{d,4})\leq 4\Delta_d^{*}-\frac{1}{q} -\frac{(1-q)f_r}{q} \leq 4\Delta_d^{*}.
\end{align}
This completes the proof of this theorem.
\end{Proof}
\section{Proof of Theorem~\ref{th:12}}\label{appen:th:12}
\begin{Proof}
We first show that $\chi_d^\pi(t)$ is uniquely determined by $\mathcal I_d^\pi(t)$. Note that, $e_{d,0}^\pi(t)$ is uniquely determined by $\mathcal I_d^\pi(t)$. Thus, from (\ref{eq:364}), we see that $c_d^\pi(t)$ is uniquely determined by $\mathcal I_d^\pi(t)$. Similarly, the sampler-transmitter pair knows whether a packet is stored at the beginning of slot $t$ and, if so, its generation time. Therefore, $b_d^\pi(t)$ is uniquely determined by $\mathcal I_d^\pi(t)$. Finally, under the fixed feedback delay $d$, at the beginning of slot $t$, the only transmission outcomes that can remain unresolved are those corresponding to the slots $t-d,t-d+1,\cdots,t-1$. The information $\mathcal I_d^\pi(t)$ contains whether a transmission occurred in each of these slots and the generation time of the corresponding transmitted packet. Thus, $u_{d,k}^\pi(t)$, $1\leq k\leq d$ are uniquely determined by $\mathcal I_d^\pi(t)$. For those indices corresponding to slots before slot $1$, no transmission has occurred and the corresponding $u_{d,k}^\pi(t)$ is zero. Thus, $\chi_d^\pi(t)$ is also uniquely determined by $\mathcal I_d^\pi(t)$.

We next prove (\ref{eq:368}). We see that $\chi_d^\pi(t)$ uniquely determines 
\begin{align}
    J_d^\pi(t),\ \{e_{d,j}^\pi(t)\}_{j=0}^{J_d^\pi(t)}, \ \{n_{d,j}^\pi(t)\}_{j=1}^{J_d^\pi(t)},\ \{p_{d,j}^\pi(t)\}_{j=1}^{J_d^\pi(t)}. 
\end{align} 
Thus, from (\ref{eq:estimated_age}) we see that $\hat v_{1,d}^\pi(t)$ is a deterministic function of $\chi_d^\pi(t)$. Since $\chi_d^\pi(t)$ is uniquely determined by $\mathcal I_d^\pi(t)$, applying the tower property of the conditional expectation, we get
\begin{align}
    \mathbb E_\pi \left[ v_1^\pi(t) \,\middle|\, \chi_d^\pi(t) \right] &= \mathbb E_\pi \left[ \mathbb E_\pi \left[ v_1^\pi(t) \,\middle|\, \mathcal I_d^\pi(t) \right] \,\middle|\, \chi_d^\pi(t) \right]\\
    &= \mathbb E_\pi \left[ \hat v_{1,d}^\pi(t) \,\middle|\, \chi_d^\pi(t) \right]\label{eq:n373} \\
    &=\hat v_{1,d}^\pi(t),\label{eq:n374}
\end{align}
where (\ref{eq:n373}) follows from (\ref{eq:estimator}), and (\ref{eq:n374}) follows from the fact that $\hat v_{1,d}^\pi(t)$ is a deterministic function of $\chi_d^\pi(t)$.

If $b_d^\pi(t)=0$, from (\ref{eq:365}) we see that $v_{2}^{\pi}(t)=\infty$. If $b_d^\pi(t)>0$, then from (\ref{eq:364}) and (\ref{eq:365}), we have $v_2^\pi(t)=c_d^\pi(t)-b_d^\pi(t)$. Thus, $\chi_d^\pi(t)$ uniquely determines $v_2^\pi(t)$. The policy $\pi$ selects its action according to $\hat v_{1,d}^\pi(t)$, whether a packet is available, and the age of that packet. Therefore, the action pair $\left(a_s^\pi(t),a_r^\pi(t)\right)$ is uniquely determined by $\chi_d^\pi(t)$.

It remains to prove (\ref{eq:369}) and the time-homogeneous Markov property. After the sampling decision in slot $t$, but before incorporating the feedback received at the end of the slot, define
\begin{align}
    \bar b^{\pi}_{d}(t) = a_s^\pi(t)c^{\pi}_{d}(t) + \left(1-a_s^\pi(t)\right)b^{\pi}_{d}(t). \label{eq:n375}
\end{align} 
At the end of slot $t$, the feedback corresponding to slot $t-d$ is received whenever $t-d\geq1$. We define,
\begin{align}
    D(t) =
    \begin{cases}
        Z(t-d)u^{\pi}_{d,1}(t), & u^{\pi}_{d,1}(t)>0,\\
        0, & u^{\pi}_{d,1}(t)=0.
    \end{cases}
\end{align}
For $t\leq d$, we have $u^{\pi}_{d,1}(t)=0$, since the corresponding slot $t-d$ precedes slot $1$, and hence $D(t)=0$. If $u^{\pi}_{d,1}(t)=0$, then $e_{d,0}^{\pi}(t+1)=e_{d,0}^{\pi}(t)$, and hence $D(t)=0$. If $u^{\pi}_{d,1}(t)>0$, then the corresponding outcome has not been revealed by the beginning of slot $t$. If that transmission failed, $D(t)=0$, whereas if it succeeded, $e_{d,0}^{\pi}(t)$ increases by exactly $u^{\pi}_{d,1}(t)$. Thus,
\begin{align}
    c_d^\pi(t+1) &= c+1-D(t), \label{eq:n377}\\
    b_d^\pi(t+1) &=\left(\bar b-D(t)\right)^+,\label{eq:n378}\\
    u_{d,k}^\pi(t+1) &= \left(u_{k+1}-D(t)\right)^+, \qquad k=1,\cdots,d-1, \label{eq:379}\\
    u_{d,d}^\pi(t+1) &= \left( a_r^\pi(t)\bar b-D(t) \right)^+. \label{eq:379d}
\end{align}

Given $\mathcal I_d^\pi(t)$, all the quantities on the right-hand sides of (\ref{eq:n377})-(\ref{eq:379d}), except the channel realization corresponding to slot $t-d$, $z(t-d)$, are known by the beginning of slot $t$. Since the channel realizations are i.i.d.~Bernoulli random variables and the policy is causal, we have
\begin{align}
    \mathbb P \left( Z(t-d)=1 \,\middle|\, \mathcal I_d^\pi(t) \right) = q,
\end{align}
almost surely. Consequently, given the current state $\chi_{d}^{\pi}(t)$ and the action selected by the policy, the distribution of $\chi_d^\pi(t+1)$ is completely determined by the current state and the action pair $(a_{s}^{\pi}(t),a_{r}^{\pi}(t))$, and  does not depend on any other part of $\mathcal I_d^\pi(t)$. We have already shown that the action pair is uniquely determined by $\chi_{d}^{\pi}(t)$. Thus, for every $B\subseteq\mathcal X_d$, (\ref{eq:369}) holds. Moreover, the transition rules (\ref{eq:n377})-(\ref{eq:379d}) and the success probability $q$ contain no explicit dependence on $t$. Thus, $\{\chi_d^\pi(t)\}_{t\geq1}$ is a time-homogeneous Markov chain on the countable state space $\mathcal X_d$.
\end{Proof}

\section{Proofs of Lemma~\ref{lemma:n7}, Lemma~\ref{lemma:n8}, and Lemma~\ref{lemma:n9}}\label{appen:lemma:n7}
\subsection{Proof of Lemma~\ref{lemma:n7}}
\begin{Proof}
We prove this lemma by contradiction. We assume that no transmission occurs in those slots. At the beginning of slot $t+d$, all feedback corresponding to transmissions prior to slot $t$ has been received, and since no transmission occurred in slots $t,\cdots,t+d-1$, we have $J_d^\pi(t+d)=0$. From (\ref{eq:estimated_age}) and (\ref{eq:364})
\begin{align}
    \hat v_{1,d}^\pi(t+d) = t+d-e_{d,0}^\pi(t+d) = c_d^\pi(t+d).
\end{align}
If $c_d^\pi(t+d)\geq W$, from the definition of the policy $\pi$, a transmission occurs in slot $t+d$, which contradicts the assumption. Hence, $c_d^\pi(t+d)<W$. For slots $t+d+\tau$, $0\leq \tau\leq W-1$, no transmission occurs as well. Thus, $e_{d,0}^\pi(\cdot)$ remains constant and we have
\begin{align}
    c_d^\pi(t+d+\tau)= c_d^\pi(t+d)+\tau.
\end{align}
Since $c_d^\pi(t+d)\geq1$, there exists an integer $0\leq \tau\leq W-1$, such that
\begin{align}
    c_d^\pi(t+d+\tau)\geq W.
\end{align}
At this time, there is no transmission for which feedback is pending. Thus, 
\begin{align}
    \hat v_{1,d}^\pi(t+d+\tau) = c_d^\pi(t+d+\tau) \geq W.
\end{align} 
From the definition of policy $\pi$ a transmission occurs at this slot, which is a contradiction of our assumption. Thus, a transmission must occur during the stated interval.
\end{Proof}
\subsection{Proof of Lemma~\ref{lemma:n8}}
\begin{Proof}
By Lemma~\ref{lemma:n7}, we see that from the beginning of time $t$ a transmission occurs no later than slot $t+d+W-1$. Let $\tau$ denote the first such transmission slot. Note that the packet transmitted in slot $\tau$ has generation time strictly larger than $e_{d,0}^\pi(\tau)$. If $e_{d,0}^{\pi}(\tau+d)>e_{d,0}^{\pi}(t)$, then we have
\begin{align}
    \bar{\tau}(t) \leq B_{d}.
\end{align}  
Otherwise, $e_{d,0}^\pi(\cdot)$ remains unchanged until the feedback for the transmission in slot $\tau$ is received. The transmission at slot $\tau$ succeeds with probability $q$. Upon success, its ACK is received at the end of slot $\tau+d$, and therefore 
\begin{align}
    \bar{\tau}(t) = \tau+d+1-t.
\end{align}
Since $\tau \leq t+d+W-1$, we have
\begin{align}
    \tau+d+1 \leq t+W+2d = t+B_{d}.
\end{align}
Therefore,
\begin{align}
    \mathbb P\!\left( \bar{\tau}(t) > B_{d} \,\middle|\, \mathcal I_d^\pi(t) \right) \leq 1-q,
\end{align}
almost surely. We denote 
\begin{align}
    \mathcal{E}_{n} = \{\bar{\tau}(t) >n B_{d}\}.
\end{align}
Note that
\begin{align}\label{eq:393}
   \mathcal{E}_{n+1} =  \mathcal{E}_{n} \cap \{\bar{\tau}(t+n B_{d}) > B_{d}\}.
\end{align}
Then,
\begin{align}
    \mathbb{P}\left(\mathcal{E}_{n+1} \,\middle|\, \mathcal{I}_{d}^{\pi}(t+nB_{d})\right) = &\mathbb{E}_{\pi}\left[\bm{1}_{\mathcal{E}_{n+1}}\,\middle|\, \mathcal{I}_{d}^{\pi}(t+nB_{d})\right] \\ 
    &= \bm{1}_{\mathcal{E}_{n}}\mathbb{P}\left(\bar{\tau}(t+n B_{d}) > B_{d} \,\middle|\, \mathcal{I}_{d}^{\pi}(t+nB_{d})\right)\label{eq:395} \\ &\leq(1-q) \bm{1}_{\mathcal{E}_{n}},
\end{align}
almost surely. Note that (\ref{eq:395}) follows from (\ref{eq:393}) and from the fact that the event $\mathcal{E}_{n}$ is determined by $\mathcal I_d^\pi(t+nB_{d})$ and hence its indicator can be taken outside the conditional expectation. Now, from the tower property of the conditional expectation, we have
\begin{align}
    \mathbb{P}\left(\mathcal{E}_{n+1} \,\middle|\, \mathcal{I}_{d}^{\pi}(t)\right) =& \mathbb{E}_{\pi} \left[\mathbb{E}_{\pi}\left[\bm{1}_{\mathcal{E}_{n+1}}\,\middle|\, \mathcal{I}_{d}^{\pi}(t+nB_{d})\right] \,\middle|\, \mathcal{I}_{d}^{\pi}(t)\right] \\ \leq& (1-q) \mathbb{E}_{\pi}\left[\bm{1}_{\mathcal{E}_{n}}\,\middle|\, \mathcal{I}_{d}^{\pi}(t)\right] \\ 
    =&(1-q) \mathbb{P}\left(\mathcal{E}_{n} \,\middle|\, \mathcal{I}_{d}^{\pi}(t)\right).
\end{align}
Now, induction on $n$ proves (\ref{eq:388}).

Since $\bar{\tau}(t)$ is a non-negative integer-valued random variable, the following  pointwise identity holds for any integer $m\geq 1$,
\begin{align}
    \bar{\tau}(t)^m = \sum_{i=0}^{\infty} \left( (i+1)^m-i^m \right) \bm{1}_{\{\bar{\tau}(t)>i\}}.
\end{align}
Thus,
\begin{align}
    \mathbb{E}_{\pi} [\bar{\tau}(t)^{m} \mid \mathcal{I}_{d}^{\pi}(t)] =& \mathbb{E}_{\pi} \left[\sum_{i=0}^{\infty} \left( (i+1)^m-i^m \right) \bm{1}_{\{\bar{\tau}(t)>i\}} \mid \mathcal{I}_{d}^{\pi}(t) \right] \\
    =& \sum_{i=0}^{\infty}((i+1)^{m}-i^m)\mathbb{E}_{\pi}[\bm{1}_{\{\bar{\tau}(t)>i\}}\mid \mathcal{I}_{d}^{\pi}(t)]\label{eq:403}\\ =&\sum_{i=0}^{\infty}((i+1)^{m}-i^m) \mathbb{P}(\bar{\tau}(t)>i \mid \mathcal{I}_{d}^{\pi}(t)),  
\end{align}
where (\ref{eq:403}) follows from Tonelli's theorem. For $nB_d\leq i<(n+1)B_d$, from (\ref{eq:388}), we have $\mathbb P(\bar\tau(t)>i\mid\mathcal I_d^\pi(t))\leq(1-q)^n$. Thus,
\begin{align}
    \mathbb E_\pi\left[\bar\tau(t)^m \,\middle|\,\mathcal I_d^\pi(t)\right] &\leq \sum_{n=0}^{\infty}(1-q)^n \sum_{i=nB_d}^{(n+1)B_d-1} \left((i+1)^m-i^m\right)\\
    &= B_d^m\sum_{n=0}^{\infty} \left((n+1)^m-n^m\right)(1-q)^n. \label{eq:404}
\end{align}
The series in (\ref{eq:404}) is finite and is independent of $t$ and $\mathcal{I}_{d}^{\pi}(t)$. We denote it by $C_{m}$, which proves (\ref{eq:389}).
In particular, for $m=1$ and $m=2$, respectively, we obtain
\begin{align}
    \mathbb E_\pi\left[\bar\tau(t) \,\middle|\,\mathcal I_d^\pi(t)\right] &\leq B_d\sum_{n=0}^{\infty}(1-q)^n =\frac{B_d}{q}, \label{eq:405}\\
    \mathbb E_\pi\left[\bar\tau(t)^2 \,\middle|\,\mathcal I_d^\pi(t)\right] &\leq B_d^2\sum_{n=0}^{\infty}(2n+1)(1-q)^n =\frac{B_d^2(2-q)}{q^2}. \label{eq:406}
\end{align}
\end{Proof}
\subsection{Proof of Lemma~\ref{lemma:n9}}
\begin{Proof}
Since $e_{d,0}^\pi({\tau})>e_{d,0}^{\pi}(\tau-1)$ a successful transmission occurred at slot
\begin{align}\label{eq:410}
    \tau{'} = \tau - d-1.
\end{align} 
Thus, at the beginning of slot $\tau-1$, this transmission is the oldest unresolved transmission. Also, as $e_{d,0}^\pi({\tau})>e_{d,0}^{\pi}(\tau-1)$, the transmitted packet at slot $\tau'$ was generated strictly after $e_{d,0}^\pi(\tau-1)$, and hence, we have $u_{d,1}^\pi(\tau-1)>0$. By the definition of $u_{d,1}^\pi(\tau-1)$, the generation time of this packet is therefore $e_{d,0}^\pi(\tau-1)+u_{d,1}^\pi(\tau-1)$. If this packet was freshly sampled in slot $\tau'$, then
\begin{align}
    \tau'-  \left( e_{d,0}^\pi(\tau-1)+u_{d,1}^\pi(\tau-1)\right) =0.
\end{align}
Otherwise, the packet was retransmitted in slot $\tau'$. By the definition of the policy $\pi$, a stored packet is retransmitted only when its age is strictly smaller than $R$. Thus, we have
\begin{align}
    \tau'- \left( e_{d,0}^\pi(\tau-1)+u_{d,1}^\pi(\tau-1) \right) \leq R-1.
\end{align}
Thus, in both cases, we have
\begin{align}
    \tau'-  \left( e_{d,0}^\pi(\tau-1)+u_{d,1}^\pi(\tau-1) \right) \leq R-1. \label{eq:414}
\end{align}
By the definition of $e_{d,0}^\pi(\tau)$, we have
\begin{align}
    e_{d,0}^\pi(\tau) \geq e_{d,0}^\pi(\tau-1)+u_{d,1}^\pi(\tau-1).
\end{align}
Consequently,
\begin{align}
    c_d^\pi(\tau)&=\tau-e_{d,0}^\pi(\tau)\\
    &\leq\tau-\left(e_{d,0}^\pi(\tau-1)+u_{d,1}^\pi(\tau-1)\right)\\
    &=\tau'+d+1-\left(e_{d,0}^\pi(\tau-1)+u_{d,1}^\pi(\tau-1)\right)\label{eq:417}\\
    &\leq R+d, \label{eq:418}
\end{align}
where (\ref{eq:417}) follows from (\ref{eq:410}) and the last inequality follows from (\ref{eq:414}).

If $b_d^\pi(\tau)>0$, then by the definition of $b_d^\pi(\tau)$, the generation time of the packet stored at the transmitter at the beginning of slot $\tau$ is $e_{d,0}^\pi(\tau)+b_d^\pi(\tau)$. Since this packet must have been generated no later than slot $\tau-1$, we have
\begin{align}
    e_{d,0}^\pi(\tau)+b_d^\pi(\tau) \leq \tau-1.
\end{align}
Thus,
\begin{align}\label{eq:n420}
    b_d^\pi(\tau) \leq \tau-1-e_{d,0}^\pi(\tau)=c_d^\pi(\tau)-1 \leq R+d-1.
\end{align}
If $b_d^\pi(\tau)=0$, the same upper bound holds trivially.

Similarly, for any $k\in\{1,\cdots,d\}$ such that $u_{d,k}^\pi(\tau)>0$, the generation time of the packet corresponding to that unresolved transmission is $e_{d,0}^\pi(\tau)+u_{d,k}^\pi(\tau)$. Since the corresponding transmission occurred before slot $\tau$, this packet must also have been generated no later than slot $\tau-1$. Therefore,
\begin{align}\label{eq:421}
    u_{d,k}^\pi(\tau) \leq \tau-1-e_{d,0}^\pi(\tau) = c_d^\pi(\tau)-1 \leq R+d-1.
\end{align}
If $u_{d,k}^\pi(\tau)=0$, the bound again holds trivially. Thus, from (\ref{eq:418}), (\ref{eq:n420}), and (\ref{eq:421}), we have (\ref{eq:409}).
\end{Proof}

\section{Proofs of Lemma~\ref{lemma:n10}, Lemma~\ref{lemma:n11}, and Lemma~\ref{lemma:n12}}\label{appen:lemma:n10}
\subsection{Proof of Lemma~\ref{lemma:n10}}
\begin{Proof}
We denote the sigma algebra generated by $\mathcal{I}_{d}^{\pi}(t)$, with $\mathcal{G}_{t}^{\pi}$, i.e., $\mathcal{G}_{t}^{\pi}=\sigma(\mathcal I_{d}^{\pi}(t))$. Since the information available at the beginning of slot $t+1$ contains all the information available at the beginning of slot $t$, we have 
\begin{align} 
    \mathcal G_t^\pi \subseteq \mathcal G_{t+1}^\pi, \qquad t\geq1. 
\end{align} 
Thus, $\{\mathcal G_t^\pi\}_{t\geq1}$ is a filtration. We first show that every $\bar{\tau}_n$ is a stopping time with respect to this filtration. Note that $\{\bar{\tau}_{n}\leq 1\}=\Phi$, where $\Phi$ denotes the empty set. Thus, the event $\{\bar{\tau}_{n}\leq 1\}$ belongs to $\mathcal{G}_{1}^{\pi}$. Now, for every deterministic integer $m\geq2$, 
\begin{align} 
    \{\bar{\tau}_n\leq m\} = \left\{ \sum_{i=2}^{m} \bm{1}_{\{ e_{d,0}^\pi(i)>e_{d,0}^\pi(i-1) \}} \geq n+1 \right\}. 
\end{align}
For every $i\leq m$, both $e_{d,0}^\pi(i)$ and $e_{d,0}^\pi(i-1)$ are determined by $\mathcal I_d^\pi(m)$. Therefore,  
\begin{align} 
    \{\bar{\tau}_n\leq m\} \in \mathcal G_m^\pi, 
\end{align} 
and hence $\bar{\tau}_n$ is a stopping time with respect to $\{\mathcal G_t^\pi\}_{t\geq1}$. We next show that 
\begin{align} 
    \bar{\tau}_n<\infty, \label{eq:n424_finite} 
\end{align} 
almost surely for every $n\geq0$. We first consider $n=0$. By Lemma~\ref{lemma:n8}, $\bar{\tau}(1)<\infty$ almost surely. Thus, from (\ref{eq:423}) we have $\bar{\tau}_0<\infty$ almost surely. We now proceed by induction. Let us assume that $\bar{\tau}_n<\infty$ almost surely. For every deterministic integer $k\geq1$, Lemma~\ref{lemma:n8} similarly gives 
\begin{align} 
    \mathbb P\left( \bar{\tau}(k)=\infty \mid \mathcal I_d^\pi(k) \right) = 0, 
\end{align}
almost surely. Since $\mathcal G_k^\pi = \sigma\left( \mathcal I_d^\pi(k) \right)$, we equivalently have 
\begin{align} 
    \mathbb P\left( \bar{\tau}(k)=\infty \mid \mathcal G_k^\pi \right) = 0, 
\end{align} 
almost surely. Since $\bar{\tau}_n$ is a stopping time, we have $\{\bar{\tau}_n=k\} \in \mathcal G_k^\pi$. Therefore,
\begin{align} 
    \mathbb P\left( \bar{\tau}_n=k, \bar{\tau}(k)=\infty \right) &= \mathbb E_\pi\left[ \bm{1}_{\{\bar{\tau}_n=k\}} \bm{1}_{\{\bar{\tau}(k)=\infty\}} \right]\\ 
    &= \mathbb E_\pi\!\left[ \bm{1}_{\{\bar{\tau}_n=k\}} \mathbb P\left( \bar{\tau}(k)=\infty \,\middle|\, \mathcal G_k^\pi \right) \right]\label{eq:nn432} \\ 
    &= 0. 
\end{align}
Since $\bar{\tau}_n<\infty$ almost surely, we have 
\begin{align} 
    \mathbb P\left( \bar{\tau}(\bar{\tau}_n)=\infty \right)  &= \sum_{k=1}^{\infty} \mathbb P\left( \bar{\tau}_n=k, \bar{\tau}(k)=\infty \right) = 0. 
\end{align}
Thus, from the recursive definition of (\ref{eq:n424}), we see that  $\bar{\tau}_{n+1}<\infty$ almost surely. Hence, by induction, (\ref{eq:n424_finite}) holds for every $n\geq0$. We define
\begin{align}\label{eq:667}
    \tilde{P}_{\pi}(x,x') = \mathbb{P}\left(\chi_{d}^{\pi}(t+\bar{\tau}(t))=x' \,\middle|\, \chi_{d}^{\pi}(t)=x\right).
\end{align}
From Theorem~\ref{th:12}, we know that $\{\chi_d^\pi(t)\}_{t\geq1}$ is a discrete-time, time-homogeneous Markov chain with transition kernel $P_{\pi}(\cdot,\cdot)$. Thus, $\tilde{P}_{\pi}(x,x')$ in (\ref{eq:667}) only depends on $x$ and $x'$ and is independent of $t$. Now, we define the stopped sigma algebra at stopping time $\bar{\tau}_{n}$ as
\begin{align} 
    \mathcal G_{\bar{\tau}_n}^\pi = \left\{ E: E\cap\{\bar{\tau}_n\leq m\} \in \mathcal G_m^\pi \text{ for every }m \right\}.
\end{align}
From the strong Markov property of $\{\chi_{d}^{\pi}(t)\}_{t\geq 1}$, at $\bar{\tau}_{n}$, we have
\begin{align}
    \mathbb P\!\left( \chi_d^\pi(\bar{\tau}_{n+1})=x' \,\middle|\, \mathcal G_{\bar{\tau}_n}^\pi \right) = \tilde P_\pi\left( \chi_d^\pi(\bar{\tau}_n),x' \right), \label{eq:nn436}
\end{align}
almost surely. Furthermore,
\begin{align}
    \sigma\!\left( \chi_d^\pi(\bar{\tau}_0), \cdots, \chi_d^\pi(\bar{\tau}_n) \right) \subseteq \mathcal G_{\bar{\tau}_n}^\pi.
\end{align}
Indeed, $\chi_d^\pi(\bar{\tau}_i)$ is $\mathcal G_{\bar{\tau}_i}^\pi$-measurable for every $i\leq n$, and $\bar{\tau}_i\leq\bar{\tau}_n$ implies $\mathcal G_{\bar{\tau}_i}^\pi \subseteq \mathcal G_{\bar{\tau}_n}^\pi$. Therefore, applying the tower property to (\ref{eq:nn436}), we obtain
\begin{align}
    \mathbb P\!\left( \chi_d^\pi(\bar{\tau}_{n+1})=x' \,\middle|\, \chi_d^\pi(\bar{\tau}_0), \cdots, \chi_d^\pi(\bar{\tau}_n) \right) &= \mathbb E_\pi\left[
    \tilde P_\pi\left( \chi_d^\pi(\bar{\tau}_n),x'\right) \,\middle|\, \chi_d^\pi(\bar{\tau}_0), \cdots, \chi_d^\pi(\bar{\tau}_n) \right]\\
    &= \tilde P_\pi\left( \chi_d^\pi(\bar{\tau}_n),x' \right).
\end{align}
Hence,
$\{\chi_d^\pi(\bar{\tau}_n)\}_{n\geq0}$ is a Markov chain with transition kernel $\tilde{P}_{\pi}(\cdot,\cdot)$. Since the transition kernel $\tilde P_\pi\left( x,x' \right)$ does not depend on $n$, the embedded chain is time homogeneous. Finally, by the definition of $\bar{\tau}_n$,
\begin{align}
    e_{d,0}^\pi(\bar{\tau}_n) > e_{d,0}^\pi(\bar{\tau}_n-1).
\end{align}
Therefore, Lemma~\ref{lemma:n9} gives
\begin{align}
    \chi_d^\pi(\bar{\tau}_n) \in \mathcal C_{W,R,d}, \quad n\geq0.
\end{align}
Since $\mathcal C_{W,R,d}$ is finite, $\{\chi_d^\pi(\bar{\tau}_n)\}_{n\geq0}$ is a finite-state, discrete-time, time-homogeneous Markov chain.
\end{Proof}

\subsection{Proof of Lemma~\ref{lemma:n11}}
\begin{Proof}
From Lemma~\ref{lemma:n10}, the sampled chain $\left\{ \chi_d^\pi(\bar{\tau}_n) \right\}_{n\geq0}$ is a finite-state, time-homogeneous Markov chain with state space contained in $\mathcal C_{W,R,d}$ and transition kernel $\tilde P_{\pi}(\cdot,\cdot)$. We first show that every state belonging to a recurrent class of $\{\chi_d^\pi(\bar{\tau}_n)\}_{n\geq0}$ is positive recurrent for the original chain $\{\chi_d^\pi(t)\}_{t\geq1}$. Let us consider a recurrent class of
$\{\chi_d^\pi(\bar{\tau}_n)\}_{n\geq0}$, $\mathcal {R}^{\textrm{sam}}$, and a state $x\in \mathcal {R}^{\textrm{sam}}$. Now, we assume that the sampled chain is initialized at $\chi_d^\pi(\bar{\tau}_0)=x$ and define its first return time to $x$ by
\begin{align}
    N_x = \inf \left\{ n\geq1: \chi_d^\pi(\bar{\tau}_n)=x \right\}.
\end{align}
Since $\mathcal R^{\textrm{sam}}$ is a recurrent class of a finite-state Markov chain, $\mathcal R^{\textrm{sam}}$ is finite, hence,
\begin{align}
    \mathbb E_\pi[N_x] < \infty. \label{eq:n441}
\end{align}
Now, we bound the expected number of slots between two successive times $\bar{\tau}_n$ and $\bar{\tau}_{n+1}$. From the definition of these times in (\ref{eq:n424}), we have
\begin{align}
    \bar{\tau}_{n+1}-\bar{\tau}_n = \bar{\tau}(\bar{\tau}_n).
\end{align}
From Lemma~\ref{lemma:n8}, with $m=1$, there exists a finite constant $C_1$, such that
\begin{align}
    \mathbb E_{\pi}\!\left[ \bar{\tau}(k) \,\middle|\, \mathcal G_k^\pi \right] \leq C_1,
\end{align}
almost surely for every deterministic $k\geq1$. We now show that the same bound holds when the deterministic time $k$ is replaced by the stopping time $\bar{\tau}_n$. For $E\in\mathcal G_{\bar{\tau}_n}^\pi$,
\begin{align}
    \mathbb E_{\pi}\!\left[ \bm{1}_E \left( \bar{\tau}_{n+1} - \bar{\tau}_n \right) \right] &= \mathbb{E}_{\pi}\left[ \sum_{k=1}^{\infty} \bm{1}_{E\cap\{\bar{\tau}_n=k\}}\bar{\tau}(k)\right]\\
    &=\sum_{k=1}^{\infty} \mathbb E_{\pi}\!\left[ \bm{1}_{E\cap\{\bar{\tau}_n=k\}} \bar{\tau}(k) \right], \label{eq:n445}
\end{align}
where in (\ref{eq:n445}), we used Tonelli's theorem. As $E\cap\{\bar{\tau}_n=k\} \in \mathcal G_k^\pi$, we get
\begin{align}
    \mathbb E\!\left[ \bm{1}_{E\cap\{\bar{\tau}_n=k\}} \bar{\tau}(k)\right] &= \mathbb E_{\pi}\!\left[ \bm{1}_{E\cap\{\bar{\tau}_n=k\}} \mathbb E_{\pi}\!\left[ \bar{\tau}(k) \,\middle|\,\mathcal G_k^\pi \right]\right]\\
    &\leq C_1 \mathbb P\!\left( E\cap\{\bar{\tau}_n=k\} \right).
\end{align}
Substituting this inequality into (\ref{eq:n445}) gives
\begin{align}
    \mathbb E_{\pi}\!\left[ \bm{1}_E \left( \bar{\tau}_{n+1}-\bar{\tau}_n \right) \right] \leq C_1\mathbb P(E).
\end{align}
Since this holds for every $E\in\mathcal G_{\bar{\tau}_n}^\pi$, we obtain
\begin{align}
    \mathbb E_\pi\!\left[\bar{\tau}_{n+1}-\bar{\tau}_n \,\middle|\, \mathcal G_{\bar{\tau}_n}^\pi \right] \leq C_1, \label{eq:n449}
\end{align}
almost surely. Now, the following holds,
\begin{align}
    \bar{\tau}_{N_x}-\bar{\tau}_0 = \sum_{n=0}^{\infty} \bm{1}_{\{N_x>n\}} \left( \bar{\tau}_{n+1}-\bar{\tau}_n \right).
\end{align}
Applying the Tonelli's theorem we get
\begin{align}
    \mathbb E_\pi\!\left[ \bar{\tau}_{N_x}-\bar{\tau}_0\right]= \sum_{n=0}^{\infty} \mathbb E_\pi\!\left[ \bm{1}_{\{N_x>n\}} \left(\bar{\tau}_{n+1}-\bar{\tau}_n\right) \right].    \label{eq:n451}
\end{align}
Note that, $\{N_x>n\} \in \mathcal G_{\bar{\tau}_n}^\pi$. Hence, using (\ref{eq:n449}) we get,
\begin{align}
    \mathbb E_\pi\!\left[ \bm{1}_{\{N_x>n\}} \left( \bar{\tau}_{n+1}-\bar{\tau}_n \right) \right] &= \mathbb E_\pi\!\left[ \bm{1}_{\{N_x>n\}} \mathbb E\!\left[ \bar{\tau}_{n+1}-\bar{\tau}_n \,\middle|\, \mathcal G_{\bar{\tau}_n}^\pi \right] \right]\\
    &\leq C_1 \mathbb P(N_x>n).
\end{align}
Thus, using (\ref{eq:n451}), we obtain
\begin{align}
    \mathbb E_\pi\!\left[ \bar{\tau}_{N_x}-\bar{\tau}_0 \right] &\leq C_1 \sum_{n=0}^{\infty} \mathbb P(N_x>n)\\
    &= C_1\mathbb E_\pi[N_x]< \infty, \label{eq:453}
\end{align}
where the last inequality follows from (\ref{eq:n441}). Now, we define
\begin{align}\label{eq:691}
    T_x^+ = \inf \left\{ t\geq1: \chi_d^\pi(\bar{\tau}_0+t)=x \right\}.
\end{align}
By the definition of $N_x$, $\chi_d^\pi(\bar{\tau}_{N_x}) = x$. Thus, $\bar{\tau}_{N_x}$ is one particular time after $\bar{\tau}_0$ at which the original chain is again in state $x$. Consequently,
\begin{align}
T_x^+ \leq \bar{\tau}_{N_x}-\bar{\tau}_0.
\end{align}
It follows from (\ref{eq:453}) that
\begin{align}
    \mathbb E_\pi[T_x^+] < \infty. \label{eq:n458}
\end{align}
The strong Markov property implies that, conditional on $\chi_d^\pi(\bar{\tau}_0)\!=\!x$, the process $\left\{ \chi_d^\pi(\bar{\tau}_0+r)
\right\}_{r\geq0}$ has the same law as the original Markov chain with transition kernel $P_{\pi}$ initialized at state $x$. Therefore, (\ref{eq:n458}) shows that the expected first return time to $x$ for the original chain initialized at $x$ is finite. Hence, $x$ is a positive recurrent state of $\{\chi_d^\pi(t)\}_{t\geq1}$. 

We next consider an arbitrary recurrent class $\mathcal R$ of the original chain $\{\chi_d^\pi(t)\}_{t\geq1}$. Fix any state $y\in\mathcal R$ and initialize the original chain at $y$. By Lemma~\ref{lemma:n8}, the next slot at which $e_{d,0}^\pi(\cdot)$ increases occurs almost surely in finite time. Since $\mathcal R$ is closed, the state of the original chain at that time still belongs to $\mathcal R$. On the other hand, Lemma~\ref{lemma:n9} implies that the state
at every such time belongs to $\mathcal C_{W,R,d}$. Therefore,
\begin{align}
    \mathcal R \cap \mathcal C_{W,R,d} \neq \Phi. \label{eq:df_eval_recurrent_intersection}
\end{align}
Now, initialize the sampled chain $\{\chi_d^\pi(\bar{\tau}_n)\}_{n\geq0}$ at any state in $\mathcal R\cap\mathcal C_{W,R,d}$. Because $\mathcal R$ is closed for the original chain, the original chain remains in $\mathcal R$ between successive times $\bar{\tau}_n$. Hence, $\chi_d^\pi(\bar{\tau}_n) \in \mathcal R$
for every $n$. Furthermore, by Lemma~\ref{lemma:n9}, $\chi_d^\pi(\bar{\tau}_n) \in \mathcal C_{W,R,d}$ for every $n$. Thus,
\begin{align}
    \chi_d^\pi(\bar{\tau}_n) \in \mathcal R\cap\mathcal C_{W,R,d},
\end{align}
for every $n$, and consequently $\mathcal R\cap\mathcal C_{W,R,d}$ is a finite closed set for the sampled chain. Every finite closed set of a finite-state Markov chain contains at least one recurrent class. Let $\mathcal R^{\textrm{sam}}$ be such a recurrent class of the sampled chain satisfying
\begin{align}
    \mathcal R^{\textrm{sam}} \subseteq \mathcal R\cap\mathcal C_{W,R,d},
\end{align}
and choose any state $x\in\mathcal R^{\mathrm{sam}}$. By the first part of the proof, $x$ is positive recurrent for the original chain. Since $x\in\mathcal R$, $\mathcal{R}$ is positive recurrent. Since $\mathcal R$ was an arbitrary recurrent class of the original chain, every recurrent class of $\{\chi_d^\pi(t)\}_{t\geq1}$ is positive recurrent.

Distinct recurrent classes are disjoint, whereas (\ref{eq:df_eval_recurrent_intersection}) shows that every recurrent class contains at least one state of the finite set $\mathcal C_{W,R,d}$. Therefore, the number of recurrent classes is at most $\left| \mathcal C_{W,R,d} \right|$, and hence is finite.

Finally, consider the original chain initialized at $x\in\mathcal{X}_{d}$. By Lemma~\ref{lemma:n8}, $\bar{\tau}_0<\infty$ almost surely. Hence, with probability one, the sampled chain $\left\{ \chi_d^\pi(\bar{\tau}_n) \right\}_{n\geq0}$ is reached and thereafter evolves according to the transition kernel $\tilde P_{\pi}(\cdot,\cdot)$. Every finite-state Markov chain eventually enters one of its recurrent classes with probability one. Therefore, with probability one, there exist a finite random index $n$ and a recurrent class $\mathcal {R}^{\textrm{sam}}$ of the sampled chain such that $\chi_d^\pi(\bar{\tau}_n) \in \mathcal{R}^{\textrm{sam}}$. We denote $ y = \chi_d^\pi(\bar{\tau}_n)$. By the first part of the proof, $y$ is positive recurrent for the original chain. At time $\bar{\tau}_n$, the original chain is in one of its recurrent classes. Thus, starting from an arbitrary state $x\in\mathcal{X}_{d}$, the original chain eventually enters one of its recurrent classes with probability one.
\end{Proof}
\subsection{Proof of Lemma~\ref{lemma:n12} }
\begin{Proof}
From Theorem~\ref{th:12}, we know that $\{\chi_d^\pi(t)\}_{t\geq1}$ is a countable-state, time-homogeneous Markov chain. From Lemma~\ref{lemma:n11}, every recurrent class is positive recurrent, and the chain eventually enters one of these classes almost surely. Fix an initial state $x\in\mathcal X_d$. Since $\mathcal R_\ell^\pi$ is irreducible and recurrent, for $y\in\mathcal R_\ell^\pi$, the probability that the chain ever visits $y$ is $\hat\psi_\ell^\pi(x)$. Thus, from \cite[Theorem 5.5.11]{durrett2019probability} and \cite[Section 5.6, (5.6.1)]{durrett2019probability}, we obtain
\begin{align}
    \lim_{T\rightarrow\infty}\bar p_T^\pi(y\mid x) = \hat\psi_\ell^\pi(x)\psi_\ell^\pi(y), \quad y\in\mathcal R_\ell^\pi.\label{eq:226}
\end{align}
For every transient state $y$, the result in \cite[Section 5.6, (5.6.1)]{durrett2019probability} also gives $\lim_{T\rightarrow\infty}\bar p_T^\pi(y\mid x)=0$. Therefore, from (\ref{eq:143}), for every $y\in\mathcal X_d$, we obtain
\begin{align}\label{eq:n488}
    \lim_{T\rightarrow\infty}\bar p_T^\pi(y\mid x) &= \sum_{\ell=1}^{M_\pi} \hat\psi_\ell^\pi(x)\psi_\ell^\pi(y)\\
    &=\bar\psi_\pi^x(y),
\end{align}
which proves (\ref{eq:467}).

Since $\bar p_T^\pi(\cdot\mid x)$ and $\bar\psi_\pi^x$ are probability mass functions, we have
\begin{align}
\sum_{y\in\mathcal X_d} \left| \bar p_T^{\pi}(y\mid x) - \bar{\psi}_{\pi}^{x}(y) \right|= 2 - 2 \sum_{y\in\mathcal X_d} \min\left\{ \bar p_T^{\pi}(y\mid x), \bar{\psi}_{\pi}^{x}(y) \right\}. \label{eq:491}
\end{align}
From (\ref{eq:467}), for every $y\in\mathcal X_d$,
\begin{align}\label{eq:n490}
    \lim_{T\rightarrow\infty}\min\left\{ \bar p_T^{\pi}(y\mid x), \bar{\psi}_{\pi}^{x}(y) \right\} = \bar{\psi}_{\pi}^{x}(y),
\end{align}
Since 
\begin{align}
    0 \leq \min\left\{ \bar p_T^{\pi}(y\mid x), \bar{\psi}_{\pi}^{x}(y) \right\} \leq \bar{\psi}_{\pi}^{x}(y), \quad\sum_{y\in\mathcal X_d} \bar{\psi}_{\pi}^{x}(y) = 1,
\end{align}
from the dominated convergence theorem and (\ref{eq:n490}) we obtain
\begin{align}
    \lim_{T\rightarrow\infty}\sum_{y\in\mathcal X_d} \min\left\{\bar p_T^{\pi}(y\mid x), \bar{\psi}_{\pi}^{x}(y) \right\} = 1.
\end{align}
Thus, from (\ref{eq:491}), we get (\ref{eq:n468}).
\end{Proof}

\section{Proof of Theorem~\ref{th:13}}\label{appen:th:13}
We first prove (\ref{eq:515}).

\begin{lemma}\label{lemma:n13}
    For every fixed initial state $x\in\mathcal{X}_{d}$, there exists a finite constant $C_c(x)$, such that
    \begin{align}\label{eq:n495}
        \sup_{t\geq1} \mathbb E_\pi\!\left[ \left( c_d^\pi(t) \right)^2 \,\middle|\, \chi_{d}^{\pi}(1)=x \right] \leq C_c(x).
    \end{align}
\end{lemma}

\begin{Proof}
We define $c_{0}= \max\left\{c_d^\pi(1), R+d \right\}$. Since the initial state is fixed $c_{0}$ is finite. Moreover, since $e_{d,0}^\pi(t)$ is non-decreasing in $t$, from (\ref{eq:364}) we have
\begin{align}
    c_d^\pi(t) &= t-e_{d,0}^\pi(t)\\
    &\leq t-e_{d,0}^\pi(1)\\
    &= c_d^\pi(1)+t-1\\
    &\leq c_0+t-1. \label{eq:495}
\end{align}
By Lemma~\ref{lemma:n9}, if we have $e_{d,0}^{\pi}(\tau)>e_{d,0}^{\pi}(\tau -1)$, then 
\begin{align}
    c_d^\pi(\tau) \leq R+d \leq c_0. \label{eq:500}
\end{align}
First, we next show that, for every $t\geq1$ and every integer $n\geq1$, the following holds,
\begin{align}
    \mathbb P_\pi\!\left( c_d^\pi(t) > c_0+nB_{d} \,\middle|\,\chi_{d}^{\pi}(1)=x \right) \leq (1-q)^n, \label{eq:501}
\end{align}
where $B_{d}$ is defined in Lemma~\ref{lemma:n8}. If $t-nB_{d}<1$, then from (\ref{eq:495}), we have $c_d^\pi(t)< c_0+nB_{d}$. Thus, (\ref{eq:501}) holds trivially. Now, we consider the case $t-nB_{d}\geq1$. If at slot $\tau\in\{t-nB_{d}+1,\cdots,t\}$, we have $e_{d,0}^{\pi}(\tau) >e_{d,0}^{\pi}(\tau-1)$, then, since $e_{d,0}^\pi(\cdot)$ is non-decreasing, from (\ref{eq:500}), we have
\begin{align}
    c_d^\pi(t)&=t-e_{d,0}^\pi(t)\\
    &\leq t-e_{d,0}^\pi(\tau)\\
    &= c_d^\pi(\tau)+(t-\tau)\\
    &\leq c_0+nB_{d}.
\end{align}
Thus,
\begin{align}
    \left\{ c_d^\pi(t) > c_0+nB_{d} \right\} \subseteq \left\{ \bar{\tau}(t-nB_{d}) > nB_{d} \right\}. \label{eq:506}
\end{align}
Let $s=t-nB_d$. Using Lemma~\ref{lemma:n8}, we get
\begin{align}
    \mathbb P\!\left( c_d^\pi(t) > c_0+nB_{d} \,\middle|\, \chi_{d}^{\pi}(1)=x \right)
    &\leq \mathbb P\!\left( \bar{\tau}(s) > nB_{d} \,\middle|\,\chi_{d}^{\pi}(1)=x \right)\\
    &= \mathbb E_\pi\!\left[ \mathbb P\!\left( \bar{\tau}(s) > nB_{d} \,\middle|\, \mathcal I_d^\pi(s) \right)\,\middle|\, \chi_{d}^{\pi}(1)=x \right]\\
    &\leq (1-q)^n.
\end{align}
Thus, (\ref{eq:501}) holds. We now apply the same argument used in the proof of Lemma~\ref{lemma:n8}. Specifically, we replace $\bar{\tau}(t)$ in that proof by $(c_{d}^{\pi}(t)-c_{0})^+$ and set $m=2$. From (\ref{eq:501}), for $nB_{d}\leq i<(n+1)B_{d}$, we have
\begin{align}
    \mathbb P\left((c_{d}^{\pi}(t)-c_0)^+>i\,\middle|\,\chi_{d}^{\pi}(1)=x\right) \leq (1-q)^n.
\end{align}
Therefore, using the identity
\begin{align}\label{eq:759}
    ((c_{d}^{\pi}(t)-c_{0})^+)^2 = \sum_{i=0}^{\infty} \left( (i+1)^2-i^2 \right) \bm{1}_{\{(c_{d}^{\pi}(t)-c_0)^+>i\}},
\end{align}
and following the same steps as in the proof of Lemma~\ref{lemma:n8}, we obtain
\begin{align}
    \mathbb E_\pi\left[((c_{d}^{\pi}(t)-c_0)^+)^2\,\middle|\, \chi_{d}^\pi(1)=x\right] \leq 2B_{d}^2 \sum_{n=0}^{\infty} (n+1)(1-q)^n. \label{eq:df_eval_Y_second_moment}
\end{align}
The series on the right-hand side is finite and independent of $t$. Hence,
\begin{align}
    \sup_{t\geq1} \mathbb E_\pi\left[((c_{d}^{\pi}(t)-c_0)^+)^2\,\middle|\,\chi_{d}^{\pi}(1)=x\right] < \infty,
\end{align}
which proves (\ref{eq:n495}).
\end{Proof}

\begin{lemma}\label{lemma:n14}
    The family $\left\{ \hat v_{1,d}^{\pi}(t): t\geq1 \right\}$ is uniformly integrable. That is,
    \begin{align}
        \lim_{K\rightarrow\infty} \sup_{t\geq1} \mathbb E_\pi\!\left[ \hat v_{1,d}^{\pi}(t) \bm{1}_{\{ \hat v_{1,d}^{\pi}(t)>K \}} \,\middle|\, \chi_{d}^{\pi}(1)=x \right ] = 0.\label{eq:514}
    \end{align}
\end{lemma}

\begin{Proof}
Since $e_{d,0}^\pi(t)$ is the generation time of a packet whose successful delivery is already known, the freshest packet actually delivered to the monitor cannot have a generation time smaller than $e_{d,0}^\pi(t)$. Thus,
\begin{align}
    0 \leq v_1^\pi(t) \leq c_d^\pi(t).
\end{align}
Taking conditional expectation with respect to $\mathcal I_d^\pi(t)$ gives
\begin{align}\label{eq:n516}
    0 \leq \hat v_{1,d}^\pi(t) \leq c_d^\pi(t).
\end{align}
Hence,
\begin{align}
    \mathbb E_\pi\!\left[ \hat v_{1,d}^\pi(t) \bm{1}_{\{ \hat v_{1,d}^\pi(t)>K
    \}}\,\middle|\, \chi_{d}^{\pi}(1)=x \right]
    &\leq \mathbb E_\pi\!\left[ c_d^\pi(t) \bm{1}_{\{ c_d^\pi(t)>K \}} \,\middle|\, \chi_{d}^{\pi}(1)=x\right]\\
    &\leq\frac{\mathbb E_\pi\left[ (c_d^\pi(t))^2\,\middle|\, \chi_{d}^{\pi}(1)=x \right]
    }{K}.
\end{align}
Taking the supremum over $t\geq 1$ and applying Lemma~\ref{lemma:n13}, we get
\begin{align}\label{eq:746}
    \sup_{t\geq1} \mathbb E_\pi\!\left[ \hat v_{1,d}^\pi(t) \bm{1}_{\{ \hat v_{1,d}^\pi(t)>K
    \}} \,\middle|\, \chi_{d}^{\pi}(1)=x \right] \leq \frac{C_c(x)}{K}.
\end{align}
Now, letting $K\rightarrow\infty$ proves (\ref{eq:514}). 
\end{Proof}

\subsection{Proof of Theorem~\ref{th:13}}
\begin{Proof}
Since $\hat v_{1,d}^\pi(t)$ is a deterministic function of $\chi_d^\pi(t)$, we have
\begin{align}
    \mathbb E_{\pi}\!\left[ \hat v_{1,d}^\pi(t)\,\middle|\, \chi_{d}^{\pi}(1)=x \right] = \sum_{y\in\mathcal X_d} p_t^\pi(y\mid x) \hat v_{1,d}(y). \label{eq:51}
\end{align}
Thus, from (\ref{eq:516}) and Tonelli's theorem, we have
\begin{align}
    \frac{1}{T} \sum_{t=1}^{T} \mathbb E_{\pi}\!\left[ v_1^\pi(t) \,\middle|\, \chi_{d}^{\pi}(1)=x \right]
    &= \sum_{y\in\mathcal X_d} \left( \frac{1}{T} \sum_{t=1}^{T} p_t^\pi(y\mid x) \right) \hat v_{1,d}(y)\\
    &= \sum_{y\in\mathcal X_d} \bar p_T^\pi(y\mid x) \hat v_{1,d}(y). \label{eq:518}
\end{align}
Now, we truncate $\hat{v}_{1,d}^{\pi}(t)$ with $K>0$. We define 
\begin{align}
    \hat v_{1,d}^{(K)}(y) = \min\left\{ \hat v_{1,d}(y),K \right\}. \label{eq:519}
\end{align}
From (\ref{eq:n468}), we have
\begin{align}
    \lim_{T\rightarrow\infty}\left| \sum_{y\in\mathcal X_d} \bar p_T^\pi(y\mid x) \hat v_{1,d}^{(K)}(y) - \sum_{y\in\mathcal X_d} \bar{\psi}_{\pi}^{x}(y) \hat v_{1,d}^{(K)}(y) \right| \leq& K \lim_{T\rightarrow\infty}\sum_{y\in\mathcal X_d} \left| \bar p_T^\pi(y\mid x) - \bar{\psi}_{\pi}^{x}(y) \right|\\
    =& 0. \label{eq:520}
\end{align}
Next, we bound the truncation error under $\bar p_T^\pi(\cdot\mid x)$, 
\begin{align}
    \sum_{y\in\mathcal X_d} \bar p_T^\pi(y\mid x) \left( \hat v_{1,d}(y) - \hat v_{1,d}^{(K)}(y) \right) &= \frac{1}{T} \sum_{t=1}^{T} \mathbb E_{\pi}\!\left[ \left(\hat v_{1,d}^\pi(t)-K  \right)^+\,\middle|\, \chi_{d}^{\pi}(1)=x \right]\label{eq:525}\\
    &\leq \frac{1}{T} \sum_{t=1}^{T} \mathbb E_{\pi}\!\left[ \hat v_{1,d}^\pi(t) \bm{1}_{\{
    \hat v_{1,d}^\pi(t)>K \}} \,\middle|\, \chi_{d}^{\pi}(1)=x\right]\\
    &\leq\sup_{t\geq1} \mathbb E_{\pi}\!\left[\hat v_{1,d}^\pi(t)\bm{1}_{\{\hat v_{1,d}^\pi(t)>K\}}\,\middle|\, \chi_{d}^{\pi}(1)=x\right], \label{eq:521}
\end{align}
where (\ref{eq:525}) follows from the definitions of $\bar p_T^\pi(y\mid x)$ and $\hat v_{1,d}^{(K)}(y)$, and from Tonelli's theorem. Now, from Lemma~\ref{lemma:n14} and (\ref{eq:521}), we have
\begin{align}
    \lim_{K\rightarrow\infty} \sup_{T\geq1} \sum_{y\in\mathcal X_d} \bar p_T^\pi(y\mid x) \left( \hat v_{1,d}(y) - \hat v_{1,d}^{(K)}(y) \right) = 0.
\label{eq:522}
\end{align}
We next bound the truncation error under $\bar{\psi}_{\pi}^{x}$. For $y= (c,b,u_1,\cdots,u_d)\in\mathcal X_d$, let $c(y)=c$ denote the first coordinate of $y$. From Lemma~\ref{lemma:n13},
\begin{align}
    \sum_{y\in\mathcal X_d} \bar p_T^\pi(y\mid x)c(y)^2 &= \frac{1}{T} \sum_{t=1}^{T} \mathbb E_{\pi}\!\left[ \left( c_d^\pi(t) \right)^2\,\middle|\, \chi_{d}^{\pi}(1)=x \right]\\
    &\leq C_c(x).\label{eq:523}
\end{align}
From (\ref{eq:467}) and Fatou's lemma we have
\begin{align}
    \sum_{y\in\mathcal X_d} \bar{\psi}_{\pi}^{x}(y)c(y)^2&\leq \liminf_{T\rightarrow\infty} \sum_{y\in\mathcal X_d} \bar p_T^\pi(y\mid x)c(y)^2\\
    &\leq C_c(x).\label{eq:524}
\end{align}
Now, 
\begin{align}
    \hat v_{1,d}(y) - \hat v_{1,d}^{(K)}(y) &\leq \hat v_{1,d}(y) \bm{1}_{\{\hat v_{1,d}(y)>K \}}\\
    &\leq c(y) \bm{1}_{\{c(y)>K\}}\label{eq:n534}\\
    &\leq \frac{c(y)^2}{K}, \label{eq:n525}
\end{align}
where (\ref{eq:n534}) follows from (\ref{eq:n516}). Now,
\begin{align}
    \sum_{y\in\mathcal X_d} \bar{\psi}_{\pi}^{x}(y) \left( \hat v_{1,d}(y) - \hat v_{1,d}^{(K)}(y) \right)
    &\leq \frac{1}{K} \sum_{y\in\mathcal X_d} \bar{\psi}_{\pi}^{x}(y)c(y)^2\\
    &\leq \frac{C_c(x)}{K},\label{eq:537}
\end{align}
where (\ref{eq:537}) follows from (\ref{eq:524}). Consequently, we get
\begin{align}
    \lim_{K\rightarrow\infty} \sum_{y\in\mathcal X_d} \bar{\psi}_{\pi}^{x}(y) \left( \hat v_{1,d}(y) - \hat v_{1,d}^{(K)}(y) \right) = 0.
\label{eq:526}
\end{align}
For every $K>0$,
\begin{align}
    &\left| \sum_{y\in\mathcal X_d} \bar p_T^\pi(y\mid x) \hat v_{1,d}(y) - \sum_{y\in\mathcal X_d} \bar{\psi}_{\pi}^{x}(y) \hat v_{1,d}(y) \right| \leq \sum_{y\in\mathcal X_d} \bar p_T^\pi(y\mid x) \left( \hat v_{1,d}(y) -\hat v_{1,d}^{(K)}(y) \right)\nonumber\\
    &\quad+\left|\sum_{y\in\mathcal X_d}\bar p_T^\pi(y\mid x)\hat v_{1,d}^{(K)}(y)-\sum_{y\in\mathcal X_d} \bar{\psi}_{\pi}^{x}(y) \hat v_{1,d}^{(K)}(y)\right|+ \sum_{y\in\mathcal X_d} \bar{\psi}_{\pi}^{x}(y)\left(\hat v_{1,d}(y)-\hat v_{1,d}^{(K)}(y)\right).\label{eq:527}
\end{align}
Now, for a fixed $\epsilon>0$, from (\ref{eq:522}) and (\ref{eq:526}), we can choose $K_{1}>0$ such that the first and third terms on the right-hand side of (\ref{eq:527}) are each smaller than $\frac{\epsilon}{3}$, with the first bound holding uniformly in $T$. For a fixed $K\geq K_{1}$, from (\ref{eq:520}), there exists $T_{1}$, such that for all $T\geq T_{1}$, the middle term is smaller than $\frac{\epsilon}{3}$. Thus, we have
\begin{align}
    \lim_{T\rightarrow\infty} \sum_{y\in\mathcal X_d} \bar p_T^\pi(y\mid x) \hat v_{1,d}(y) = \sum_{y\in\mathcal X_d} \bar{\psi}_{\pi}^{x}(y) \hat v_{1,d}(y).
\label{eq:528}
\end{align}
Thus, from (\ref{eq:518}) and (\ref{eq:528}), we have (\ref{eq:515}).

From (\ref{eq:nn4}), (\ref{eq:nn5}), and (\ref{eq:nnn145}), we have $\nu_{k,d}(\pi;T)=\sum_{y\in\mathcal X_d}\bar p_T^\pi(y\mid x)a_k^\pi(y)$. Since $0\leq a_k^\pi(y)\leq1$, we obtain
\begin{align}
    \left|\nu_{k,d}(\pi;T) -\sum_{y\in\mathcal X_d}\bar\psi_\pi^x(y)a_k^\pi(y)\right| \leq \sum_{y\in\mathcal X_d} \left|\bar p_T^\pi(y\mid x)-\bar\psi_\pi^x(y)\right|.
\end{align}
Thus, from (\ref{eq:n468}), we have (\ref{eq:52}).
\end{Proof}

\section{Proof of Theorem~\ref{th:15}}\label{appen:th:15}
\begin{Proof}
We prove Theorem~\ref{th:15} in five steps.
\subsection{Step 1: Selection of reference states and the time required to reach them.}
Let $\bar{\mathcal R}_{\pi}$ denote the union of all recurrent classes of the finite embedded chain $\{\chi_d^\pi(\bar\tau_n)\}_{n\geq0}$. The proof of Lemma~\ref{lemma:n11} shows that every recurrent class $\mathcal R_\ell^\pi$ of the original chain contains at least one recurrent class of the embedded chain. For each $\ell\in\{1,\cdots,M_\pi\}$, select one state
\begin{align}
    r_\ell\in\mathcal R_\ell^\pi\cap \bar{\mathcal{R}}_{\pi}.
\end{align}
We first show that the embedded chain reaches $\bar{\mathcal R}_\pi$ in a finite expected number of embedded transitions. We define
\begin{align}
    N_{\bar{{R}}_{\pi}} = \inf\left\{n\geq0: \chi_d^\pi(\bar\tau_n)\in \bar{\mathcal{R}}_\pi \right\}.
\end{align}
If the embedded chain has no transient states, then $N_{\bar{{R}}_{\pi}}=0$ almost surely, and hence
\begin{align}
    \mathbb E_\pi\left[ N_{\bar{{R}}_{\pi}} \,\middle|\, \chi_d^\pi(1)=x \right] =0.
\end{align}

Making every state in $\bar{\mathcal R}_\pi$ absorbing gives an absorbing Markov chain with absorption time $N_{\bar R_\pi}$. From Lemma~\ref{lemma:n10}, we know that the state space of the embedded chain is finite. Thus, applying the mean absorption time results in \cite[Theorems 11.4 and 11.5]{grinstead_snell}, we obtain
\begin{align}\label{eq:782}
    \mathbb E_\pi\left[N_{\bar R_\pi}\,\middle|\,\chi_d^\pi(1)=x\right]<\infty.
\end{align}

From Tonelli's theorem, we have
\begin{align}
    \mathbb E_\pi\left[ \bar\tau_{N_{\bar{R}_{\pi}}} \!\,\middle|\,\! \chi_d^\pi(1)=x \right] &\!=\! \mathbb E_\pi\left[\bar\tau_0 \,\middle|\, \chi_d^\pi(1)=x\right] \!+\! \sum_{n=0}^{\infty} \mathbb E_\pi\left[ \bm 1_{\{N_{\bar{R}_{\pi}}>n\}} (\bar\tau_{n+1}-\bar\tau_n)
    \,\middle|\,\chi_d^\pi(1)=x \right] \\ 
    &\leq \mathbb E_\pi\left[ \bar\tau_0 \,\middle|\, \chi_d^\pi(1)=x \right] + C_1 \mathbb E_\pi\left[ N_{\bar{R}_{\pi}} \,\middle|\, \chi_d^\pi(1)=x \right]\label{eq:785} \\ 
    &< \infty,\label{eq:786}
\end{align}
where (\ref{eq:785}) follows from (\ref{eq:n449}), and the fact that $\{N_{\bar{R}_{\pi}}>n\}$ is measurable with respect to $\mathcal G_{\bar\tau_n}^\pi$ and (\ref{eq:786}) follows from Lemma~\ref{lemma:n8} and (\ref{eq:782}). 

We next show that, once the chain reaches a state $z\in\bar{\mathcal R}_\pi\cap\mathcal R_\ell^\pi$, it reaches $r_\ell$ in finite expected time. When the
chain is initialized at $z$, we define
\begin{align}
    T_{r_\ell} =\inf\left\{t\geq0:\chi_d^\pi(1+t)=r_\ell\right\}.
\end{align}
From Lemma~\ref{lemma:n11}, we know that $\mathcal R_\ell^\pi$ is positive recurrent. The chain restricted to $\mathcal R_\ell^\pi$ is countable and irreducible. Thus, from \cite[Proposition 21.11]{levin_peres}, we have
\begin{align}\label{eq:800}
    \mathbb E_\pi\left[ T_{r_\ell} \,\middle|\, \chi_d^\pi(1)=z \right]<\infty.
\end{align}
Since $\bar{\mathcal R}_\pi$ is finite, from (\ref{eq:800}), we obtain
\begin{align}\label{eq:801}
    \max_{\substack{ 1\leq\ell\leq M_\pi\\
    z\in\bar{\mathcal R}_\pi\cap\mathcal R_\ell^\pi}} \mathbb E_\pi\left[ T_{r_\ell} \,\middle|\, \chi_d^\pi(1)=z \right]<\infty.
\end{align}
We define 
\begin{align}\label{eq:803}
    H = \inf\left\{ t\geq1: \chi_d^\pi(t)\in \{r_1,\cdots,r_{M_\pi}\} \right\}.
\end{align}
Note that, $\{N_{\bar{R}_{\pi}}=n\}\in \mathcal{G}_{\bar{\tau}_{n}}^{\pi}$ for every $n$, thus for every integer $m\geq 1$,
\begin{align}
    \{\bar\tau_{N_{\bar{R}_{\pi}}}\leq m\} = \bigcup_{n=0}^{\infty} \left( \{N_{\bar{R}_{\pi}}=n\}\cap\{\bar\tau_n\leq m\} \right) \in \mathcal G_m^\pi.
\end{align}
Thus, the random time $\bar\tau_{N_{\bar R_\pi}}$ is a stopping time with respect to $\{\mathcal G_t^\pi\}_{t\geq1}$. Now, note that the following pathwise relation holds
\begin{align}
    H\leq \bar{\tau}_{N_{\bar{R}_{\pi}}} + \sum_{\ell=1}^{M_{\pi}}\sum_{z\in \mathcal R_\ell^\pi\cap \bar{\mathcal{R}}_{\pi}}\bm{1}_{\left\{\chi_{d}^{\pi}\left(\bar{\tau}_{N_{\bar{R}_{\pi}}}\right)=z\right\}} \inf\{t\geq 0 : \chi_{d}^{\pi}(\bar{\tau}_{N_{\bar{R}_{\pi}}}+t) = r_{\ell}\}.
\end{align}
From (\ref{eq:786}), we have $\bar\tau_{N_{\bar{R}_{\pi}}}<\infty$ almost surely. Now, applying strong Markov property at $\bar{\tau}_{N_{\bar{R}_{\pi}}}$and from the time-homogeneity of the Markov chain $\{\chi_{d}^{\pi}(t)\}_{t\geq 1}$, we get
\begin{align}
    \mathbb{E}_{\pi} \left[H\,\middle|\chi_{d}^{\pi}(1)=x\,\right] \leq& \mathbb{E}_{\pi} \left[\bar{\tau}_{N_{\bar{R}_{\pi}}} \,\middle|\chi_{d}^{\pi}(1)=x\,\right] + \sum_{\ell=1}^{M_{\pi}} \sum_{z\in \mathcal R_\ell^\pi\cap \bar{\mathcal{R}}_{\pi}} \mathbb{E}_{\pi}\Big[ \bm{1}_{\left\{\chi_{d}^{\pi}\left(\bar{\tau}_{N_{\bar{R}_{\pi}}}\right)=z\right\}} \nonumber\\&\ \ \mathbb{E}_{\pi}\left[ \inf\{t\geq 0: \chi_{d}^{\pi}(\bar{\tau}_{N_{\bar{R}_{\pi}}}+t) = r_{\ell}\}\,\middle|\mathcal{G}_{\bar{\tau}_{N_{\bar{R}_{\pi}}}}\,\right]\Big |\chi_{d}^{\pi}(1)=x\,\Big]\label{eq:805} \\ 
    = & \mathbb{E}_{\pi} \left[\bar{\tau}_{N_{\bar{R}_{\pi}}} \,\middle|\chi_{d}^{\pi}(1)=x\,\right] + \sum_{\ell=1}^{M_{\pi}} \sum_{z\in \mathcal R_\ell^\pi\cap \bar{\mathcal{R}}_{\pi}} \mathbb{E}_{\pi}\Big[ \bm{1}_{\left\{\chi_{d}^{\pi}\left(\bar{\tau}_{N_{\bar{R}_{\pi}}}\right)=z\right\}} \nonumber\\&\ \ \mathbb{E}_{\pi}\left[ \inf\{t\geq 0: \chi_{d}^{\pi}(\bar{\tau}_{N_{\bar{R}_{\pi}}}\!\!+\!t) = r_{\ell}\}\,\middle|\chi_{d}^{\pi}\left(\bar{\tau}_{N_{\bar{R}_{\pi}}}\right)=z\,\right]\Big |\chi_{d}^{\pi}(1)=x\,\Big] \label{eq:806} \\ 
    =& \mathbb{E}_{\pi} \left[\bar{\tau}_{N_{\bar{R}_{\pi}}} \,\middle|\chi_{d}^{\pi}(1)=x\,\right] + \sum_{\ell=1}^{M_{\pi}} \sum_{z\in \mathcal R_\ell^\pi\cap \bar{\mathcal{R}}_{\pi}} \mathbb{E}_{\pi}\Big[ \bm{1}_{\left\{\chi_{d}^{\pi}\left(\bar{\tau}_{N_{\bar{R}_{\pi}}}\right)=z\right\}} \nonumber\\&\ \ \mathbb{E}_{\pi}\left[T_{r_{\ell}}\,\middle|\chi_{d}^{\pi}(1)=z\,\right] \Big| \chi_{d}^{\pi}(1)=x\Big] \label{eq:807} \\ =&\mathbb{E}_{\pi} \left[\bar{\tau}_{N_{\bar{R}_{\pi}}} \,\middle|\chi_{d}^{\pi}(1)=x\,\right] + \sum_{\ell=1}^{M_{\pi}} \sum_{z\in \mathcal R_\ell^\pi\cap \bar{\mathcal{R}}_{\pi}} \mathbb{P}\left(\chi_{d}^{\pi}\left(\bar{\tau}_{N_{\bar{R}_{\pi}}}\right)=z\,\middle|\chi_{d}^{\pi}(1)=x\,\right) \nonumber\\& \ \ \mathbb{E}_{\pi}\left[T_{r_{\ell}}\,\middle|\chi_{d}^{\pi}(1)=z\,\right] \\ 
    \leq&\mathbb{E}_{\pi} \left[\bar{\tau}_{N_{\bar{R}_{\pi}}} \,\middle|\chi_{d}^{\pi}(1)=x\,\right] +  \max_{ 1\leq\ell\leq M_\pi,  \  z\in\bar{\mathcal{R}}_\pi\cap\mathcal R_\ell^\pi} \mathbb E_\pi\left[ T_{r_\ell} \,\middle|\, \chi_d^\pi(1)=z \right],
\end{align}
where (\ref{eq:805}) follows from the fact that the event $\{\chi_{d}^{\pi}\left(\bar{\tau}_{N_{\bar{R}_{\pi}}}\right)=z\}$ is measurable with respect to the stopped sigma algebra $\mathcal{G}_{\bar{\tau}_{N_{\bar{R}_{\pi}}}}^{\pi}$, (\ref{eq:806}) and (\ref{eq:807}) follow from the strong Markov property of $\{\chi_{d}^{\pi}(t)\}_{t\geq 1}$ at $\bar{\tau}_{N_{\bar{R}_{\pi}}}$ and the time homogeneity of $\{\chi_{d}^{\pi}(t)\}_{t\geq 1}$, respectively. Now, from (\ref{eq:786}) and (\ref{eq:801}), we get
\begin{align}\label{eq:811}
     \mathbb{E}_{\pi} \left[H\,\middle|\chi_{d}^{\pi}(1)=x\,\right] < \infty. 
\end{align}
\subsection{Step 2: Second moment of the return time to each reference state is finite.}
Let us consider a fixed $\ell\in\{1,2,\cdots,M_\pi\}$. We define
\begin{align}\label{eq:n787}
    \bar T_{r_\ell}^{+} =\inf\left\{t\geq1: \chi_d^\pi(1+t)=r_\ell\right\}.
\end{align}
Let
\begin{align}\label{eq:812}
    \mathcal A_\ell =\mathcal R_\ell^\pi\cap\mathcal C_{W,R,d}.
\end{align}
The set $\mathcal A_\ell$ is finite and nonempty, since $\mathcal C_{W,R,d}$ is finite and $r_\ell\in\mathcal A_\ell$.

When the chain is initialized at any $z\in\mathcal A_\ell$, we define
\begin{align}
    T_{\mathcal A_\ell}^{+} =\inf\left\{t\geq1: \chi_d^\pi(1+t)\in\mathcal A_\ell\right\}.
\end{align}
From Lemma~\ref{lemma:n9}, we know that $\chi_d^\pi(1+\bar\tau(1))$ belongs to $\mathcal C_{W,R,d}$ almost surely, and since $\mathcal R_\ell^\pi$ is closed, this state also belongs to $\mathcal R_\ell^\pi$. Thus, $T_{\mathcal A_\ell}^{+}\leq\bar\tau(1)$, almost surely. Consequently, from Lemma~\ref{lemma:n8}, for every $z\in\mathcal A_\ell$, we have
\begin{align}\label{eq:817}
    \mathbb E_\pi\left[ \left(T_{\mathcal A_\ell}^{+}\right)^2 \,\middle|\, \chi_d^\pi(1)=z \right] &\leq \mathbb E_\pi\left[ \bar\tau(1)^2 \,\middle|\,
    \chi_d^\pi(1)=z \right]\\
    &\leq C_2<\infty.
\end{align}
The chain restricted to $\mathcal R_\ell^\pi$ is countable, time-homogeneous, and irreducible. Applying \cite[Lemma 2.6.1]{menshikov2017non} to this chain with the finite set $\mathcal A_\ell$ and second moment, we obtain
\begin{align}\label{eq:827}
    \mathbb E_\pi\left[ \left(\bar T_{r_\ell}^{+}\right)^2 \,\middle|\, \chi_d^\pi(1)=r_\ell  \right]<\infty.
\end{align}

\subsection{Step 3: Convergence when the chain is initialized by $r_\ell$.}
Let us assume that $\chi_d^\pi(1)=r_\ell$. Let $S_k$ denote the elapsed time until the $k$th return to $r_\ell$. We initialize $S_0=0$, thus for every $k\geq1$,
\begin{align}
    S_k = \inf\left\{ t>S_{k-1}: \chi_d^\pi(1+t)=r_\ell \right\}.
\end{align}
Thus, $S_k+1$ is the slot of the $k$th positive return to $r_\ell$. We denote the length of the $k$th return cycle and the occupation of state $y$ during that cycle by
\begin{align}
    \bar T_{k,r_\ell}^{+} &= S_k-S_{k-1},\label{eq:920}\\
    O_k(y) &= \sum_{t=S_{k-1}+1}^{S_k} \bm1_{\{\chi_d^\pi(t)=y\}}. \label{eq:612}
\end{align}
By the strong Markov property at the return slots $S_k+1$, the sequence
\begin{align}
    \left\{ \left( \bar T_{k,r_\ell}^{+}, \{O_k(y):y\in\mathcal X_d\} \right) \right\}_{k\geq1}
\end{align}
is i.i.d. Note that, $\bar{T}_{1,r_{\ell}}^{+}= \bar{T}_{r_{\ell}}^{+}$. Since $r_{\ell}$ is a positive recurrent state, we have
\begin{align}
    \mathbb E_\pi\left[ \bar T_{1,r_\ell}^{+} \,\middle|\, \chi_d^\pi(1)=r_\ell \right] <\infty.
\end{align}
Moreover, $0\leq O_1(y)\leq \bar T_{1,r_\ell}^{+}$, and hence
\begin{align}
    \mathbb E_\pi\left[ O_1(y) \,\middle|\, \chi_d^\pi(1)=r_\ell \right] <\infty.
\end{align}
Therefore, the renewal-reward theorem gives
\begin{align}
    \lim_{T\rightarrow\infty} \frac{1}{T} \sum_{t=1}^{T} \bm1_{\{\chi_d^\pi(t)=y\}} = \frac{ \mathbb E_\pi\left[ O_1(y) \,\middle|\, \chi_d^\pi(1)=r_\ell
    \right] }{ \mathbb E_\pi\left[ \bar T_{1,r_\ell}^{+}\,\middle|\, \chi_d^\pi(1)=r_\ell \right]},
\end{align}
almost surely. Since $\mathcal R_\ell^\pi$ is an irreducible positive recurrent class, the Markov-chain ergodic theorem gives
\begin{align}
    \lim_{T\rightarrow\infty} \frac{1}{T} \sum_{t=1}^{T} \bm1_{\{\chi_d^\pi(t)=y\}} = \psi_\ell^\pi(y),
\end{align}
almost surely. Thus, we get
\begin{align}
    \mathbb E_\pi\left[ O_1(y) \,\middle|\, \chi_d^\pi(1)=r_\ell \right] = \mathbb E_\pi\left[ \bar T_{1,r_\ell}^{+} \,\middle|\, \chi_d^\pi(1)=r_\ell \right]
    \psi_\ell^\pi(y). \label{eq:613}
\end{align}
We define the index of the cycle containing slot $T$ by
\begin{align}
    \eta_{T} = \inf\left\{ k\geq1: S_{k}> T-1 \right\}.
\end{align}
Since every cycle has length at least one, $\eta_{T}\leq T$. Furthermore,
\begin{align}
    \{\eta_{T}\geq k\} = \{S_{k-1}<T\}.
\end{align}
Since $\eta_T$ is a bounded stopping time with respect to the filtration 
\begin{align}
    \left\{ \sigma\left( \left(\bar T_{j,r_\ell}^{+},  \{O_j(z):z\in\mathcal X_d\} \right): 1\leq j\leq k \right) \right\}_{k\geq0},
\end{align} 
Wald's identity gives, for every $y\in\mathcal X_d$,
\begin{align}
    \mathbb E_\pi\left[ \sum_{k=1}^{\eta_T}O_k(y) \,\middle|\, \chi_d^\pi(1)=r_\ell \right] &= \mathbb E_\pi\left[ \eta_T \,\middle|\, \chi_d^\pi(1)=r_\ell \right] \mathbb E_\pi\left[ O_1(y) \,\middle|\, \chi_d^\pi(1)=r_\ell \right], \label{eq:840} \\
    \mathbb E_\pi\left[ S_{\eta_T} \,\middle|\, \chi_d^\pi(1)=r_\ell \right]  &= \mathbb E_\pi\left[ \sum_{k=1}^{\eta_T}\bar T_{k,r_\ell}^{+} \,\middle|\, \chi_d^\pi(1)=r_\ell \right]\\
    &= \mathbb E_\pi\left[ \eta_T \,\middle|\, \chi_d^\pi(1)=r_\ell \right] \mathbb E_\pi\left[\bar T_{1,r_\ell}^{+} \,\middle|\, \chi_d^\pi(1)=r_\ell \right].
    \label{eq:842}
\end{align}
Note that, the following pathwise identity holds
\begin{align}
    \sum_{k=1}^{\eta_{T}}O_{k}(y)  = \sum_{t=1}^{S_{\eta_{T}}} \bm 1_{\{\chi_d^\pi(t)=y\}}.
\end{align}
Now, from (\ref{eq:613}), (\ref{eq:840}), and (\ref{eq:842}), we get
\begin{align}\label{eq:844}
    \mathbb E_\pi\left[\sum_{t=1}^{S_{\eta_{T}}} \bm 1_{\{\chi_d^\pi(t)=y\}} \,\middle|\, \chi_d^\pi(1)=r_\ell \right] &= \mathbb E_\pi\left[S_{\eta_{T}} \,\middle|\, \chi_d^\pi(1)=r_\ell \right] \psi_\ell^\pi(y).
\end{align}
Since $S_{\eta_{T}}\geq T$, the following pathwise identity holds
\begin{align}
    \sum_{t=1}^{T} \bm 1_{\{\chi_d^\pi(t)=y\}} &= \sum_{t=1}^{S_{\eta_{T}}} \bm 1_{\{\chi_d^\pi(t)=y\}} - \sum_{t=T+1}^{S_{\eta_{T}}} \bm 1_{\{\chi_d^\pi(t)=y\}}.
\end{align}
Thus, from (\ref{eq:844}), we have
\begin{align}
    \sum_{t=1}^{T} \mathbb P\left(\chi_d^\pi(t)=y \,\middle|\, \chi_d^\pi(1)=r_\ell \right) - T\psi_\ell^\pi(y) =& \left(\mathbb E_\pi\left[S_{\eta_{T}} \,\middle|\, \chi_d^\pi(1)=r_\ell \right] -T \right) \psi_\ell^\pi(y)\nonumber\\&-\mathbb E_\pi\left[\sum_{t=T+1}^{S_{\eta_{T}}}\bm 1_{\{\chi_d^\pi(t)=y\}}
    \,\middle|\,\chi_d^\pi(1)=r_\ell\right].
\end{align}
Thus,
\begin{align}
    \sum_{y\in\mathcal X_d}\left|\sum_{t=1}^{T}\mathbb P\left(\chi_d^\pi(t)=y\,\middle|\,\chi_d^\pi(1)=r_\ell\right)- T\psi_\ell^\pi(y) \right| \leq & \left( \mathbb E_\pi\left[S_{\eta_{T}}\,\middle|\,\chi_d^\pi(1)=r_\ell\right]-T\right)\sum_{y\in\mathcal X_d}\psi_\ell^\pi(y)\nonumber\\&+ \mathbb E_\pi\left[\sum_{t=T+1}^{S_{\eta_{T}}} \sum_{y\in\mathcal X_d} \bm 1_{\{\chi_d^\pi(t)=y\}} \,\middle|\, \chi_d^\pi(1)=r_\ell\right] \\ 
    \leq & 2\mathbb E_\pi\left[S_{\eta_{T}}-T\,\middle|\,\chi_d^\pi(1)=r_\ell\right].\label{eq:848}
\end{align}
The sequence $\{\bar T_{k,r_\ell}^{+}\}_{k\geq1}$ consists of i.i.d.~integer-valued random variables, and
\begin{align}
    S_k=\sum_{j=1}^{k}\bar T_{j,r_\ell}^{+}.
\end{align}
Moreover, from (\ref{eq:827}), we have
\begin{align}
    \mathbb E_\pi\left[\left(\bar T_{1,r_\ell}^{+}\right)^2\,\middle|\,\chi_d^\pi(1)=r_\ell\right]<\infty.
\end{align}
Applying Lorden's inequality \cite{lorden1970excess} at level $T-1$ gives
\begin{align}\label{eq:851}
    \mathbb E_\pi\left[S_{\eta_T}-(T-1)\,\middle|\,\chi_d^\pi(1)=r_\ell\right]\leq\frac{ \mathbb E_\pi\left[\left(\bar T_{1,r_\ell}^{+}\right)^2 \,\middle|\,
    \chi_d^\pi(1)=r_\ell\right]}{\mathbb E_\pi\left[\bar T_{1,r_\ell}^{+}\,\middle|\,\chi_d^\pi(1)=r_\ell\right]}.
\end{align}
Combining (\ref{eq:848}) with (\ref{eq:851}) gives
\begin{align}\label{eq:852}
    \sum_{y\in\mathcal X_d}\left|\bar p_T^\pi(y\mid r_\ell)-\psi_\ell^\pi(y)\right|\leq\frac{2}{T}\left(\frac{\mathbb E_\pi\left[\left(\bar T_{1,r_\ell}^{+}\right)^2\,\middle|\,\chi_d^\pi(1)=r_\ell\right]}{\mathbb E_\pi\left[\bar T_{1,r_\ell}^{+}\,\middle|\,\chi_d^\pi(1)=r_\ell\right]}-1\right).
\end{align}

\subsection{Proof of (\ref{eq:600}).}
Recall the hitting time $H$ in (\ref{eq:803}), from (\ref{eq:811}), we see that $H$ is finite almost surely.  For $1\leq\ell\leq M_\pi$ and $m\geq1$, we define
\begin{align}
    A_{\ell,m}&=\{H=m,\chi_d^\pi(H)=r_\ell\},\\
    B_{\ell,T}&=  \{H>T,\chi_d^\pi(H)=r_\ell\}.
\end{align} 
Note that 
\begin{align}
    \mathbb P\left(\chi_d^\pi(H)=r_\ell\,\middle|\,\chi_d^\pi(1)=x\right)=\hat\psi_\ell^\pi(x).
\end{align}
Consequently from the definition of $\bar{\psi}_{\pi}^{x}(y)$ in (\ref{eq:143}), we get
\begin{align}
    \bar\psi_\pi^x(y)=\sum_{\ell=1}^{M_\pi}\left[\sum_{m=1}^{T}\mathbb P\left(A_{\ell,m}\,\middle|\,\chi_d^\pi(1)=x\right)+\mathbb P\left(B_{\ell,T}\,\middle|\,\chi_d^\pi(1)=x\right)\right]\psi_\ell^\pi(y).\label{eq:mixture-decomposition}
\end{align}
Similarly,
\begin{align}
    \sum_{t=1}^{T}p_t^\pi(y\mid x) =& \sum_{\ell=1}^{M_\pi}\sum_{m=1}^{T}\mathbb E_\pi\left[\bm1_{A_{\ell,m}}\sum_{t=1}^{T}\bm1_{\{\chi_d^\pi(t)=y\}}
    \,\middle|\,\chi_d^\pi(1)=x\right]\nonumber\\&+\sum_{\ell=1}^{M_\pi}\mathbb E_\pi\left[\bm1_{B_{\ell,T}}\sum_{t=1}^{T}\bm1_{\{\chi_d^\pi(t)=y\}} \,\middle|\,\chi_d^\pi(1)=x\right].
\end{align}
Thus, we have
\begin{align}\label{eq:858}
    \sum_{t=1}^{T}p_t^\pi(y\mid x)- T\bar\psi_\pi^x(y) =&\sum_{\ell=1}^{M_\pi}\sum_{m=1}^{T}\mathbb E_\pi\left[\bm1_{A_{\ell,m}}\left(\sum_{t=1}^{T} \bm1_{\{\chi_d^\pi(t)=y\}}-T\psi_\ell^\pi(y)\right)\,\middle|\,\chi_d^\pi(1)=x\right]\nonumber\\&+\sum_{\ell=1}^{M_\pi}\mathbb E_\pi\left[\bm1_{B_{\ell,T}}
    \left( \sum_{t=1}^{T}\bm1_{\{\chi_d^\pi(t)=y\}}-T\psi_\ell^\pi(y)\right)\,\middle|\,\chi_d^\pi(1)=x\right].
\end{align}
Now,
\begin{align}
    &\sum_{y\in\mathcal X_d}\left|\mathbb E_\pi\left[\bm1_{A_{\ell,m}}\left(\sum_{t=1}^{m-1}\bm1_{\{\chi_d^\pi(t)=y\}} -(m-1)\psi_\ell^\pi(y)\right)
    \,\middle|\,\chi_d^\pi(1)=x\right]\right|\nonumber\\&\leq \mathbb E_\pi\left[\bm1_{A_{\ell,m}}\sum_{y\in\mathcal X_d}\left|\sum_{t=1}^{m-1} \bm1_{\{\chi_d^\pi(t)=y\}}-(m-1)\psi_\ell^\pi(y)\right|\,\middle|\,\chi_d^\pi(1)=x\right]\\
    &\leq 2(m-1)\mathbb P\left(A_{\ell,m} \,\middle|\,\chi_d^\pi(1)=x\right).\label{eq:860}
\end{align}
From the strong Markov property at $H$, the following relation holds
\begin{align}
    &\mathbb E_\pi\left[\bm1_{A_{\ell,m}}\left(\sum_{t=m}^{T}\bm1_{\{\chi_d^\pi(t)=y\}}-(T-m+1)\psi_\ell^\pi(y)\right)\,\middle|\,\chi_d^\pi(1)=x\right]
    \nonumber\\&=\mathbb P\left(A_{\ell,m}\,\middle|\,\chi_d^\pi(1)=x\right)\left(\sum_{s=1}^{T-m+1} p_s^\pi(y\mid r_\ell) - (T-m+1)\psi_\ell^\pi(y) \right).
\end{align}
Now, applying (\ref{eq:852}), we obtain
\begin{align}
    &\sum_{y\in\mathcal X_d} \left| \mathbb E_\pi\left[ \bm1_{A_{\ell,m}} \left( \sum_{t=m}^{T} \bm1_{\{\chi_d^\pi(t)=y\}} - (T-m+1)\psi_\ell^\pi(y) \right)\,\middle|\,\chi_d^\pi(1)=x\right]\right|\nonumber\\&\leq 2\mathbb P\left(A_{\ell,m}\,\middle|\,\chi_d^\pi(1)=x\right)\left(\frac{\mathbb E_\pi\left[\left(\bar T_{1,r_\ell}^{+}\right)^2\,\middle|\,\chi_d^\pi(1)=r_\ell\right]}{\mathbb E_\pi\left[\bar T_{1,r_\ell}^{+}\,\middle|\, \chi_d^\pi(1)=r_\ell\right]}-1\right).\label{eq:862}
\end{align}
Thus, combining (\ref{eq:860}) and (\ref{eq:862}), we get
\begin{align}\label{eq:863}
    &\sum_{y\in\mathcal X_d}\left|\mathbb E_\pi\left[\bm1_{A_{\ell,m}}\left(\sum_{t=1}^{T}\bm1_{\{\chi_d^\pi(t)=y\}}-T\psi_\ell^\pi(y)\right)\,\middle|\, \chi_d^\pi(1)=x\right]\right|\nonumber\\&\leq 2\mathbb P\left(A_{\ell,m}\,\middle|\,\chi_d^\pi(1)=x\right) \left[m-1 + \frac{\mathbb E_\pi\left[ \left(\bar T_{1,r_\ell}^{+}\right)^2 \,\middle|\,\chi_d^\pi(1)=r_\ell \right]}{\mathbb E_\pi\left[\bar T_{1,r_\ell}^{+}\,\middle|\,\chi_d^\pi(1)=r_\ell
    \right]}-1 \right].
\end{align}
Similarly, 
\begin{align}
    &\sum_{y\in\mathcal X_d} \left|\mathbb E_\pi\left[\bm1_{B_{\ell,T}}\left(\sum_{t=1}^{T}\bm1_{\{\chi_d^\pi(t)=y\}} - T\psi_\ell^\pi(y) \right) \,\middle|\, \chi_d^\pi(1)=x \right]\right|\nonumber\\&\leq\mathbb E_\pi\left[\bm1_{B_{\ell,T}}\left(\sum_{t=1}^{T} \sum_{y\in\mathcal X_d} \bm1_{\{\chi_d^\pi(t)=y\}}+ T\sum_{y\in\mathcal X_d}\psi_\ell^\pi(y)\right)\,\middle|\,\chi_d^\pi(1)=x\right],\\
    &=2T\mathbb P\left(B_{\ell,T}\,\middle|\,\chi_d^\pi(1)=x\right).\label{eq:865}
\end{align}
From (\ref{eq:858}), (\ref{eq:863}) and (\ref{eq:865}), we get
\begin{align}
    &\sum_{y\in\mathcal X_d} \left|\sum_{t=1}^{T}p_t^\pi(y\mid x)-T\bar\psi_\pi^x(y)\right|\leq 2\sum_{\ell=1}^{M_\pi}\sum_{m=1}^{T} \mathbb P\left(
    A_{\ell,m}\,\middle|\,\chi_d^\pi(1)=x\right)\nonumber\\&\left[m-1 +\frac{\mathbb E_\pi\left[ \left(\bar T_{1,r_\ell}^{+}\right)^2 \,\middle|\, \chi_d^\pi(1)=r_\ell\right]}{\mathbb E_\pi\left[\bar T_{1,r_\ell}^{+}\,\middle|\,\chi_d^\pi(1)=r_\ell\right]}-1\right]+2T\sum_{\ell=1}^{M_\pi} \mathbb P\left(B_{\ell,T}\,\middle|\,\chi_d^\pi(1)=x\right).\label{eq:866}
\end{align}
Note that
\begin{align}
    &\sum_{\ell=1}^{M_\pi}\sum_{m=1}^{T}(m-1) \mathbb P\left(A_{\ell,m}\,\middle|\,\chi_d^\pi(1)=x\right) + T\sum_{\ell=1}^{M_\pi} \mathbb P\left( B_{\ell,T}
    \,\middle|\,\chi_d^\pi(1)=x\right)\nonumber\\ &\leq \mathbb E_\pi\left[H\bm1_{\{H\leq T\}}\,\middle|\, \chi_d^\pi(1)=x\right] + \mathbb E_\pi\left[ H\bm1_{\{H>T\}}\,\middle|\,\chi_d^\pi(1)=x\right]\\
    &=\mathbb E_\pi\left[H\,\middle|\,\chi_d^\pi(1)=x\right].\label{eq:868}
\end{align}
Moreover, for every $\ell$,
\begin{align}
    \sum_{m=1}^{T}\mathbb P\left(A_{\ell,m}\,\middle|\,\chi_d^\pi(1)=x\right)&\leq\mathbb P\left(\chi_d^\pi(H)=r_\ell \,\middle|\, \chi_d^\pi(1)=x \right)\\
    &= \hat\psi_\ell^\pi(x).\label{eq:869}
\end{align}
Thus, from (\ref{eq:868}), (\ref{eq:869}), we have
\begin{align}\label{eq:871}
    \sum_{y\in\mathcal X_d}\left|\sum_{t=1}^{T} p_t^\pi(y\mid x) - T\bar\psi_\pi^x(y) \right| \leq&  2\mathbb E_\pi\left[H \,\middle|\, \chi_d^\pi(1)=x
    \right]\nonumber\\&+ 2\sum_{\ell=1}^{M_\pi} \hat\psi_\ell^\pi(x) \left( \frac{ \mathbb E_\pi\left[ \left(\bar T_{1,r_\ell}^{+}\right)^2 \,\middle|\,
    \chi_d^\pi(1)=r_\ell \right]}{\mathbb E_\pi\left[\bar T_{1,r_\ell}^{+} \,\middle|\, \chi_d^\pi(1)=r_\ell \right]} -1 \right).
\end{align}
Now, we define
\begin{align}\label{eq:1025}
    C_{\pi}(x) = 2\mathbb E_\pi\left[H\,\middle|\,\chi_d^\pi(1)=x\right]+ 2\sum_{\ell=1}^{M_\pi} \hat\psi_\ell^\pi(x)\left(\frac{\mathbb E_\pi\left[\left(\bar T_{1,r_\ell}^{+}\right)^2\,\middle|\,\chi_d^\pi(1)=r_\ell\right]}{\mathbb E_\pi\left[\bar T_{1,r_\ell}^{+}\,\middle|\,\chi_d^\pi(1)=r_\ell\right]}-1 \right).
\end{align}
From (\ref{eq:811}), (\ref{eq:827}), and from Lemma~\ref{lemma:n11}, we see that $C_{\pi}(x)<\infty$. Thus, from (\ref{eq:871}), we have (\ref{eq:600}).
\subsection{Proof of (\ref{eq:601}).}
Recall the decomposition in (\ref{eq:527}),
\begin{align}
    \left|\Delta_d(\pi;T)-\Delta_d(\pi)\right|\leq&\sum_{y\in\mathcal X_d}\bar p_T^\pi(y\mid x)\left(\hat v_{1,d}(y)-\hat v_{1,d}^{(K)}(y)\right)+
    \left|\sum_{y\in\mathcal X_d}\left(\bar p_T^\pi(y\mid x)-\bar\psi_\pi^x(y)\right)\hat v_{1,d}^{(K)}(y)\right|\nonumber\\&+\sum_{y\in\mathcal X_d} \bar\psi_\pi^x(y) \left(\hat v_{1,d}(y)-\hat v_{1,d}^{(K)}(y)\right).\label{eq:619}
\end{align}
By (\ref{eq:746}) and (\ref{eq:521}),
\begin{align}\label{eq:874}
    \sum_{y\in\mathcal X_d}\bar p_T^\pi(y\mid x)\left(\hat v_{1,d}(y)-\hat v_{1,d}^{(K)}(y)\right)\leq\frac{C_c(x)}{K}.
\end{align}
From (\ref {eq:537}) 
\begin{align}\label{eq:875}
    \sum_{y\in\mathcal X_d} \bar{\psi}_{\pi}^{x}(y) \left(\hat v_{1,d}(y) - \hat v_{1,d}^{(K)}(y)\right) \leq \frac{C_c(x)}{K}.
\end{align}
Finally, because $0\leq\hat v_{1,d}^{(K)}(y)\leq K$, (\ref{eq:600}) gives
\begin{align}\label{eq:876}
    \left| \sum_{y\in\mathcal X_d} \left(\bar p_T^\pi(y\mid x)-\bar\psi_\pi^x(y)\right)\hat v_{1,d}^{(K)}(y)\right|\leq \frac{KC_\pi(x)}{T}.
\end{align}
Substituting (\ref{eq:874}), (\ref{eq:875}) and (\ref{eq:876}) into (\ref{eq:619}), gives 
\begin{align}\label{eq:877}
    \left|\Delta_d(\pi;T)-\Delta_d(\pi)\right|\leq\frac{2C_c(x)}{K}+\frac{KC_\pi(x)}{T}.
\end{align}
Minimizing the right-hand side over $K>0$ gives
\begin{align}
    K=\sqrt{\frac{2C_c(x)T}{C_\pi(x)}},
\end{align}
and substituting it into (\ref{eq:877}) proves (\ref{eq:601}).
\end{Proof}

\section{Proof of Theorem~\ref{th:16}}\label{appen:th:16}
\begin{Proof}
From the proof of Lemma~\ref{lemma:n13}, we have
\begin{align}
    \mathbb P\left((c_d^\pi(t)-c_0)^+>nB_d\,\middle|\,\chi_d^\pi(1)=x\right)\leq(1-q)^n,\quad n\geq0,\label{eq:nn173}
\end{align}
where $c_0=\max\{c(x),R+d\}$.  Now, from the tail-sum representation of expectation, we have
\begin{align}
    \mathbb E_\pi\left[(c_d^\pi(t)-c_0)^+\,\middle|\,\chi_d^\pi(1)=x\right]\leq B_d\sum_{n=0}^\infty(1-q)^n=\frac{B_d}{q}.
\end{align}
Now, from the identity in (\ref{eq:759}) in Appendix~\ref{appen:th:13}, we have
\begin{align}
    \mathbb E_\pi\left[((c_{d}^{\pi}(t)-c_0)^+)^2\,\middle|\, \chi_{d}^\pi(1)=x\right] \leq B_{d}^2 \sum_{n=0}^{\infty} (2n+1)(1-q)^n=\frac{B_d^2(2-q)}{q^2}.
\end{align}
As $c_d^\pi(t)\leq c_0+(c_d^\pi(t)-c_0)^+$, we get
\begin{align}\label{eq:nn176}
    \sup_{t\geq1}\mathbb E_\pi\left[(c_d^\pi(t))^2\,\middle|\,\chi_d^\pi(1)=x\right]\leq c_0^2+\frac{2c_0B_d}{q}+\frac{B_d^2(2-q)}{q^2}.
\end{align}
Now, from the statement of Lemma~\ref{lemma:n13}, we see that the right-sided quantity of (\ref{eq:nn176}) is a valid choice for $C_{c}(x)$. 

The proof of Lemma~\ref{lemma:n11} establishes that every recurrent class of the Markov chain $\{\chi_{d}^{\pi}(t)\}_{t\geq 1}$ contains at least one recurrent class of the embedded chain. Now, we show that every recurrent class of the original chain only contains one recurrent class of the embedded chain. We show this by contradiction. Let us consider that the recurrent class $\mathcal R_{\ell} ^\pi$ contains two distinct embedded recurrent classes $\mathcal{R}_{1}^{\mathrm{sam}}$ and $\mathcal{R}_{2}^{\mathrm{sam}}$. Since each $\mathcal R_i^{\mathrm{sam}}$ is closed for the embedded chain
\begin{align}
    \sum_{y\in\mathcal R_i^{\mathrm{sam}}} \tilde P_\pi(z,y)=1, \quad z\in\mathcal R_i^{\mathrm{sam}}, \quad i\in\{1,2\}. \label{eq:nnn177}
\end{align}
Let us assume that $\chi_d^\pi(1)=x$, where  $x\in\mathcal R_\ell^\pi$. We define 
\begin{align}
    N_{\mathrm{sam},1}&=\inf\left\{t\geq1:\chi_d^\pi(t)\in\mathcal R_1^{\mathrm{sam}}\right\},\label{eq:nnn178}\\
    N_{\mathrm{sam},2}&=\inf\left\{t\geq N_{\mathrm{sam},1}+\bar\tau(N_{\mathrm{sam},1}):\chi_d^\pi(t)\in\mathcal R_2^{\mathrm{sam}}\right\}.\label{eq:nnn179}
\end{align}
Note that, these are stopping times with respect to $\{\mathcal G_t^\pi\}_{t\geq1}$. Irreducibility and recurrence of $\mathcal R_\ell^\pi$ imply that the chain visits every state in each $\mathcal R_i^{\mathrm{sam}}$ infinitely often almost surely. Therefore, $N_{\mathrm{sam},1}<\infty$ almost surely, moreover, from Lemma~\ref{lemma:n8}, we have $N_{\mathrm{sam},1}+\bar\tau(N_{\mathrm{sam},1})<\infty$ almost surely, and recurrence of $\mathcal{R}_{\ell}^{\pi}$ again
implies $N_{\mathrm{sam},2}<\infty$ almost surely.

By the definition of $N_{\mathrm{sam},1}$, $\chi_d^\pi(N_{\mathrm{sam},1})\in\mathcal R_1^{\mathrm{sam}}$ almost surely. Now, applying strong Markov property of $\{\chi_{d}^{\pi}(t)\}_{t\geq 1}$, at $N_{\mathrm{sam},1}$ and the definition of $\tilde{P}_{\pi}(\cdot,\cdot)$ in (\ref{eq:667}) yields
\begin{align}
    \mathbb P\left(\chi_d^\pi(N_{\mathrm{sam},1}+\bar\tau(N_{\mathrm{sam},1}))\in\mathcal R_1^{\mathrm{sam}}\,\middle|\,\mathcal G_{N_{\mathrm{sam},1}}^\pi
    \right)&=\sum_{z\in\mathcal R_1^{\mathrm{sam}}}\bm1_{\{\chi_d^\pi(N_{\mathrm{sam},1})=z\}} \sum_{y\in\mathcal R_1^{\mathrm{sam}}} \tilde P_\pi(z,y)\\
    &=\sum_{z\in\mathcal R_1^{\mathrm{sam}}}\bm1_{\{\chi_d^\pi(N_{\mathrm{sam},1})=z\}}=1, \label{eq:nnn181}
\end{align}
where (\ref{eq:nnn181}) follows from (\ref{eq:nnn177}). Thus, $\chi_d^\pi(N_{\mathrm{sam},1}+\bar\tau(N_{\mathrm{sam},1}))$ belongs to $\mathcal R_1^{\mathrm{sam}}$ almost surely. From the definition of $\bar{\tau}(\cdot)$, we know that there exists an integer $k\geq0$, such that
\begin{align}
    \bar{\tau}_{k} = N_{\mathrm{sam},1} + \bar{\tau}(N_{\mathrm{sam},1}). 
\end{align}
Thus, closedness of $\mathcal R_1^{\mathrm{sam}}$ for the embedded chain implies that, almost surely,
\begin{align}\label{eq:nnn183}
    \chi_d^\pi(\bar\tau_n)\in\mathcal R_1^{\mathrm{sam}} \quad\text{for every }n\geq0\text{ such that } \bar\tau_n\geq\bar{\tau}_{k}.
\end{align}
Now,
\begin{align}
    N_{\mathrm{sam},2} + \bar{\tau}(N_{\mathrm{sam},2}) > N_{\mathrm{sam},2} \geq N_{\mathrm{sam},1} + \bar{\tau}(N_{\mathrm{sam},1}).
\end{align}
Again from the definition of $\bar{\tau}(\cdot)$, there exists $k_{1}> k$, such that
\begin{align}
    \bar{\tau}_{k_1} = N_{\mathrm{sam},2} + \bar{\tau}(N_{\mathrm{sam},2}).
\end{align}
Therefore, (\ref{eq:nnn183}) implies $\chi_d^\pi(N_{\mathrm{sam},2}+\bar\tau(N_{\mathrm{sam},2})) \in\mathcal R_1^{\mathrm{sam}}$ almost surely. On the other hand, the definition of $N_{\mathrm{sam},2}$ gives $\chi_d^\pi(N_{\mathrm{sam},2})\in\mathcal R_2^{\mathrm{sam}}$ almost
surely. Similar to (\ref{eq:nnn181}), we also have
\begin{align}
    \mathbb P\left(\chi_d^\pi(N_{\mathrm{sam},2}+\bar\tau(N_{\mathrm{sam},2}))\in\mathcal R_2^{\mathrm{sam}}\,\middle|\,\mathcal G_{N_{\mathrm{sam},2}}^\pi \right)=1.
\end{align}
Therefore,
\begin{align}
    1=\mathbb P\left(\chi_d^\pi(N_{\mathrm{sam},2}+\bar\tau(N_{\mathrm{sam},2}))\in\mathcal R_1^{\mathrm{sam}}\cap\mathcal R_2^{\mathrm{sam}} \,\middle|\,\chi_d^\pi(1)=x\right)=0,
\end{align}
where the second equality follows from the fact that the sets $\mathcal{R}_{1}^{\mathrm{sam}}$ and $\mathcal{R}_{2}^{\mathrm{sam}}$ are disjoint. This contradiction proves that an original recurrent class contains at most one embedded recurrent class. In other words, each recurrent class of the embedded chain contains exactly one reference state, i.e.,
\begin{align}
    \left|\mathcal R^{\mathrm{sam}}\cap\{r_1,\cdots,r_{M_\pi}\} \right|=1.
\end{align}

Since every embedded recurrent class contains a reference state, making $r_1,\ldots,r_{M_\pi}$ absorbing states gives a finite absorbing Markov chain whose transition matrix among the remaining states is $\bar P_\pi$. By the argument in \cite[proof of Theorem 11.3]{grinstead_snell}, there exist $L_{\mathrm{ref}}\in\mathbb N$ and $\epsilon_{\mathrm{ref}}\in(0,1)$, such that 
\begin{align}\label{eq:nn195}
    \sum_{z'\in\mathcal C_\pi^{nr}} (\bar P_\pi^n)_{z,z'} \leq (1-\epsilon_{\mathrm{ref}})^{ \left\lfloor n/L_{\mathrm{ref}}\right\rfloor},\quad z\in\mathcal C_\pi^{nr},\quad n\geq0.
\end{align}
Moreover, \cite[Theorem 11.4]{grinstead_snell} gives 
\begin{align}\label{eq:nn196}
    (I-\bar P_\pi)^{-1} = \sum_{n=0}^{\infty}\bar P_\pi^n,
\end{align}
and all entries of this matrix are finite. Now, we define 
\begin{align}
    N_{\mathrm{ref}}=\inf\left\{n\geq0:\chi_d^\pi(\bar\tau_n)\in\{r_1,\cdots,r_{M_\pi}\}\right\}. \label{eq:205}
\end{align}
If $\chi_d^\pi(\bar\tau_0)$ is a reference state, then $N_{\mathrm{ref}}=0$. If it belongs to $\mathcal C_\pi^{nr}$, then (\ref{eq:nn195}) gives $N_{\mathrm{ref}}<\infty$ almost surely.

For $z\in\mathcal C_\pi^{nr}$, we initialize the chain at $\chi_d^\pi(1)=z$ and consider the states at slots $1,\bar\tau_0,\bar\tau_1,\ldots$. By the strong Markov property and time homogeneity, the joint law of the next observed state and the duration until that observation depends only on the current observed state. Each duration is at least one slot and has finite first and second moments by Lemma~\ref{lemma:n8}. Combine the reference states into one absorbing state with a one-slot self-transition. The time to reach the absorbing state is $\bar\tau_{N_{\mathrm{ref}}}-1$, which is finite almost surely.
Thus, from \cite[Theorem 4.2]{warr_cordeiro} and the definitions in (\ref{eq:nn159}) and (\ref{eq:nn160}), we obtain
\begin{align}\label{eq:nn241}
    \mathbb E_\pi\left[(\bar\tau_{N_{\mathrm{ref}}}-1)^j\,\middle|\,\chi_d^\pi(1)=z\right]=v_\pi^{(j)}(z)<\infty, \quad z\in\mathcal C_\pi^{nr},\quad j\in\{1,2\}.
\end{align}
Now, 
\begin{align}
    \mathbb E_\pi\left[ \bar\tau(1)(\bar\tau_{N_{\mathrm{ref}}}-\bar\tau_0) \,\middle|\,\chi_d^\pi(1)=r_\ell \right] &= \mathbb E_\pi\Bigg[ \bar\tau(1) \mathbb E_\pi\left[ \bar\tau_{N_{\mathrm{ref}}}-\bar\tau_0 \,\middle|\,\mathcal G_{\bar\tau_0}^\pi \right] \,\Big|\,\chi_d^\pi(1)=r_\ell \Bigg] \\
    &=\sum_{z'\in\mathcal C_\pi^{nr}} \mathbb E_\pi\left[ \bar\tau(1) \bm1_{\{\chi_d^\pi(\bar\tau_0)=z'\}} \,\middle|\,\chi_d^\pi(1)=r_\ell \right]v_\pi^{(1)}(z')\label{eq:nn243}\\
    &= \sum_{z'\in\mathcal C_\pi^{nr}} \tilde P_{\pi,1}(r_\ell,z')v_\pi^{(1)}(z'), \label{eq:nn244}
\end{align}
where (\ref{eq:nn243}) follows from (\ref{eq:nn241}). Note that 
\begin{align}
    \bar\tau_{N_{\mathrm{ref}}}-1 =\bar\tau(1) +(\bar\tau_{N_{\mathrm{ref}}}-\bar\tau_0).\label{eq:nn228}
\end{align}
Applying strong Markov property at $\bar{\tau}_0$, we get
\begin{align}
    \mathbb E_\pi\left[ (\bar\tau_{N_{\mathrm{ref}}}-\bar\tau_0)^j \,\middle|\,\chi_d^\pi(1)=x \right]= \sum_{z'\in\mathcal C_\pi^{nr}} \tilde P_\pi(x,z') \mathbb E_\pi\left[ (\bar\tau_{N_{\mathrm{ref}}}-1)^j \,\middle|\,\chi_d^\pi(1)=z' \right] <\infty. \label{eq:nn224}
\end{align}

From (\ref{eq:155}), (\ref{eq:nn228}), (\ref{eq:nn244}) and (\ref{eq:nn224}), we have
\begin{align}
    \mathbb E_\pi\left[ (\bar\tau_{N_{\mathrm{ref}}}-1)^2 \,\middle|\,\chi_d^\pi(1)=r_\ell\right] &= \sum_{z'\in\mathcal C_{W,R,d}} \tilde P_{\pi,2}(r_\ell,z') + 2\sum_{z'\in\mathcal C_\pi^{nr}} \tilde P_{\pi,1}(r_\ell,z')v_\pi^{(1)}(z') \nonumber\\ &\quad+ \sum_{z'\in\mathcal C_\pi^{nr}} \tilde P_\pi(r_\ell,z')v_\pi^{(2)}(z')= \bar m_{\pi,\ell}^{(2)}, \label{eq:nn245}
\end{align}
where the last equality follows from the definition of $\bar m_{\pi,\ell}^{(2)}$ in (\ref{eq:nn164}). Now, let us assume that $\chi_d^\pi(1)=r_\ell$. From the definition of $\bar T_{1,r_\ell}^{+}$ in (\ref{eq:920}), we have
\begin{align}
    \bar T_{1,r_\ell}^{+} \leq \bar\tau_{N_{\mathrm{ref}}}-1,
\end{align}
almost surely. Thus,
\begin{align}\label{eq:nn247}
    \mathbb E_\pi\left[ (\bar T_{1,r_\ell}^{+})^2\,\middle|\,\chi_d^\pi(1)=r_\ell \right] \leq \bar m_{\pi,\ell}^{(2)}.
\end{align}
If $\mathcal C_{\pi}^{nr}$ is empty then we have $\bar{\tau}_{N_{\mathrm{ref}}} = \bar{\tau}_{0}$. Thus, (\ref{eq:nn247}) still holds. 

We next show that the sequence defined in (\ref{eq:nn161}) and (\ref{eq:nn162}) satisfies
\begin{align}
    \bar\zeta_{\pi,\ell,n}(y) &= \mathbb P\left( \chi_d^\pi(n+1)=y,\ \bar T_{1,r_\ell}^{+}>n \,\middle|\, \chi_d^\pi(1)=r_\ell \right), \quad n\geq0,\quad y\in\mathcal X_d. \label{eq:nn248}
\end{align}
We prove (\ref{eq:nn248}) by induction on $n$. Since $\bar T_{1,r_\ell}^{+}\geq1$, the initialization in (\ref{eq:nn161}) gives
\begin{align}
    \bar\zeta_{\pi,\ell,0}(y)= \mathbb P\left( \chi_d^\pi(1)=y,\ \bar T_{1,r_\ell}^{+}>0 \,\middle|\, \chi_d^\pi(1)=r_\ell \right).
\end{align}
Suppose that (\ref{eq:nn248}) holds for every state at some index $n\geq0$. From the definition of $\bar{T}_{1,r_\ell}^{+}$, we see that $\{\bar T_{1,r_\ell}^{+}>n\}\in\mathcal G_{n+1}^\pi$ and
\begin{align}
    \{\bar T_{1,r_\ell}^{+}>n+1\}= \{\bar T_{1,r_\ell}^{+}>n\} \cap\{\chi_d^\pi(n+2)\ne r_\ell\}.
\end{align}
Now,
\begin{align}
    \bar\zeta_{\pi,\ell,n+1}(y)&= \bm1_{\{y\ne r_\ell\}} \sum_{z\in\mathcal X_d} \bar\zeta_{\pi,\ell,n}(z)P_\pi(z,y),\\
    &= \bm1_{\{y\ne r_\ell\}} \sum_{z\in\mathcal X_d} P_\pi(z,y) \mathbb P\left( \chi_d^\pi(n+1)=z,\ \bar T_{1,r_\ell}^{+}>n \,\middle|\, \chi_d^\pi(1)=r_\ell \right)\\
    &= \bm1_{\{y\ne r_\ell\}} \mathbb E_\pi\left[ \bm1_{\{\bar T_{1,r_\ell}^{+}>n\}}
    \sum_{z\in\mathcal X_d} \bm1_{\{\chi_d^\pi(n+1)=z\}}P_\pi(z,y) \,\middle|\, \chi_d^\pi(1)=r_\ell \right]\label{eq:nn253}\\
    &= \bm1_{\{y\ne r_\ell\}} \mathbb E_\pi\left[ \bm1_{\{\bar T_{1,r_\ell}^{+}>n\}} P_\pi\bigl(\chi_d^\pi(n+1),y\bigr) \,\middle|\, \chi_d^\pi(1)=r_\ell \right]\\
    &= \bm1_{\{y\ne r_\ell\}} \mathbb E_\pi\left[ \bm1_{\{\bar T_{1,r_\ell}^{+}>n\}} \mathbb E_\pi\left[ \bm1_{\{\chi_d^\pi(n+2)=y\}} \,\middle|\, \mathcal G_{n+1}^\pi \right] \,\middle|\, \chi_d^\pi(1)=r_\ell \right]\label{eq:nn255}\\
    &=\bm1_{\{y\ne r_\ell\}} \mathbb E_\pi\left[\bm1_{\{\bar T_{1,r_\ell}^{+}>n\}} \bm1_{\{\chi_d^\pi(n+2)=y\}} \,\middle|\, \chi_d^\pi(1)=r_\ell \right]\label{eq:nn256}\\
    &= \mathbb P\left( \chi_d^\pi(n+2)=y,\ \bar T_{1,r_\ell}^{+}>n+1 \,\middle|\, \chi_d^\pi(1)=r_\ell \right),
\end{align}
where the first equality follows from (\ref{eq:nn162}), and the second uses the $n$th induction step. Equation (\ref{eq:nn253}) follows from Tonelli's theorem. The sixth follows from the fact that $\{\bar T_{1,r_\ell}^{+}>n\}\in\mathcal G_{n+1}^\pi$ and $\sigma(\chi_d^\pi(1))\subseteq\mathcal G_{n+1}^\pi$. This completes the induction. Now, from the definition of $\bar{m}_{\pi,\ell}(\tilde{K})$ and (\ref{eq:nn248}) we have
\begin{align}
    \bar m_{\pi,\ell}(\tilde K) &= \sum_{n=0}^{\tilde K-1} \mathbb P\left(\bar T_{1,r_\ell}^{+}>n \,\middle|\,\chi_d^\pi(1)=r_\ell\right)\\
    &\leq \mathbb E_\pi\left[ \bar T_{1,r_\ell}^{+} \,\middle|\,\chi_d^\pi(1)=r_\ell\right].
\end{align}
Together with (\ref{eq:nn247}) this proves
\begin{align}
    &\frac{ \mathbb E_\pi\left[(\bar T_{1,r_\ell}^{+})^2 \,\middle|\,\chi_d^\pi(1)=r_\ell\right]} {\mathbb E_\pi\left[\bar T_{1,r_\ell}^{+} \,\middle|\,\chi_d^\pi(1)=r_\ell\right]}-1\leq \frac{\bar m_{\pi,\ell}^{(2)}} {\bar m_{\pi,\ell}(\tilde K)}-1. \label{eq:nn261}
\end{align}
We define
\begin{align}
    \tilde H =\inf\left\{\bar\tau_n:n\geq0,\ \chi_d^\pi(\bar\tau_n)\in\{r_1,\ldots,r_{M_\pi}\}\right\}.
\end{align}
Recall the definition of $H$ in (\ref{eq:803}). Thus, we have
\begin{align}
    \mathbb E_\pi[H\mid\chi_d^\pi(1)=x] &\leq\mathbb E_\pi[\tilde H\mid\chi_d^\pi(1)=x]\\
    &=1+ \mathbb{E}_{\pi}\left[\bar{\tau}(  1)\,\middle|\chi_{d}^{\pi}(1)=x\,\right] + \mathbb{E}_{\pi}\left[\bar{\tau}_{N_{\textrm{ref}}} -\bar{\tau}_{0}\,\middle|\chi_{d}^{\pi}(1)=x\,\right]\\
    & =1+ \sum_{z\in\mathcal C_{W,R,d}} \tilde P_{\pi,1}(x,z)+ \sum_{z\in\mathcal C_{W,R,d}}
    \tilde P_\pi(x,z)v_\pi^{(1)}(z),\label{eq:nn264}
\end{align}
where (\ref{eq:nn264}) follows from the definition of $\tilde{P}_{\pi,1}(x,z)$, (\ref{eq:nn224}), and (\ref{eq:nn241}). From the definition of $C_{\pi}(x)$ in (\ref{eq:1025}), from (\ref{eq:nn261}) and (\ref{eq:nn264}), we have (\ref{eq:nn167}). From the definition of $\hat{\psi}_{\ell}^{\pi}(r_{j})$, (\ref{eq:nn168}) follows immediately. Now, let us assume that $z\in\mathcal C_\pi^{nr}$. Since $\mathcal R_\ell^\pi$ is closed, if the chain enters $\mathcal R_\ell^\pi$ before $\bar\tau_0$, it remains in $\mathcal R_\ell^\pi$ at $\bar\tau_0$. Therefore,
\begin{align}
    \left\{ \chi_d^\pi(t)\in\mathcal R_\ell^\pi \text{ for some }t\geq1 \right\}= \left\{\chi_d^\pi(\bar\tau_0+k)\in\mathcal R_\ell^\pi \text{ for some }k\geq0 \right\}
\end{align}
almost surely. Applying the strong Markov property at $\bar\tau_0$ and using time homogeneity gives
\begin{align}
    \mathbb P\left( \chi_d^\pi(t)\in\mathcal R_\ell^\pi \text{ for some }t\geq1 \,\middle|\,\mathcal G_{\bar\tau_0}^\pi \right)&= \mathbb P\left( \chi_d^\pi(\bar\tau_0+k)\in\mathcal R_\ell^\pi \text{ for some }k\geq0 \,\middle|\,\mathcal G_{\bar\tau_0}^\pi \right)\\
    &= \sum_{z'\in\mathcal C_{W,R,d}} \bm1_{\{\chi_d^\pi(\bar\tau_0)=z'\}} \hat\psi_\ell^\pi(z')\\
    &= \hat\psi_\ell^\pi\left( \chi_d^\pi(\bar\tau_0) \right).
\end{align}
Taking expectations conditional on $\chi_d^\pi(1)=z$ and using the tower property, we obtain
\begin{align}
    \hat\psi_\ell^\pi(z) &= \mathbb P\left( \chi_d^\pi(t)\in\mathcal R_\ell^\pi \text{ for some }t\geq1 \,\middle|\, \chi_{d}^{\pi}(1)=z \right)\\
    &=\mathbb{E}_{\pi}\left[\mathbb P\left( \chi_d^\pi(t)\in\mathcal R_\ell^\pi \text{ for some }t\geq1 \,\middle|\,\mathcal G_{\bar\tau_0}^\pi \right)\,\middle|\,\chi_{d}^{\pi}(1)=z\right]\\
    &=  \mathbb E_\pi\left[ \hat\psi_\ell^\pi\left( \chi_d^\pi(\bar\tau_0) \right) \,\middle|\, \chi_d^\pi(1)=z \right]\\
    &= \sum_{z'\in\mathcal C_{W,R,d}} \hat\psi_\ell^\pi(z') \mathbb P\left( \chi_d^\pi(\bar\tau_0)=z' \,\middle|\, \chi_d^\pi(1)=z \right)\\
    &= \sum_{z'\in\mathcal C_{W,R,d}} \tilde P_\pi(z,z')\hat\psi_\ell^\pi(z').\label{eq:nn274}
\end{align}
Note that
\begin{align}
    \hat\psi_\ell^\pi(r_j)=
    \begin{cases}
        1, & j=\ell,\\
        0, & j\ne\ell.
    \end{cases}
\end{align}
Thus, for all $z\in\mathcal C_{\pi}^{nr}$,
\begin{align}
    \hat\psi_\ell^\pi(z)
    &=\sum_{z'\in\mathcal C_\pi^{nr}} \tilde P_\pi(z,z')\hat\psi_\ell^\pi(z') + \sum_{j=1}^{M_\pi} \tilde P_\pi(z,r_j)\hat\psi_\ell^\pi(r_j)\\
    &= \tilde P_\pi(z,r_\ell) + \sum_{z'\in\mathcal C_\pi^{nr}} \tilde P_\pi(z,z')\hat\psi_\ell^\pi(z'),\label{eq:nn277}
\end{align}
where (\ref{eq:nn277}) is equivalent to (\ref{eq:nn169}). For $x\in\mathcal X_d$, following the same steps used to obtain (\ref{eq:nn274}), we get (\ref{eq:nnn171}). From (\ref{eq:600}), (\ref{eq:601}), and (\ref{eq:nn166}), for $T\geq 1$, we have
\begin{align}
    &\sum_{y\in\mathcal X_d} \left| \bar p_T^\pi(y\mid x) - \bar\psi_\pi^x(y) \right| \leq \frac{C_\pi(x;\tilde K)}{T}, \\
    &\left| \Delta_d(\pi;T)-\Delta_d(\pi)\right|\leq 2\sqrt{ \frac{2C_c(x)C_\pi(x;\tilde K)}{T}}.
\end{align}
From the definition of $T_\epsilon$ in (\ref{eq:nn170}), for all $T\geq T_{\epsilon}$, we get (\ref{eq:nn171}) and (\ref{eq:nn172}). This completes the proof.
\end{Proof}

\section{Proof of Theorem~\ref{th:17}}\label{appen:th:17}
\begin{Proof}
Let us assume that $\chi_d^\pi(1)=z$. From the definition of $\bar{\tau}(1)$, for every $n\geq1$, we have
\begin{align}
    \{\bar{\tau}(1)>n\} &= \{\bar{\tau}(1)>n-1\} \cap \{e_{d,0}^\pi(n+1)=e_{d,0}^\pi(n)\}, \label{eq:nn301}\\
    \{\bar{\tau}(1)=n\} &= \{\bar{\tau}(1)>n-1\} \cap \{e_{d,0}^\pi(n+1)>e_{d,0}^\pi(n)\}. \label{eq:nn302}
\end{align}
Since $\bar\tau(1)\geq1$ and $\chi_d^\pi(1)=z$, the initialization in (\ref{eq:nn303}) follows.

Recall the states $x^{(0)}$ and $x^{(1)}$ defined in (\ref{eq:543}) and (\ref{eq:544}), respectively. Also, recall that the quantity $e_{d,0}^\pi(t)$ denotes the sampling time of the freshest packet whose successful delivery is known to the transmitter at the beginning of slot $t$. We refer to this quantity as the confirmed generation time at the beginning of slot $t$. Suppose that $\chi_d^\pi(t)=x=(c,b,u_1,\cdots,u_d)$. If $u_1>0$, the transition to $x^{(0)}$ occurs with probability $1-q$ and gives $e_{d,0}^\pi(t+1)=e_{d,0}^\pi(t)$. The transition to $x^{(1)}$ occurs with probability $q$ and gives $e_{d,0}^\pi(t+1)=e_{d,0}^\pi(t)+u_1$. If $u_1=0$, the two successor states are equal, and $e_{d,0}^\pi(t+1)=e_{d,0}^\pi(t)$ in both cases. Thus, conditional on $\chi_d^\pi(t)=x$, the probabilities that $e_{d,0}^\pi(t+1)=e_{d,0}^\pi(t)$ and $e_{d,0}^\pi(t+1)>e_{d,0}^\pi(t)$ are $1-q\bm1_{\{u_1>0\}}$ and $q\bm1_{\{u_1>0\}}$, respectively. For every $n\geq1$, the event $\{\bar{\tau}(1)>n-1\}$ and the state $\chi_d^\pi(n)$ are $\mathcal{G}_n^\pi$-measurable. Therefore, applying the Markov property at slot $n$ to (\ref{eq:nn301}) gives (\ref{eq:nn304}). Similarly, applying the same properties to (\ref{eq:nn302}) gives (\ref{eq:nn305}). Further, (\ref{eq:nn303}) assigns probability one to the initial state $z$. In (\ref{eq:nn304}), the probability at a state $x$ is assigned only to $x^{(0)}$. Thus, by induction, for every $n$, at most one state has positive probability in this recursion. Also, (\ref{eq:nn305}) assigns probability only to $x^{(1)}$ for the same state $x$. Thus, at each index $n\geq1$, each of the sums in (\ref{eq:nn304}) and (\ref{eq:nn305}) contains at most one nonzero term.

From Lemma~\ref{lemma:n8}, we know that $\bar{\tau}(1)$ is finite almost surely. Thus, for $y\in\mathcal{C}_{W,R,d}$,
\begin{align}
    \tilde{P}_\pi(z,y) &= \sum_{n=1}^{\infty} \mathbb{P}\left(\bar{\tau}(1)=n,\ \chi_d^\pi(n+1)=y\,\middle|\,\chi_d^\pi(1)=z\right),\label{eq:nn306}\\
    \tilde{P}_{\pi,j}(z,y)&=\sum_{n=1}^{\infty}n^j \mathbb{P}\left(\bar{\tau}(1)=n,\ \chi_d^\pi(n+1)=y\,\middle|\,\chi_d^\pi(1)=z\right), \quad j\in\{1,2\}.
    \label{eq:nn307}
\end{align}
The second relation follows from the definition in (\ref{eq:155}) and Tonelli's theorem. If $q=1$, Lemma~\ref{lemma:n8} gives
\begin{align}
    \mathbb{P}\left(\bar{\tau}(1)>B_d\,\middle|\,\chi_d^\pi(1)=z\right)&=0.
\end{align}
Therefore, the sums in (\ref{eq:nn306}) and (\ref{eq:nn307}) contain terms only for $n\leq B_d$, which gives (\ref{eq:nn330}) and (\ref{eq:nn331}). This proves the statement for $q=1$.

Now, let us consider $0<q<1$. Under fixed feedback delay $d$, the outcomes of at most $d$ transmissions are unknown at the beginning of a slot. Thus,
\begin{align}
    \sum_{h=1}^{J_d^\pi(t)}n_{d,h}^\pi(t) &\leq d.
\end{align}
From the definition of $p_{d,h}^\pi(t)$, we get
\begin{align}
    \prod_{h=1}^{J_d^\pi(t)}(1-p_{d,h}^\pi(t))&=(1-q)^{\sum_{h=1}^{J_d^\pi(t)}n_{d,h}^\pi(t)}\geq(1-q)^d.\label{eq:nn308}
\end{align}
From (\ref{eq:110}) and (\ref{eq:111}), we have
\begin{align}
    \prod_{h=1}^{J_d^\pi(t)}(1-p_{d,h}^\pi(t))+\sum_{h=1}^{J_d^\pi(t)} p_{d,h}^\pi(t) \prod_{k=1}^{h-1}(1-p_{d,k}^\pi(t)) &=1.
\end{align}
Using this identity in (\ref{eq:estimated_age}), we have
\begin{align}
    \hat{v}_{1,d}^\pi(t)&=c_d^\pi(t)\prod_{h=1}^{J_d^\pi(t)}(1-p_{d,h}^\pi(t))+\sum_{h=1}^{J_d^\pi(t)} (t-e_{d,h}^\pi(t))p_{d,h}^\pi(t)\prod_{k=1}^{h-1}(1-p_{d,k}^\pi(t))\\
    &\geq (1-q)^d c_d^\pi(t).\label{eq:nn309}
\end{align}
The inequality follows from (\ref{eq:nn308}) and the fact that all packet ages are nonnegative.

On the event $\{\bar{\tau}(1)>t-1\}$, the confirmed generation time has not changed between slots $1$ and $t$. Therefore,
\begin{align}
    c_d^\pi(t) &=c(z)+t-1. \label{eq:nn310}
\end{align}
For every integer $t\geq\bar{n}_\pi(z)-d-1$, the definition of $\bar{n}_\pi(z)$ gives, on the same event,
\begin{align}
    c_d^\pi(t) &\geq c(z)+\max\left\{ 0,\left\lceil\frac{W}{(1-q)^d}\right\rceil-c(z) \right\}\\
    &\geq\left\lceil\frac{W}{(1-q)^d}\right\rceil.
\end{align}
Thus, from (\ref{eq:nn309}) and the definition of the policy $\pi$, we get
\begin{align}
    \{\bar{\tau}(1)>t-1\} &\subseteq\{a_r^\pi(t)=1\}, \quad t\geq\bar{n}_\pi(z)-d-1. \label{eq:nn311}
\end{align}

We next consider the packet stored at the transmitter during these slots. If no packet is stored, or its age is at least $R$, the sampler samples a fresh packet, which is transmitted in that slot. Otherwise, the stored packet is retransmitted and its age belongs to $\{1,\cdots,R-1\}$. An ACK for the currently stored packet would increase $e_{d,0}^\pi(\cdot)$, since the generation time of that packet is strictly greater than the confirmed generation time. An ACK for a packet that has already been replaced does not remove the currently stored packet. Thus, on $\{\bar{\tau}(1)>t\}$, the packet transmitted in slot $t$
remains stored at the beginning of slot $t+1$. Its age is $1$ if a fresh packet was sampled and is at most $R$ otherwise. Consequently, for every $t\geq\bar{n}_\pi(z)-d$, on $\{\bar{\tau}(1)>t-1\}$, we have  $1\leq v_2^\pi(t)\leq R$. For these slots, on $\{\bar{\tau}(1)>t\}$, we have
\begin{align}
    v_2^\pi(t+1) &=
    \begin{cases}
        v_2^\pi(t)+1,&1\leq v_2^\pi(t)<R,\\
        1,&v_2^\pi(t)=R.
    \end{cases} \label{eq:nn313}
\end{align}
The recursion in (\ref{eq:nn313}) returns to its initial value after $R$ transitions. Therefore, for $t\geq\bar{n}_\pi(z)-d$, on $\{\bar{\tau}(1)>t+R-1\}$, we get
\begin{align}
    v_2^\pi(t+R)&=v_2^\pi(t), \label{eq:nn314}\\
    c_d^\pi(t+R)&=c_d^\pi(t)+R, \label{eq:nn315}\\
    b_d^\pi(t+R)&=b_d^\pi(t)+R, \label{eq:nn316}
\end{align}
where (\ref{eq:nn315}) follows from (\ref{eq:nn310}), and (\ref{eq:nn316}) follows from (\ref{eq:nn314}), (\ref{eq:nn315}), and the fact that $b_d^\pi(t)=c_d^\pi(t)-v_2^\pi(t)$ when a packet is stored. Moreover, by (\ref{eq:nn311}), the transmission actions at slots $t$ and $t+R$ satisfy $a_r^\pi(t)=a_r^\pi(t+R)=1$. The sampling actions are equal on both slots, as the ages of the stored packets are equal. Thus, from the definition of $\bar{b}(\cdot)$,
\begin{align}
    \bar{b}\left(\chi_d^\pi(t+R)\right) &= \bar{b}\left(\chi_d^\pi(t)\right)+R. \label{eq:nn317}
\end{align}

For $t\geq\bar{n}_\pi(z)+1$, all slots $t-d,\cdots,t-1$ are strictly greater than $\bar{n}_\pi(z)-d$. On $\{\bar{\tau}(1)>t-1\}$, a transmission occurs in each
of these slots. The generation time of each transmitted packet is strictly greater than the unchanged confirmed generation time. Therefore, from the definition in (\ref{eq:366}) and the successive shifts in (\ref{eq:543}), we have
\begin{align}
    u_{d,k}^\pi(t) &= \bar{b}\left(\chi_d^\pi(t-d+k-1)\right)>0, \quad k=1,\cdots,d. \label{eq:nn318}
\end{align}
On the event $\{\bar{\tau}(1)>t+R-1\}$, applying (\ref{eq:nn317}) at slots $t-d+k-1$ gives 
\begin{align}
    u_{d,k}^\pi(t+R) &=u_{d,k}^\pi(t)+R, \quad k=1,\cdots,d. \label{eq:nn319}
\end{align}

Let us consider $n\geq\bar{n}_\pi(z)+1$. From (\ref{eq:nn304}), conditional on $\{\bar{\tau}(1)>n-1,\chi_d^\pi(1)=z\}$, the state at slot $n$ has only one possible value. We denote that state by $x=(c,b,u_1,\cdots,u_d)$. The conditioning event has positive probability, because the probability of a transition without an increase of the confirmed generation time is at least $1-q$. In particular,
\begin{align}
    \mathbb{P}\left(\bar{\tau}(1)>n-1 \,\middle|\, \chi_d^\pi(1)=z \right) \geq(1-q)^{n-1}>0.
\end{align}
Equations (\ref{eq:nn315}), (\ref{eq:nn316}), and (\ref{eq:nn319}) imply that, on $\{\bar{\tau}(1)>n+R-1\}$,
\begin{align}
    \chi_d^\pi(n+R) &=(c+R,b+R,u_1+R,\cdots,u_d+R). \label{eq:nn320}
\end{align}
Thus, on the event $\{\bar\tau(1)>n+R-1\}$, (\ref{eq:544}) shows that the successor state corresponding to an increase in the confirmed generation time is the same state $x^{(1)}$ at both slots $n$ and $n+R$. From (\ref{eq:nn318}), on the event $\{\bar{\tau}(1)>s-1\}$ for every $s\geq\bar{n}_\pi(z)+1$, we have $u_{d,1}^\pi(s)>0$. Hence, the conditional probability that the confirmed generation time increases in the transition from slot $s$ to slot $s+1$ is $q$. Equivalently,
\begin{align}
    \mathbb{P}\left(\bar{\tau}(1)>s \,\middle|\, \mathcal{G}_s^\pi \right) &= (1-q)\bm{1}_{\{\bar{\tau}(1)>s-1\}}.
\end{align}
Taking expectations and applying this relation through $R$ transitions gives
\begin{align}
    &\mathbb{P}\left(\bar{\tau}(1)>n+R-1\,\middle|\,\chi_d^\pi(1)=z\right)=(1-q)^R\mathbb{P}\left(\bar{\tau}(1)>n-1\,\middle|\, \chi_d^\pi(1)=z \right). \label{eq:nn321}
\end{align}
Using (\ref{eq:nn305}), the equality of the resulting states, and (\ref{eq:nn321}), we obtain
\begin{align}
    \mathbb{P}\left(\bar{\tau}(1)=n+R,\ \chi_d^\pi(n+R+1)=y \,\middle|\, \chi_d^\pi(1)=z \right) &=q\bm{1}_{\{y=x^{(1)}\}} \mathbb{P}\left(
    \bar{\tau}(1)>n+R-1\,\middle|\,\chi_d^\pi(1)=z\right)\\
    &= q\bm{1}_{\{y=x^{(1)}\}} (1-q)^R \mathbb{P}\left(\bar{\tau}(1)>n-1
    \,\middle|\,\chi_d^\pi(1)=z\right)\\
    &=(1-q)^R\mathbb{P}\left(\bar{\tau}(1)=n,\ \chi_d^\pi(n+1)=y\,\middle|\,\chi_d^\pi(1)=z\right).\label{eq:nn322}
\end{align}
Thus, by induction, for every $n\geq\bar{n}_\pi(z)+1$ and every $k\geq0$, we have
\begin{align}
    \mathbb{P}\left(\bar{\tau}(1)=n+kR,\ \chi_d^\pi(n+kR+1)=y \,\middle|\,\chi_d^\pi(1)=z\right)=(1-q)^{kR}\mathbb{P}\left(\bar{\tau}(1)=n,\ \chi_d^\pi(n+1)=y\,\middle|\,\chi_d^\pi(1)=z\right).\label{eq:nn323}
\end{align}
For every $y\in\mathcal C_{W,R,d}$, we have
\begin{align}
    \tilde P_\pi(z,y)&=\sum_{n=1}^{\bar n_\pi(z)} \mathbb P\left(\bar\tau(1)=n,\ \chi_d^\pi(n+1)=y\,\middle|\,\chi_d^\pi(1)=z\right)\nonumber\\&\quad+
    \sum_{n=\bar n_\pi(z)+1}^{\bar n_\pi(z)+R}\sum_{k=0}^{\infty}\mathbb P\left(\bar\tau(1)=n+kR,\ \chi_d^\pi(n+kR+1)=y\,\middle|\,\chi_d^\pi(1)=z
    \right)\label{eq:316}\\
    &=\sum_{n=1}^{\bar n_\pi(z)}\mathbb P\left(\bar\tau(1)=n,\ \chi_d^\pi(n+1)=y\,\middle|\,\chi_d^\pi(1)=z\right)\nonumber\\&\quad+\sum_{n=\bar n_\pi(z)+1}^{\bar n_\pi(z)+R}\left(\sum_{k=0}^{\infty}(1-q)^{kR}\right)\mathbb P\left(\bar\tau(1)=n,\ \chi_d^\pi(n+1)=y\,\middle|\,\chi_d^\pi(1)=z \right).
\label{eq:nn324}
\end{align}
Equation (\ref{eq:316}) follows from (\ref{eq:nn306}), (\ref{eq:nn324}) follows from (\ref{eq:nn323}). Similarly, using (\ref{eq:nn307}) and (\ref{eq:nn323}), we obtain
\begin{align}
    \tilde P_{\pi,1}(z,y) &= \sum_{n=1}^{\bar n_\pi(z)} n\,\mathbb P\left( \bar\tau(1)=n,\ \chi_d^\pi(n+1)=y \,\middle|\, \chi_d^\pi(1)=z \right)+ \sum_{n=\bar n_\pi(z)+1}^{\bar n_\pi(z)+R} \sum_{k=0}^{\infty}(n+kR)\nonumber\\&\quad \mathbb P\left(\bar\tau(1)=n+kR,\ \chi_d^\pi(n+kR+1)=y \,\middle|\,\chi_d^\pi(1)=z\right)\\
    &=\sum_{n=1}^{\bar n_\pi(z)} n\,\mathbb P\left(\bar\tau(1)=n,\ \chi_d^\pi(n+1)=y\,\middle|\,\chi_d^\pi(1)=z\right)+ \sum_{n=\bar n_\pi(z)+1}^{\bar n_\pi(z)+R}\Bigg[n\sum_{k=0}^{\infty}(1-q)^{kR}\nonumber\\&\quad+R\sum_{k=0}^{\infty}k(1-q)^{kR}\Bigg]\mathbb P\left(\bar\tau(1)=n,\ \chi_d^\pi(n+1)=y
    \,\middle|\,\chi_d^\pi(1)=z\right),\label{eq:nn325}\\
    \tilde P_{\pi,2}(z,y)&=\sum_{n=1}^{\bar n_\pi(z)}n^2\,\mathbb P\left(\bar\tau(1)=n,\ \chi_d^\pi(n+1)=y\,\middle|\,\chi_d^\pi(1)=z\right)+\sum_{n=\bar n_\pi(z)+1}^{\bar n_\pi(z)+R}\sum_{k=0}^{\infty}(n+kR)^2 \nonumber\\&\quad\mathbb P\left(\bar\tau(1)=n+kR,\ \chi_d^\pi(n+kR+1)=y \,\middle|\,
    \chi_d^\pi(1)=z\right)\\
    &\!=\!\sum_{n=1}^{\bar n_\pi(z)} n^2\,\mathbb P\left(\bar\tau(1)\!=\!n,\ \chi_d^\pi(n+1)=y \,\middle|\, \chi_d^\pi(1)=z \right) \!+\!\!\!\! \sum_{n=\bar n_\pi(z)+1}^{\bar n_\pi(z)+R} \Bigg[n^2\sum_{k=0}^{\infty}(1-q)^{kR}+2nR \nonumber\\&\ \ \sum_{k=0}^{\infty}k(1\!-\!q)^{kR}\!+\! R^2\sum_{k=0}^{\infty}k^2(1\!-\!q)^{kR}\Bigg]\mathbb P\left(\bar\tau(1)=n,\ \chi_d^\pi(n\!+\!1)\!=\!y\,\middle|\,\chi_d^\pi(1)\!=\!z\right). \label{eq:nn326}
\end{align}
Evaluating the geometric series in (\ref{eq:nn324}), (\ref{eq:nn325}) and (\ref{eq:nn326}) gives (\ref{eq:nn327}), (\ref{eq:nn328}) and (\ref{eq:nn329}), respectively. Only the probabilities with $1\leq n\leq\bar n_\pi(z)+R$ are required in these expressions. They are computed using the initialization in
(\ref{eq:nn303}) and at most $\bar n_\pi(z)+R$ iterations of the recursions in (\ref{eq:nn304}) and (\ref{eq:nn305}). This completes the proof.
\end{Proof}

\section{Proof of Theorem~\ref{th:18}}\label{appen:th:18}
\begin{Proof}
We consider a fixed $\ell\in\{1,\cdots,M_\pi\}$ and initialize the chain at $\chi_d^\pi(1)=r_\ell$. We set $\bar\tau_{-1}=1$. From (\ref{eq:423}) and (\ref{eq:667}), the transition from slot $\bar\tau_{-1}$ to slot $\bar\tau_0$ has kernel $\tilde P_\pi$. Since $\tilde{\mathcal R}_\ell^\pi$ is closed for the embedded chain and contains exactly one reference state, $r_\ell$, from (\ref{eq:205}) we see that $\bar\tau_{N_{\mathrm{ref}}}$ is the slot of the first return to $r_\ell$ at an embedded epoch. Thus, the first return cycle consists of slots $1,\cdots,\bar\tau_{N_{\mathrm{ref}}}-1$. By the strong Markov property at successive embedded returns to $r_\ell$, the return cycles are i.i.d. From (\ref{eq:155}), (\ref{eq:nn228}) and (\ref{eq:nn224}), we have
\begin{align}\label{eq:1138}
    1\leq \mathbb E_\pi\left[ \bar\tau_{N_{\mathrm{ref}}}-1 \,\middle|\,\chi_d^\pi(1)=r_\ell \right]<\infty.
\end{align}
Now, applying the same renewal-reward and Markov chain ergodic theorem argument used to obtain (\ref{eq:613}), for every $y\in\mathcal X_d$, we have
\begin{align}\label{eq:1129}
    \mathbb E_\pi\left[ \sum_{t=1}^{\bar\tau_{N_{\mathrm{ref}}}-1} \bm1_{\{\chi_d^\pi(t)=y\}} \,\middle|\,\chi_d^\pi(1)=r_\ell \right]= \mathbb E_\pi\left[ \bar\tau_{N_{\mathrm{ref}}}-1 \,\middle|\,\chi_d^\pi(1)=r_\ell \right]\psi_\ell^\pi(y).
\end{align}
Multiplying (\ref{eq:1129}) by $g(y)$, summing over $y\in\mathcal X_d$, we obtain
\begin{align}\label{eq:1123}
    \sum_{y\in\mathcal X_d}\psi_\ell^\pi(y)g(y) &= \frac{\mathbb E_\pi\left[ \sum_{t=1}^{\bar\tau_{N_{\mathrm{ref}}}-1} g(\chi_d^\pi(t)) \,\middle|\,\chi_d^\pi(1)=r_\ell \right]} {\mathbb E_\pi\left[ \bar\tau_{N_{\mathrm{ref}}}-1 \,\middle|\,\chi_d^\pi(1)=r_\ell \right]}.
\end{align}

The same cycle has $N_{\mathrm{ref}}+1$ embedded transitions, with starting slots $\bar\tau_{-1},\cdots,\bar\tau_{N_{\mathrm{ref}}-1}$. Since each transition takes at least one slot, $N_{\mathrm{ref}}+1\leq\bar\tau_{N_{\mathrm{ref}}}-1$. Thus, the number of embedded transitions has finite expectation. Applying the same argument used to obtain (\ref{eq:613}) to the embedded chain, for every $z\in\tilde{\mathcal R}_\ell^\pi$, we have
\begin{align}\label{eq:1124}
    \mathbb E_\pi\left[ \sum_{n=-1}^{N_{\mathrm{ref}}-1} \bm1_{\{\chi_d^\pi(\bar\tau_n)=z\}} \,\middle|\,\chi_d^\pi(1)=r_\ell \right]&= \mu_\ell^\pi(z) \mathbb E_\pi\left[ N_{\mathrm{ref}}+1 \,\middle|\,\chi_d^\pi(1)=r_\ell \right].
\end{align}
Now,
\begin{align}\label{eq:1130}
    \sum_{t=1}^{\bar\tau_{N_{\mathrm{ref}}}-1} g(\chi_d^\pi(t)) = \sum_{n=-1}^{\infty} \bm1_{\{N_{\mathrm{ref}}>n\}} \sum_{t=\bar\tau_n}^{\bar\tau_{n+1}-1} g(\chi_d^\pi(t)).
\end{align}
By the strong Markov property at $\bar\tau_n$ and the definition of $F_\pi(z;g)$ in (\ref{eq:1121}), we have
\begin{align}
    \mathbb E_\pi\left[ \bm1_{\{N_{\mathrm{ref}}>n\}} \sum_{t=\bar\tau_n}^{\bar\tau_{n+1}-1} g(\chi_d^\pi(t)) \,\middle|\,\mathcal G_{\bar\tau_n}^\pi \right] &= \bm1_{\{N_{\mathrm{ref}}>n\}} \mathbb E_\pi\left[ \sum_{t=\bar\tau_n}^{\bar\tau_{n+1}-1} g(\chi_d^\pi(t)) \,\middle|\,\mathcal G_{\bar\tau_n}^\pi \right]\\
    &= \bm1_{\{N_{\mathrm{ref}}>n\}} F_\pi(\chi_d^\pi(\bar\tau_n);g).\label{eq:1131}
\end{align}

From (\ref{eq:1130}) and (\ref{eq:1131}), we get
\begin{align}
    \mathbb E_\pi\left[ \sum_{t=1}^{\bar\tau_{N_{\mathrm{ref}}}-1} g(\chi_d^\pi(t)) \,\middle|\,\chi_d^\pi(1)=r_\ell \right] &= \sum_{n=-1}^{\infty} \mathbb E_\pi\left[ \bm1_{\{N_{\mathrm{ref}}>n\}} F_\pi(\chi_d^\pi(\bar\tau_n);g) \,\middle|\,\chi_d^\pi(1)=r_\ell \right]\label{eq:1125}\\ 
    &= \mathbb E_\pi\left[ \sum_{n=-1}^{N_{\mathrm{ref}}-1} F_\pi(\chi_d^\pi(\bar\tau_n);g)\,\middle|\,\chi_d^\pi(1)=r_\ell \right]\\
    &=\sum_{z\in\tilde{\mathcal R}_\ell^\pi}F_\pi(z;g) \mathbb E_\pi\left[ \sum_{n=-1}^{N_{\mathrm{ref}}-1} \bm1_{\{\chi_d^\pi(\bar\tau_n)=z\}} \,\middle|\,\chi_d^\pi(1)=r_\ell \right]\\
    &= \mathbb E_\pi\left[ N_{\mathrm{ref}}+1 \,\middle|\,\chi_d^\pi(1)=r_\ell \right] \sum_{z\in\tilde{\mathcal R}_\ell^\pi} \mu_\ell^\pi(z)F_\pi(z;g).\label{eq:n1125}
\end{align}
Applying the same argument as in (\ref{eq:1125})-(\ref{eq:n1125}), for $g\equiv1$, we obtain
\begin{align}\label{eq:1126}
    \mathbb E_\pi\left[ \bar\tau_{N_{\mathrm{ref}}}-1 \,\middle|\,\chi_d^\pi(1)=r_\ell \right]= \mathbb E_\pi\left[ N_{\mathrm{ref}}+1 \,\middle|\,\chi_d^\pi(1)=r_\ell \right] \sum_{z\in\tilde{\mathcal R}_\ell^\pi} \mu_\ell^\pi(z)L_\pi(z).
\end{align}
Substituting (\ref{eq:n1125}) and (\ref{eq:1126}) into (\ref{eq:1123}), we obtain
\begin{align}\label{eq:1132}
    \sum_{y\in\mathcal X_d}\psi_\ell^\pi(y)g(y) &=\frac{\sum_{z\in\tilde{\mathcal R}_\ell^\pi} \mu_\ell^\pi(z)F_\pi(z;g)} {\sum_{z\in\tilde{\mathcal R}_\ell^\pi} \mu_\ell^\pi(z)L_\pi(z)}.
\end{align}
Multiplying (\ref{eq:1132}) by $\hat\psi_\ell^\pi(x)$ and summing over $\ell=1,\cdots,M_\pi$, gives (\ref{eq:1122}) by (\ref{eq:143}).

By Lemma~\ref{lemma:n9}, the state at slot $1+\bar\tau(1)$ belongs to $\mathcal C_{W,R,d}$. Thus, from (\ref{eq:155}) and (\ref{eq:1120}), we have
\begin{align}\label{eq:1133}
    L_\pi(z)=\sum_{y\in\mathcal C_{W,R,d}}\tilde P_{\pi,1}(z,y).
\end{align}
Theorem~\ref{th:17} provides each entry in this sum in finitely many recursion steps.

For $n\geq1$, let $s_{\pi,n}(z)=\mathbb P(\bar\tau(1)>n-1\mid\chi_d^\pi(1)=z)$. When $s_{\pi,n}(z)>0$, let $z_{\pi,n}(z)$ denote the unique state at slot $n$ conditional on $\{\bar\tau(1)>n-1\}$ and $\chi_d^\pi(1)=z$. The uniqueness follows from (\ref{eq:nn303}) and (\ref{eq:nn304}), which also provide $s_{\pi,n}(z)$ and $z_{\pi,n}(z)$. From (\ref{eq:1121}), we get
\begin{align}
    F_\pi(z;g)  &=\sum_{n=1}^{\infty} \mathbb E_\pi\left[ \bm1_{\{\bar\tau(1)>n-1\}}g(\chi_d^\pi(n)) \,\middle|\,\chi_d^\pi(1)=z \right]\\
    &=\sum_{n:\,s_{\pi,n}(z)>0} s_{\pi,n}(z)g(z_{\pi,n}(z)).\label{eq:1127}
\end{align}
If $q=1$, (\ref{eq:388}) gives $s_{\pi,n}(z)=0$ for $n>B_d$. Thus, using (\ref{eq:nn303}) and (\ref{eq:nn304}) for $1\leq n\leq B_d-1$, we obtain all terms in (\ref{eq:1127}). For $0<q<1$, following the same procedure of using (\ref{eq:nn320}) in Appendix~\ref{appen:th:17}, we get
\begin{align}\label{eq:1128}
    \sum_{n=\bar n_\pi(z)+1}^{\infty} s_{\pi,n}(z)g(z_{\pi,n}(z))= \sum_{n=\bar n_\pi(z)+1}^{\bar n_\pi(z)+R} s_{\pi,n}(z) \left[ \frac{g(z_{\pi,n}(z))}{1-(1-q)^R} +\frac{R(1-q)^{d+R}\bm1_{\{g=\hat v_{1,d}\}}} {(1-(1-q)^R)^2} \right].
\end{align}
Thus, to evaluate $F_\pi(z;g)$, we use (\ref{eq:nn303}) for initialization and apply (\ref{eq:nn304}) for $1\leq n\leq\bar n_\pi(z)+R-1$. We evaluate the terms with $1\leq n\leq\bar n_\pi(z)$ in (\ref{eq:1127}) directly and use (\ref{eq:1128}) for the sum of the remaining terms. Starting with the row $\tilde P_\pi(x,\cdot)$, we use Theorem~\ref{th:17} to compute the rows for all states reached with positive probability from the rows already computed. By Lemma~\ref{lemma:n9}, every successor belongs to the finite set $\mathcal C_{W,R,d}$, thus this procedure terminates after at most $1+|\mathcal C_{W,R,d}|$ rows. From the positive entries of these rows, we identify the recurrent classes reachable from $x$. For each such class $\tilde{\mathcal R}_\ell^\pi$, we obtain
$\mu_\ell^\pi$, by solving
\begin{align}\label{eq:1137}
    \mu_\ell^\pi(y) &=\sum_{z\in\tilde{\mathcal R}_\ell^\pi} \mu_\ell^\pi(z)\tilde P_\pi(z,y), \quad y\in\tilde{\mathcal R}_\ell^\pi,\nonumber\\
    \sum_{y\in\tilde{\mathcal R}_\ell^\pi} \mu_\ell^\pi(y)&=1.
\end{align}
For the initial state $x=(1,0,\cdots,0)$, we compute $\hat\psi_\ell^\pi(x)$ from (\ref{eq:nnn171}), restricted to the reachable embedded states. Thus, every term required in (\ref{eq:1122}) is obtained using finitely many recursion steps and finite linear systems.
\end{Proof}
\section{Proofs of Lemma~\ref{lemma:16}, Lemma~\ref{lemma:17}, and Theorem~\ref{th:8}}\label{appen:th:8}
\subsection{Proof of Lemma~\ref{lemma:16}}
\begin{Proof}
From (\ref{eq:df_exact_age_evolution}), the age of the monitor at slot $t+1$ is $\tilde v_2^\pi(t)+1$ if $a_r^\pi(t)Z(t)=1$ and $v_1^\pi(t)+1$ if $a_r^\pi(t)Z(t)=0$. Thus,
\begin{align}
    &(v_1^\pi(t)+1)^2-(v_1^\pi(t+1))^2= a_r^\pi(t)Z(t) \left((v_1^\pi(t)+1)^2-(\tilde v_2^\pi(t)+1)^2\right).
\end{align}
Since $a_r^\pi(t)$ and $\tilde v_2^\pi(t)$ are $\mathcal I_d^\pi(t)$-measurable and $Z(t)$ is independent of $v_1^\pi(t)$ and $\mathcal{I}_d^\pi(t)$, we have
\begin{align}
    \mathbb E_\pi\left[(v_1^\pi(t)+1)^2-(v_1^\pi(t+1))^2 \,\middle|\,\mathcal I_d^\pi(t)\right]&= q a_r^\pi(t)\left(\mathbb E_\pi\left[(v_1^\pi(t)+1)^2\,\middle|\,\mathcal I_d^\pi(t)\right]-(\tilde v_2^\pi(t)+1)^2\right)\label{eq:n116}\\
    &\geq q a_r^\pi(t)\left((\hat v_{1,d}^\pi(t)+1)^2 -(\tilde v_2^\pi(t)+1)^2\right)\label{eq:n117}\\
    &\geq (W+2)q a_r^\pi(t) \bigl(\hat v_{1,d}^\pi(t)-\tilde v_2^\pi(t)\bigr)\label{eq:n118}\\
    &\geq (W+2)\mathbb E_\pi\left[v_1^\pi(t)+1-v_1^\pi(t+1)\,\middle|\, \mathcal{I}_d^\pi(t)\right],\label{eq:n119}
\end{align}
where (\ref{eq:n117}) follows from Jensen's inequality and the definition of $\hat v_{1,d}^\pi(t)$ and (\ref{eq:n118}) follows from the fact that when $a_r^\pi(t)=1$, we have $\hat v_{1,d}^\pi(t)\geq W$ and $0\leq\tilde v_2^\pi(t)\leq\hat v_{1,d}^\pi(t)$. From (\ref{eq:df_exact_age_evolution}) and the definition of $\hat v_{1,d}^\pi(t)$, we have (\ref{eq:n119}). Taking expectations, summing over $t=1,\ldots,T$ and dividing by $T$, we obtain
\begin{align}\label{eq:n120}
    &2\Delta_d(\pi;T)+1+ \frac{\mathbb E_\pi[(v_1^\pi(1))^2] -\mathbb E_\pi[(v_1^\pi(T+1))^2]}{T}\geq (W+2)\left(1+ \frac{\mathbb E_\pi[v_1^\pi(1)] -\mathbb E_\pi[v_1^\pi(T+1)]}{T}\right).
\end{align}
Now, from the definition of $c_{d}^\pi(t)$ and from Lemma~\ref{lemma:n13}, we have 
\begin{align}
     \mathbb E_\pi[(v_1^\pi(t))^2] \leq \mathbb E_\pi[(c_d^\pi(t))^2]\leq C_c(x).
\end{align}
Thus, taking $T\rightarrow\infty$ in (\ref{eq:n120}), we get (\ref{eq:n114}).
\end{Proof}

\subsection{Proof of Lemma~\ref{lemma:17}}
\begin{Proof}
First, we assume that $q=1$. Every fresh packet is successfully delivered in its sampling slot. The corresponding ACK is available at the beginning of the slot in which its age reaches $d+1$. Thus, for a given $W$, all packet replacement thresholds $R\geq d+1$ give the same actions from the initial state. We choose $R_*=\max\{R_0,d+1\}$, which gives (\ref{eq:1141}).

Now, we assume that $0<q<1$. Let $z_*=(d+1,0,\cdots,0)$ and $B=2K+4d$. We first consider a policy $\pi=(W,R)_d$ with $W<K$ and $R\geq d+1$. Suppose that
\begin{align}\label{eq:130}
    Z(t)=\cdots=Z(t+B-1)=1.
\end{align}
By Lemma~\ref{lemma:n7}, a transmission occurs by slot $t+W+d-1$. If a fresh packet has been sampled by the beginning of the slot when the ACK for this transmission becomes available, we consider the first such sample. Otherwise, the transmitter is empty at that slot, and the next transmission samples a fresh packet. Applying Lemma~\ref{lemma:n7} again, we see that
a fresh packet is sampled by slot $t+2W+3d-1$.

Due to (\ref{eq:130}), this fresh packet succeeds in its sampling slot. Since $R\geq d+1$, the packet remains stored until its ACK arrives.
At the beginning of the slot when this ACK becomes
available, the confirmed age of the monitor at the sampler-transmitter pair is $d+1$ and the transmitter is empty. Any unresolved transmission at that slot is of the same packet, so the remaining coordinates of $\chi_d^\pi$ are zero. Thus, the chain visits $z_*$ by the beginning of slot $t+2W+4d$, which is no later than slot $t+B$.

Let $\tau_*^\pi(t)$ denote the number of slots from the beginning of slot $t$ to the next visit to $z_*$. Applying a similar argument as in the proof of Lemma~\ref{lemma:n8}, we get
\begin{align}
    \mathbb E_\pi\left[ \tau_*^\pi(t)\,\middle|\,\mathcal G_t^\pi \right] &\leq \frac{B}{q^B} = M_1,\label{eq:n1144}\\
    \mathbb E_\pi\left[ (\tau_*^\pi(t))^2\,\middle|\,\mathcal G_t^\pi
    \right] &\leq \frac{B^2(2-q^B)}{q^{2B}} = M_2.\label{eq:1144}
\end{align}
These bounds hold for all $W<K$ and $R\geq d+1$. We now initialize the chain at $\chi_d^\pi(1)=z_*$, and define 
\begin{align}\label{eq:1145}
    U^\pi &=\sum_{t=1}^{\tau_*^\pi(1)} \left(v_1^\pi(t)+\lambda_s a_s^\pi(t) +\lambda_r a_r^\pi(t)\right).
\end{align}
The chain reaches $z_*$ from every initial state and from (\ref{eq:1144}), the finite mean return time to $z_*$ is finite. Thus, the chain has a unique positive recurrent class, which contains $z_*$. Applying the same return-cycle argument as in (\ref{eq:1123}) gives
\begin{align}\label{eq:1146}
    G_\lambda(\pi) &=\frac{\mathbb E_\pi\left[ U^\pi\,\middle|\,\chi_d^\pi(1)=z_*\right]} {\mathbb E_\pi\left[ \tau_*^\pi(1)\,\middle|\,\chi_d^\pi(1)=z_*\right]}.
\end{align}
At $z_*$, the age of the monitor is $d+1$, and it increases by at most one per slot. Thus, using $\lambda_s+\lambda_r\leq\frac{B_0}{\gamma}$,
\begin{align}\label{eq:1147}
    U^\pi &\leq \left(d+1+\frac{B_0}{\gamma}\right)\tau_*^\pi(1) +\frac{\tau_*^\pi(1)(\tau_*^\pi(1)-1)}{2}.
\end{align}
Using (\ref{eq:1144}), (\ref{eq:1146}) and $\tau_*^\pi(1)\geq1$, we obtain
\begin{align}\label{eq:1148}
    0\leq G_\lambda(\pi) &\leq d+1+\frac{B_0}{\gamma}+\frac{M_2}{2} \triangleq J_{\max}.
\end{align}
We choose
\begin{align}\label{eq:1149}
    R_*=\max\Bigg\{ R_0,\ d+2,\ \left\lceil\frac{K}{(1-q)^d}\right\rceil,\ \left\lceil \frac{M_2+\left(\frac{B_0}{\gamma}+J_{\max}\right)M_1}{q} \right\rceil \Bigg\}.
\end{align}

Next we fix $W<K$ and $R>R_*$ and compare $\pi=(W,R)_d$ and $\pi_*=(W,R_*)_d$ using the same channel realizations, with both the corresponding chains initialized at $z_*$. Let $\tau$ be the first slot at which either the two policies select different actions or both the corresponding chains return to $z_*$. The two policies have the same information at the beginning of slot $\tau$, and the event that their actions differ at $\tau$ is determined by this information.

Suppose that their actions differ at $\tau$. Then, $\pi_*$ samples and transmits a fresh packet, whereas $\pi$ retransmits a packet of age at least $R_*$. Let $S=\inf\{t\geq\tau:Z(t)=1\}$. All packets sampled before slot $\tau$ were generated no later than slot $\tau-R_*$. Since $Z(t)=0$ for $\tau\leq t<S$, no packet generated at or after $\tau$ has been delivered by the beginning of any slot $t\leq S$. Thus, for $\pi'\in\{\pi,\pi_*\}$ and $\tau\leq t\leq S$, we have $c_d^{\pi'}(t)\geq R_*$. From (\ref{eq:nn309}) and (\ref{eq:1149}), we obtain
\begin{align}\label{eq:1150}
    \hat v_{1,d}^{\pi'}(t) &\geq (1-q)^d c_d^{\pi'}(t) \geq (1-q)^dR_*\geq K>W.
\end{align}
Hence, both policies transmit in every slot from $\tau$ through $S$. Since $c_d^\pi(t)>d+1$ and $c_d^{\pi_*}(t)>d+1$ for $t=\{\tau,\cdots,S\}$, both return cycles contain these slots. The age of the monitor under two policies are equal through slot $S$, since they are the same at $\tau$ and $Z(t)=0$ for $t\in\{\tau,\cdots,S-1\}$. Under $\pi_*$, the packet delivered in slot $S$ was generated at or after $\tau$. Thus, the age of the monitor under $\pi_*$, for $t\geq S+1$ is therefore at most $t-\tau$. Under $\pi$, when $S=\tau$, the monitor age in slot $\tau+1$ is at least $R_*+1$. Thus, $c_d^\pi(\tau+1)\geq v_1^\pi(\tau+1) \geq R_*+1>d+1$. Thus, the return cycle under $\pi$ also contains slot $\tau+1$. From (\ref{eq:1145}) and the above analysis, we have
\begin{align}
    \left(U^{\pi_*}-G_\lambda(\pi)\tau_*^{\pi_*}(1)\right) -\left(U^\pi-G_\lambda(\pi)\tau_*^\pi(1)\right) \leq & \sum_{t=S+1}^{\tau_*^{\pi_*}(1)}(t-\tau)
    +(\lambda_s+\lambda_r)\tau_*^{\pi_*}(\tau)\nonumber\\
    &  +G_\lambda(\pi)\left(\tau_*^\pi(\tau)-\tau_*^{\pi_*}(\tau)\right) -(R_*+1)\bm1_{\{S=\tau\}}\\
    \leq &\left(\tau_*^{\pi_*}(\tau)\right)^2 \!+\!\frac{B_0}{\gamma}\tau_*^{\pi_*}(\tau)
    \!+\!G_\lambda(\pi)\tau_*^\pi(\tau)
    \!-\!(R_*\!+\!1)\bm1_{\{S=\tau\}}.\label{eq:1151}
\end{align}
On the event that the two policies select different actions at $\tau$, we have $\mathbb P(S=\tau\mid\mathcal G_\tau^\pi)=q$. Taking the conditional expectation of (\ref{eq:1151}) and using (\ref{eq:n1144}), (\ref{eq:1144}) and (\ref{eq:1148}), gives
\begin{align}\label{eq:1152}
    \mathbb E\left[ (U^{\pi_*}-G_\lambda(\pi) \tau_*^{\pi_*}(1))-(U^\pi-G_\lambda(\pi)\tau_*^\pi(1)) \,\middle|\,\mathcal G_\tau^\pi \right] \leq M_2+\left(\frac{B_0}{\gamma}+J_{\max}\right)M_1-q(R_*+1)<0.
\end{align}
If the two corresponding chains return to $z_*$ before their actions differ, the two return cycles coincide. Thus, the difference inside the conditional expectation in (\ref{eq:1152}) is zero on
this event. 

From (\ref{eq:1146}), we have $\mathbb{E}[U^{\pi} - G_{\lambda}(\pi) \tau_*^{\pi}(1)] = 0$, thus from (\ref{eq:1152})
\begin{align}
    \mathbb E\left[U^{\pi_*}-G_\lambda(\pi) \tau_*^{\pi_*}(1)\right] \leq 0, 
\end{align}
which proves (\ref{eq:1141}) from (\ref{eq:1146}).
\end{Proof}
\subsection{Proof of Theorem~\ref{th:8}}
\begin{Proof}
From (\ref{eq:1113}), we have $G_{\lambda^K}\geq\zeta_K$ for $\pi\in\bar\Pi_{d,K}$. For $W<K$ and $R>R_*$, from Lemma~\ref{lemma:17} and (\ref{eq:n129}), we have $G_{\lambda^K}((W,R)_d)\geq G_{\lambda^K}((W,R_*)_d)$. Since $(W,R_*)_d\in\bar\Pi_{d,K}$, from (\ref{eq:1113}), we have $G_{\lambda^K}((W,R_*)_d)\geq\zeta_K$. For $W>\lfloor 2\zeta_K-1\rfloor$, from (\ref{eq:n114}), we have $G_{\lambda^K}((W,R)_d)>\zeta_K$. Combining these inequalities with (\ref{eq:1117}), we obtain $G_{\lambda^K}(\pi)\geq\zeta_K$ for every $\pi\in\bar\Pi_d$.

As in the definition of $\pi_w$, let $\bar{\pi}_w$ denote the policy that selects $\pi\in\bar\Pi_d$ with probability $w_\pi$ at $t=1$ and follows the selected policy thereafter. Assume that $\bar{\pi}_w$ is feasible, i.e., $\nu_{k,d}(\bar{\pi}_w)\leq f_k$ for $k\in\{s,r\}$. For each $\pi\in\bar\Pi_d$, Theorem~\ref{th:13} gives $\Delta_d(\pi;T)\rightarrow\Delta_d(\pi)$ and $\nu_{k,d}(\pi;T)\rightarrow\nu_{k,d}(\pi)$ for $k\in\{s,r\}$. Now, conditioning on the selected policy, we obtain the following by Fatou's lemma and the bounded convergence theorem,
\begin{align}
    \Delta_d(\bar\pi_w) &\geq \sum_{\pi\in\bar\Pi_d}w_\pi G_{\lambda^K}(\pi)-\lambda_s^K f_s-\lambda_r^K f_r.
\end{align}
Thus, from (\ref{eq:n115}), we obtain
\begin{align}\label{eq:n127}
    \Delta_d(\bar\pi_w)\geq V_K+\sum_{\pi\in\bar\Pi_d}w_\pi\bigl(G_{\lambda^K}(\pi)-\zeta_K\bigr) \geq V_K.
\end{align}
An optimal solution of (\ref{eq:1108}) attains $V_K$. If $G_{\lambda^K}(\pi)>\zeta_K$ is strict for a policy $\pi\in\bar{\Pi}_d$, then assigning positive probability to $\pi$ makes the last inequality in (\ref{eq:n127}) strict.
\end{Proof}

\end{appendices}

\bibliographystyle{unsrt}
\bibliography{references}

\end{document}